\documentclass[usenatbib]{mn2e}
\usepackage{epsf}

\def\apj{ApJ}

\def\apjl{ApJL}
\def\apjs{ApJS}
\def\mnras{MNRAS}
\def\aj{AJ}

\def\nat{Nature}

\newcommand{\tvir}{T_{\rm vir}} 
\newcommand{\tvirmin}{T_{\rm vir,min}} 

\newcommand{\vcirc}{v_{\rm circ}}
\newcommand{\Tvir}{T_{\rm vir}} 
\newcommand{\tmax}{T_{\rm max}} 
\newcommand{\Tmax}{T_{\rm max}} 
\newcommand{\hmpc}{h^{-1}\,{\rm Mpc}}
\newcommand{\K}{\,{\rm K}} 
\newcommand{\kms}{\,{\rm km}\,{\rm s}^{-1}}
\newcommand{\msun}{M_{\odot}} 

\newcommand{\lya}{Ly$\alpha$}
\newcommand{\lcdm}{$\Lambda$CDM}
\newcommand{\hubunits}{{\rm km}\;{\rm s}^{-1}\;{\rm Mpc}^{-1}}
\newcommand{\hkpc}{h^{-1}\;{\rm kpc}}
\newcommand{\msph}{m_{\rm SPH}}
\newcommand{\mgal}{M_{\rm gal}}
\newcommand{\mgalmin}{M_{\rm gal,min}}
\newcommand{\mhalomin}{M_{\rm halo,min}}
\newcommand{\mhalo}{M_{\rm halo}}
\newcommand{\rhovir}{\rho_{\rm vir}}

\newcommand{\rvir}{R_{\rm vir}}
\newcommand{\Rvir}{R_{\rm vir}}
\newcommand{\mvir}{M_{\rm vir}}
\newcommand{\surfdunits}{\;h_{75}^2\;{\rm Mpc}^{-2}}
\newcommand{\dunits}{\;h^3\;{\rm Mpc}^{-3}}
\newcommand{\accgalunits}{\;M_\odot\;{\rm yr}^{-1}\;{\rm gal}^{-1}}
\newcommand{\cosij}{{\rm cos}[{\bf r}^g_i,{\bf r}^g_j]}
\newcommand{\tcool}{t_{\rm cool}}
\newcommand{\tdyn}{t_{\rm dyn}}
\newcommand{\tff}{t_{\rm ff}}
\newcommand{\rcool}{r_{\rm cool}}

\title{How Do Galaxies Get Their Gas?}

\begin{document}
\author[D. Kere\v{s} et al.]{ Du\v{s}an Kere\v{s}$^1$, Neal Katz$^1$
David H. Weinberg$^2$, Romeel Dav\'e$^3$ \\
\\
$^1$Astronomy Department, University of Massachusetts at Amherst, MA 01003; 
keres@nova.astro.umass.edu, nsk@kaka.astro.umass.edu \\
$^2$Ohio State University, Department of Astronomy, Columbus, OH 43210;
dhw@astronomy.ohio-state.edu \\
$^3$University of Arizona, Steward Observatory, Tucson, AZ 85721;
rad@astro.as.arizona.edu
}

\maketitle

\begin{abstract}
We examine the temperature history of gas accreted by forming galaxies in
smoothed particle hydrodynamics (SPH) simulations.  About half of the gas 
follows the track expected in the conventional picture of galaxy formation,
shock heating to roughly the virial temperature of the galaxy potential
well ($T \sim 10^6\K$ for a Milky Way type galaxy) before cooling, condensing,
and forming stars.  However, the other half radiates its acquired gravitational
energy at much lower temperatures, typically $T < 10^5\K$, and the
histogram of maximum gas temperatures is clearly bimodal.  The ``cold mode''
of gas accretion dominates for low mass galaxies (baryonic mass
$\mgal \la 10^{10.3}\msun$ or halo mass $\mhalo \la 10^{11.4} \msun$), 
while the conventional
``hot mode'' dominates the growth of high mass systems.  Cold accretion is
often directed along filaments, allowing galaxies to efficiently draw gas
from large distances, while hot accretion is quasi-spherical.  The galaxy 
and halo mass dependence leads to redshift and environment dependence of 
cold and hot accretion rates, with cold mode dominating at high redshift and
in low density regions today, and hot mode dominating in group and cluster
environments at low redshift.  The simulations reproduce an important
feature of the observed relation between galaxy star formation rate  (SFR) and
environment, namely a break in star formation rates at surface densities
$\Sigma \sim 1 h_{75}^2{\rm Mpc}^{-2}$, outside the virial radii of large
groups and clusters.  
The cosmic SFR tracks the overall history of gas accretion, and its
decline at low redshift follows the combined decline of cold and hot
accretion rates.  The drop in cold accretion is driven by the decreasing infall
rate onto halos, while for hot accretion this slower mass growth is
further modified by the longer cooling times within halos.
If we allowed hot accretion 
to be suppressed by conduction or AGN feedback, then the simulation
predictions would change in interesting ways, perhaps resolving conflicts with
the colours of ellipticals and the cutoff of the galaxy luminosity function.
The transition at $\mhalo \sim 10^{11.4}\msun$ between cold mode domination and
hot mode domination is similar to that found by \cite{birnboim03}
using 1-d simulations and analytic arguments.  The corresponding baryonic
mass is tantalisingly close to the scale at which \cite{kauffmann03} find a
marked shift in galaxy properties, and we speculate on possible connections
between these theoretical and observational transitions.
\end{abstract}


\section{Introduction \label{sec:intro}}

The conventional sketch of galaxy formation has its roots in classic
papers of the late '70s and early '80s, with initial discussions
of collapse and cooling criteria by \cite{rees77} and \cite{silk77},
the addition of dark matter halos by \cite{white78},
and the disk formation model of \cite{fall80}.
According to this sketch, gas falling into a dark matter
potential well is shock heated to approximately the halo virial
temperature, $\tvir = 10^6 (v_{\rm circ}/167\kms)^2\, \K$,
putting it in quasi-hydrostatic equilibrium with the dark matter.
Gas in the dense, inner regions of
this shock heated halo radiates its thermal energy, 
loses its pressure support, settles into a
centrifugally supported disk, and forms stars.  Mergers of disks can
scatter stars onto disordered orbits, producing spheroidal systems,
which may regrow disks if they experience subsequent gas accretion.
Over the last decade, the ideas of these seminal papers have been
updated and extended into a powerful ``semi-analytic'' framework for
galaxy formation calculations 
(e.g., \citealt{white91,kauffmann93,cole94,avila98,mo98,somerville99}).

The geometry seen in N-body and hydrodynamic
cosmological simulations, where the densest structures often form at
the nodes of a filamentary network, is clearly more complicated than
the spherical geometry underlying semi-analytic calculations.
Nonetheless, a substantial
fraction of the gas in these simulations does shock heat to $T\sim
\tvir$, and some of this gas does cool and settle into galaxies.  The
approximate agreement between semi-analytic models
and smoothed particle hydrodynamics
(SPH) calculations of galaxy masses 
(e.g., \citealt{benson01,yoshida02,helly03})
has therefore been taken as evidence that the
conventional sketch, while idealised, captures most of the essential physics.
In this paper, we use SPH simulations of cosmological volumes to 
argue that this sketch requires an important revision:
roughly half of the gas accreted by the simulated galaxies is
never shock heated close to the halo virial temperature 
($T \sim 10^6\K$ for a Milky Way type galaxy) but instead
radiates its acquired gravitational energy from 
$T \la 2.5 \times 10^5\K$ (often $T \la 5\times 10^4\K$).
This ``cold mode'' of gas accretion dominates for lower mass
galaxies (baryon mass $\mgal \la 2\times 10^{10}M_\odot$), while the
conventional, ``hot mode'' of gas accretion dominates the growth
of high mass systems.  As a result, ``cold mode'' accretion 
dominates at high redshift ($z \ga 3$) and in low density
environments today, while ``hot mode'' accretion dominates in group
and cluster environments at low redshift.

There is, in fact, a long history of results suggesting that 
cold accretion could be an important element of galaxy formation.
Binney (1977), using analytic models of proto-galaxy collapse, argued that
the amount of shock heating could be small for plausible physical
conditions, with only a fraction of the gas reaching temperatures
$T \sim \tvir$.  In the first SPH simulations of forming galaxies (Katz
\& Gunn 1991), which had idealised initial conditions but included
small scale power leading to hierarchical formation, most of the gas
never heated above $T\sim 3\times 10^4\K$, with much of the cooling
radiation therefore emerging in the Ly$\alpha$ line.  \citet{katz93}
and \citet{katz94}
showed the importance of filamentary structures as channels for
gas accretion in simulations with cold dark matter (CDM) initial
conditions.  Recent studies based on SPH simulations of cosmological volumes
reveal the situation even more starkly.  Fardal et al.\ (2001)
showed that most of the cooling radiation in their simulations comes
from gas with $T < 2\times 10^4\K$, again implying that a significant
fraction emerges in the Ly$\alpha$ line.  Since gas starting at 
$T\sim 10^6\K$ {\it must} radiate 90\% of its thermal energy by the time it
cools to $T\sim 10^5\K$, \cite{fardal01} concluded that the majority
of the gas entering galaxies (indeed, the majority of the gas
experiencing any significant cooling) must not be heated to the virial
temperature of any dark matter halo resolved by the simulation.
Kay et al.\ (2000) directly tracked the temperature histories of
particles that ended up in their simulated galaxies and found
that only 11\% of these particles were ever heated to a temperature
above $10^5\K$.

Motivated by these results, we here investigate the temperature histories
of accreted gas particles as a function of galaxy mass, redshift,
and environment, thoroughly quantifying the relative importance of
the cold and hot modes of gas accretion in our SPH simulations.
We use several simulations to demonstrate the insensitivity of our
primary conclusions to numerical resolution over a wide dynamic
range.  We reported initial results from our study in \cite{katz03}.
\cite{birnboim03} have investigated similar issues with a complementary
approach based on high resolution, spherically symmetric collapse
calculations.  They find that a virial shock fails to develop if the
gas cooling time is shorter than the local dynamical time, so in these
cases gas shells fall far inside the halo virial radius without ever
being heated to high temperature.  
\cite{birnboim03} show that the cooling time condition corresponds
approximately to a threshold in the galaxy's halo mass, with little
dependence on redshift.  As we will show in \S\ref{sec:results} below,
our results are in quite good agreement with Birnboim \& Dekel's,
despite the radically different approaches. 
Traditional semi-analytic models also distinguish between halos with rapid
post-shock cooling and halos with slow post-shock cooling 
(e.g., \citealt{white91}), though the distinction has received relatively
little attention in discussions of these models, and heating to the
virial temperature is assumed in either case. 
In quantitative terms, we find that cold accretion plays a much larger
role in our simulations than rapid-cooling accretion plays in 
standard semi-analytic calculations (see \S\ref{sec:disc_sf} and
the Appendix).

In addition to being an important aspect of the physics of galaxy formation,
the existence of distinct cold and hot modes of gas accretion could have
interesting observational implications.  \cite{fardal01} emphasised one
of these implications: cold accretion allows much of the cooling radiation
associated with galaxy formation to emerge in the \lya\ line instead of
the X-ray continuum (see also \citealt{haiman00}).  A reduced role for
hot accretion might help explain why diffuse X-ray emission from late-type
galaxy halos is well below the predictions of standard semi-analytic
calculations \citep{benson00}.  A second class of implications
relates to the cosmic star formation history.
Murali et al. (\citeyear{murali02}; hereafter MKHWD) show that
galaxies in SPH simulations, like the ones analysed here, gain most of their
mass through smooth accretion of gas, not through mergers with pre-existing
galaxies (at least not galaxies above the simulation's resolution limit).
They further find that the global history of star formation tracks
the global history of gas accretion rather than the merger history.
In subsequent analysis, we have found that these generalisations
hold fairly well on a galaxy-by-galaxy basis (Maller et al., in preparation).
Thus, understanding gas accretion is nearly tantamount to understanding 
the history of star formation, at least in the simulations.

The cosmic star formation history inferred from near-UV luminosity
functions declines sharply between $z \sim 1$ and $z=0$ 
(e.g., \citealt{madau96}).  While semi-analytic
models and hydrodynamic simulations both predict a drop in the
global star formation rate (SFR) over this redshift interval,
it is difficult to explain the full order-of-magnitude
reduction implied by the data \citep{baugh.etal04}.  
Without an {\it ad hoc} fix, semi-analytic and numerical calculations
also predict continuing gas accretion and star formation in old,
massive galaxies, so they have difficulty reproducing the
``red envelope'' of the observed galaxy population as a function
of redshift (e.g., \citealt{cole00}).
Of course, partitioning the accretion into cold and hot modes
does not, in itself, change the simulation predictions, but it
may illuminate the physics behind the predicted drop in the SFR
and explain differences between numerical and semi-analytic
results.  The declining SFR is frequently attributed to the
longer cooling times in the hotter, lower density halos that
prevail at low redshift (e.g., \citealt{blanton00}), but the
efficiency of cold accretion could be strongly affected by other
factors.
Furthermore, stellar or AGN feedback, or heat conduction, could have
different effects on the cold and hot accretion modes, since the incoming
gas has different geometry and density.  Allowing for such a difference
in theoretical models could have interesting observational consequences,
since the relative importance of cold and hot accretion depends on
redshift, environment, and galaxy mass.

The sensitivity of cold and hot accretion rates to environment
could also play a role in explaining the well known morphology-density
relation (e.g., \citealt{hubble36,dressler80,postman84}) and the
associated correlation between galaxy SFR and local density
(e.g., \citealt{lewis02,gomez03,kauffmann04}).  
Mergers, ram pressure stripping,
\citep{gunn72}, galaxy ``harassment'' by weak perturbations
in clusters \citep{moore96,moore98}, 
truncation of gas supplies \citep{larson80,somerville99}, and longer cooling
times in hotter environments \citep{blanton00} may all contribute
to the origin of these correlations.  However, environmental 
effects that shut off cold accretion flows could suppress star formation
and disk growth, and they might explain why transitions in galaxy
properties appear to start well beyond the virial radii of
groups and clusters \citep{lewis02,gomez03}.  \cite{kauffmann03} find
a clear transition in galaxy properties at a baryonic mass
$M_b \sim 3\times 10^{10} M_\odot$, with lower mass galaxies having
active star formation, low surface mass density, and a disk morphology
while higher mass galaxies have old stellar populations, high surface
mass density, and a bulge-dominated morphology.
The transition that we find between
galaxies dominated by cold accretion and galaxies dominated by
hot accretion occurs at a similar mass scale and could 
be connected to this broader transition in galaxy properties.

After briefly describing our simulations and analysis methods in
\S\ref{sec:methods}, we present our basic results on the global
significance and mass dependence of cold and hot accretion 
in \S\ref{sec:results}.
We assess the numerical robustness of these results in \S\ref{sec:numerical},
in particular comparing different simulations to show that cold mode
gas accretion is found in simulations with a large dynamic range in
mass resolution.  We investigate the dependence of the cold and hot accretion
rates and the corresponding star formation rates
on galaxy environment in \S\ref{sec:environment}, including a
comparison of the predicted correlations between SFR and environment
to the \cite{gomez03} observations.
In \S\ref{sec:discussion} we describe our current physical understanding
of the cold and hot accretion modes and discuss some of the potential
implications mentioned above.  We summarise our conclusions in 
\S\ref{sec:conclusions}.

\section{Simulations and Numerical Methods}
\label{sec:methods}

\subsection{Simulation parameters \label{sec:param}}

We adopt an inflationary cold dark matter model dominated by a cosmological
constant, \lcdm, with
$\Omega_m=0.4$, $\Omega_{\Lambda}=0.6$, 
$h\equiv H_0/(100\;\hubunits)=0.65$,
and a primordial power spectrum index $n=0.93$. For the
amplitude of mass fluctuations we use $\sigma_8=0.8$, which for our
adopted parameters is consistent both with COBE normalisation using CMBFAST 
\citep{seljak96,zaldarriaga98} and with the observed abundance
of rich clusters \citep{white93}.
For the baryonic density we adopt $\Omega_b=0.02 h^{-2}$, a value
consistent both with the deuterium abundance in high redshift Lyman limit
systems \citep{burles98} and the value derived from cosmic microwave
background (CMB) anisotropy measurements \citep{debernardis02}.
Our values of $\Omega_b$, $n$, and $\sigma_8$ are close to 
those inferred by recent joint analysis of CMB anisotropy measurements
from WMAP (\citealt{bennett03})
and galaxy clustering data from the 2dF Galaxy Redshift Survey
(2dFGRS; \citealt{colless01}) and the Sloan Digital Sky Survey
(SDSS; \citealt{york00}), while our assumed $\Omega_m$ is higher
by about $1.5\sigma$.  We have recently repeated one of our runs
using the parameter values implied by the WMAP analysis of
\cite{spergel03}, and 
preliminary investigation shows results similar
to those reported here.

Our primary results are derived from a simulation that models a 
22.222$\hmpc$ comoving periodic cube using $128^3$ dark matter
particles and $128^3$ gas particles.  Gravitational forces are
softened using a cubic spline kernel of 
comoving radius $5\hkpc$, approximately equivalent to a Plummer
force softening of $\epsilon_{\rm grav} = 3.5\hkpc$.
Our baryonic mass threshold for resolved galaxies 
(see \S\ref{sec:skid}) is $6.8 \times 10^9 M_\odot$, the mass
of 64 gas particles, and there are 1120 galaxies in the box
above this threshold at $z=0$.  To approximately match our
galaxy mass scale to an observed luminosity scale, we note that
the \cite{blanton03} $r$-band luminosity function yields a space
density of $0.0032\dunits$ for galaxies brighter than the
characteristic luminosity $L_*$ of a \cite{schechter76} function fit.
The baryonic mass threshold that yields the same space density in
our simulation is $2.45 \times 10^{11} M_\odot$, so if we identify
this mass with $L_*$ and assume that luminosity is approximately 
proportional to mass, our resolution threshold corresponds roughly to
$L_*/36$.

We draw on six additional simulations to investigate the influence of
mass resolution and the presence of a UV background field on our 
results.  Parameters of all the simulations are listed in 
Table~\ref{tbl:sims}.  A simulation of comoving box length $x~\hmpc$
and $N^3$ particles is designated L$x/N$; thus, our primary simulation
is L22/128.  Most of the simulations were run with a photoionizing
UV background (see below); those that were not are designated ``nb''
for ``no background.''  The full suite of simulations ---
L50/144nb, L22/128, L22/64nb, L11/64, L11/64nb, L11/128, L5.5/128 ---
spans a factor of 512 in mass resolution.  However, the highest
resolution simulations (L11/128 and L5.5/128) have been evolved
only to $z=3$, and the lowest resolution simulations have no UV background.

\begin{table*}
\begin{tabular}{ccccccc}

\hline Name&$L (\hmpc)$&$N$&$z_{\rm fin}$& UV &
$M_{\rm res}$($\msun$)&\\ 
\hline 
L50/144nb&$50$&$2\times 144^3$&$0$&
No &$5.4\times 10^{10}$&\\ 
\bf{L22/128}&$22.22$&$2\times
128^3$&$0$& Yes &$6.8\times 10^9$&\\ 
L22/64nb&$22.22$&$2\times
64^3$&$0$& No &$5.4\times 10^{10}$&\\ 
L11/64&$11.11$&$2\times
64^3$&$3$& Yes &$6.8\times 10^9$&\\ 
L11/64nb&$11.11$&$2\times
64^3$&$3$& No &$6.8\times 10^9$&\\ 
L11/128&$11.11$&$2\times
128^3$&$3$& Yes &$8.5\times 10^8$&\\ 
L5.5/128&$5.55$&$2\times
128^3$&$3$& Yes &$1.1\times 10^8$&\\ 
\hline
\end{tabular}
\caption{Parameters of the simulations used in this paper. $L$
is the comoving box size, $N$ is the total number of particles
(dark+baryonic), $z_{\rm fin}$ is the final redshift to which the
simulation has been evolved, UV indicates whether or not a UV 
background is included in the calculation of cooling and heating rates,
and $M_{\rm res}$ is the baryonic mass resolution threshold, corresponding
to the mass of 64 gas particles.
}
\label{tbl:sims}
\end{table*}

\subsection{The Simulation Code \label{sec:simulations}}

Our simulations are performed using the parallel version of TreeSPH
(\citealt{hernquist89}; \citealt{katz96}, hereafter KWH; \citealt{dave97}).
This code combines smoothed
particle hydrodynamics (SPH; \citealt{lucy77,gingold77})
with the hierarchical tree algorithm for computation of gravitational
forces \citep{barnes86,hernquist87}.
TreeSPH is a completely Lagrangian code, adaptive both in space and in time. 
In our simulations, gas properties are estimated by smoothing over 32 nearby
particles.  There are three kinds of particles in our simulations:
dark matter, stars and gas. Collisionless particles (dark matter and
stars) are influenced only by gravity, while gas particles are
influenced by pressure gradients and shocks in addition to
gravity.

We use the geometric averaging
form of the energy equation \citep{hernquist89}.
Gas particles experience adiabatic heating and cooling, shock heating,
inverse Compton cooling off the microwave background, and radiative
cooling via free-free emission, collisional ionization and recombination,
and collisionally excited line cooling.  We assume primordial abundances
(since we are primarily interested in following gas before it gets
into galaxies), and we include only atomic cooling processes, so gas
cannot cool below $T \sim 10^4\K$.
In most of our simulations (see above), we include photoionization by a
spatially uniform UV background, 
which heats low temperature gas and suppresses cooling processes
involving neutral atoms at low gas densities.
For the spectral shape and intensity of the UV
background we use the calculations of Haardt \& Madau (1996).
Our calculations of cooling and heating rates are discussed
in detail by KWH, who also illustrate the influence of photoionization
on these rates.

We heuristically include star formation and its associated supernova feedback,
as described by KWH.
In brief, gas with physical density $\rho_{\rm gas} > 0.1m_H {\rm cm}^{-3}$
(hydrogen number density $n_H > 0.1{\rm cm}^{-3}$) is assumed to convert
into stars on a timescale set by the dynamical time or the cooling
time, whichever is longer.  We also require star-forming gas to
be Jeans unstable, part of a converging flow ($\nabla\cdot {\bf v} < 0$),
and above the virial overdensity 
($\rho_{\rm gas}/\bar{\rho}_{\rm bar} > 55.7$), but gas that
satisfies the physical density criterion usually satisfies the other
three criteria as well.  Gas reaches this high density only
after cooling to $T \sim 10^4\K$, and we implicitly include the
subsequent molecular and metal-line cooling to lower temperatures
as part of the star formation process.  Our formula for the star 
formation rate leads to a relation with gas surface density 
similar to a Schmidt law \citep{schmidt59,kennicutt98}.
At a technical level, we convert gas particles to collisionless star
particles by means of intermediate particles that feel reduced
gas forces (see KWH). This allows us to trace each star particle
back to a unique gas progenitor.

During the formation of stars, supernova feedback energy is added to
the surrounding gas particles in the form of heat. This energy is
added gradually with an exponential time decay of $2\times 10^7$ years.
We calculate this energy assuming that stars with masses above $8\msun$ 
explode as supernovae,
which for our adopted Miller-Scalo initial mass function (Miller \& Scalo 1979)
gives $7.35\times 10^{-3}$ supernovae per solar mass of formed stars.
Each supernova deposits $10^{51}$ ergs of energy into the surrounding medium.
The surrounding medium is usually dense, so the deposited energy is
typically radiated away before it can drive a galactic scale wind.

\subsection{Identifying Galaxies and Halos \label{sec:skid}}

Cosmological simulations that incorporate cooling and star formation produce
dense groups of baryonic particles with the sizes and masses of 
observed galaxies \citep{katz92,evrard94}.
To identify these dense groups we use the group finding algorithm
Spline Kernel Interpolative DENMAX (SKID)\footnote{
\tt http://www-hpcc.astro.washington.edu/tools/skid.html} (\citealt{gelb94};
KWH).
This algorithm involves four basic steps: (1)
determine the smoothed baryonic density field; (2) move baryonic particles
towards higher density along the initial gradient of the baryonic
density field; (3) define the initial group to be the set of particles that
aggregate at a particular density peak; (4) link together initial groups 
that are very close together; (5) remove particles from the
group that do not satisfy a negative energy binding criterion relative
to the group's centre of mass. 
We apply SKID to the population of all star particles and those gas
particles that have temperatures $T < 3 \times 10^4\K$
and overdensities $\rho_{\rm gas}/{\bar\rho}_{\rm gas} > 10^3$,
and we henceforth refer to the groups of stars and cold gas that
SKID identifies simply as ``galaxies.''
Tests on simulations with varying mass resolution show that 
the simulated galaxy population becomes substantially incomplete
below a baryonic mass corresponding to $\sim 64 m_{\rm SPH}$
but is fairly robust above this limit (see, e.g., MKHWD).
We therefore adopt $64 m_{\rm SPH}$ ($6.8 \times 10^9 M_\odot$ in 
the L22/128 simulation) as our resolution threshold and ignore
lower mass galaxies in our analysis.
Because of our high overdensity threshold for star formation,
essentially all star formation in the simulation takes place
in galaxies, though some of these are below the resolution limit,
and some stars are tidally stripped from galaxies
during dynamical interactions.

We identify dark matter halos using
a friends-of-friends (FOF) algorithm \citep{davis85},
which selects groups of particles in which every particle has
at least one neighbour within a specified linking length.
We choose the linking length to correspond to the interparticle
separation at $1/3$ of the virial overdensity $\rhovir/\bar{\rho}$,
which is calculated for the value of $\Omega_m$ at each redshift
using the fitting formula of \cite{kitayama96}.
We refine the halos and assign a virial mass, virial radius,
and virial temperature using a spherical overdensity (SO)
criterion.  Specifically, we set the centre of the group to be at
the position of the most bound FOF particle, and we go out in radius
until the mean enclosed overdensity (dark matter plus baryons)
equals $\rhovir(z)/\bar{\rho}$.  We define
this radius to be $\rvir$, the mass within it to be $\mvir$, and the
halo circular velocity to be $\vcirc = (G\mvir/\rvir)^{1/2}$.
We define the halo virial temperature by 
$k\tvir = \frac{1}{2} \mu m_p \vcirc^2$, so it represents
the temperature at which the gas would be in hydrostatic equilibrium
if the potential well were isothermal.
Since gas at this temperature would be fully ionized, we adopt
$\mu = 0.59$, appropriate to fully ionized, primordial composition gas.
We associate galaxies with these, refined,  SO halos.

Our minimum galaxy mass $\mgalmin = 64 m_{\rm SPH}$ corresponds
to a minimum host halo mass $\mhalomin \approx (\Omega_m/\Omega_b)\mgalmin$,
since the fraction of cold gas in a halo never exceeds the universal
baryon fraction by a large factor (though it can sometimes be slightly
larger).  The corresponding minimum virial temperature is
$\tvirmin(z) \approx 80,000 (1+z)\K$ for the L22/128 simulation,
where the $(1+z)$ factor
arises from the increasing physical density at higher $z$ for
fixed virial {\rm over}density. 
Including the redshift dependence of the virial overdensity makes
the pre-factor
slightly lower at high redshifts ($\sim$ 75000 K) 
and slightly higher at low redshifts (88000 K at $z=0$).
Any resolved galaxy in the simulation 
resides in a dark matter halo with $\tvir \geq \tvirmin(z)$.

\subsection{Determining Gas Accretion Histories  \label{sec:acchistory}}

At each of our output redshifts (listed below), we use SKID to identify
all galaxies above our $64 m_{\rm SPH}$ baryonic mass resolution
threshold ($6.8 \times 10^9 M_\odot$ for L22/128).  
We discard from consideration any galaxy whose progenitor was 
not resolved at the previous output redshift, since we do not want to 
count simply
passing across the resolution threshold as ``accretion.''  For the 
remaining galaxies, resolved in both outputs, we identify any gas
particles that are in the galaxy at the later redshift but were not 
in any resolved galaxy at the earlier redshift as smoothly accreted
gas.  For each of these accreted gas particles, we trace back the
entire temperature history since the beginning of the simulation
and record the particle's maximum temperature $\Tmax$.  These $\Tmax$
values will be used to distinguish the cold and hot accretion modes.

To determine the global accretion properties at a given redshift, we sum
over all the particles that have been smoothly accreted between our
previous analysis output and that redshift.  By ``smoothly accreted''
we simply mean particles that were not in resolved galaxies at
the previous output.  For most of our calculations, we include only
accreted {\it gas} particles in the statistics.  Star particles can
also be smoothly accreted either because they were in under-resolved
galaxies at the previous output or because they were tidally
stripped ``field'' stars. (Often these ``field'' stars are effectively 
part of the galactic halo or the intragroup or
intracluster environment, but SKID still attaches them to 
a galaxy.)
However, since all the stars in our simulation form within
galaxies, the gas that formed these smoothly accreted stars was itself
accreted onto a (perhaps under-resolved) galaxy at a higher 
redshift, making the $\Tmax$ values representative of gas accretion
at higher redshifts.  Including accreted star particles in our
statistics has only a small impact, increasing the apparent importance
of cold accretion at low redshift 
(see the dotted line in Figure~\ref{fig:diff_acc_ns} below).

Our measured rates of smooth accretion really represent the sum of
genuinely smooth gas flow and mergers with galaxies below our resolution
limit, including systems that are not even under-resolved bound objects
in the simulation but would nevertheless exist in the real universe.  In 
\S\ref{sec:submerge} we extrapolate the measured distribution of
merger mass ratios to show that this sub-resolution merging is
unlikely to be an important correction, so that our measured smooth
accretion rates indeed correspond mainly to smooth gas flows.

We use the following redshift spacing for determining $\Tmax$ values
and calculating accretion rates: for 
$0 \le z \le 1$ we use $\Delta z\sim 0.125$, for $1 < z \le 3.5$ we use 
$\Delta z\sim 0.25$, for $3.5 < z \le 5$ we use $\Delta z\sim 0.5$, and for
higher redshifts we use $\Delta z\sim 1$. In the cosmology we consider here,
these redshift intervals correspond to time
intervals of roughly $\Delta t=0.25$ Gyr for $z > 3.5$, 
$\Delta t\sim 0.5$ Gyr at $z=1.5$ and $\Delta t \sim 1.5$ Gyr at $z=0$.  
These time intervals are close to the typical infall time scales at the 
corresponding outputs, except at $z=0$ where they are slightly longer.
We have checked the effect of using smaller redshift intervals
(see \S\ref{sec:tint}) and find that it makes only a small difference
to our statistical results.

\section{Cold and Hot Accretion} 
\label{sec:results}

\begin{figure*}
\centerline{
\epsfxsize=1.0\columnwidth
\epsfbox[35 25 565 690]{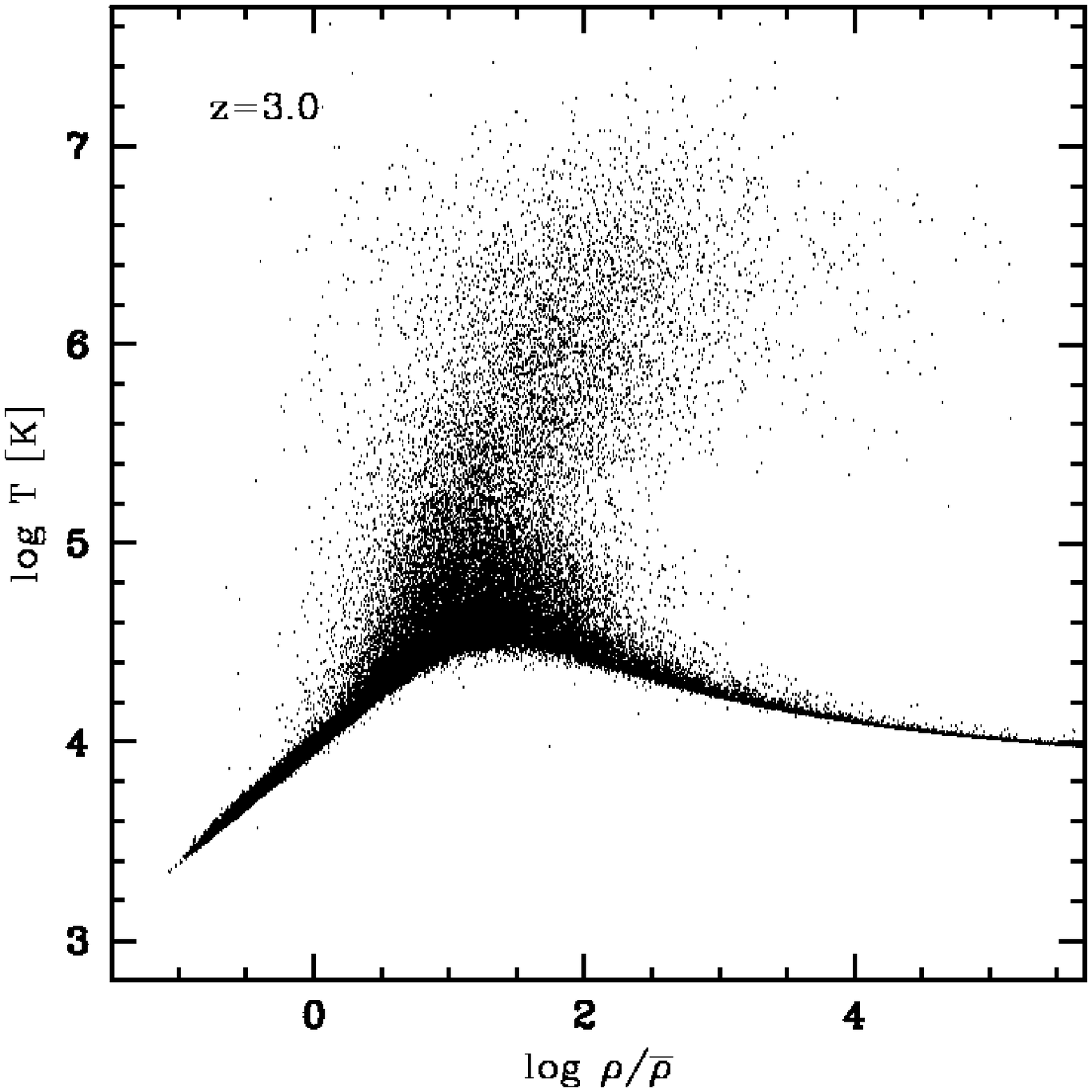}
\hfill
\epsfxsize=1.0\columnwidth
\epsfbox[35 150 565 690]{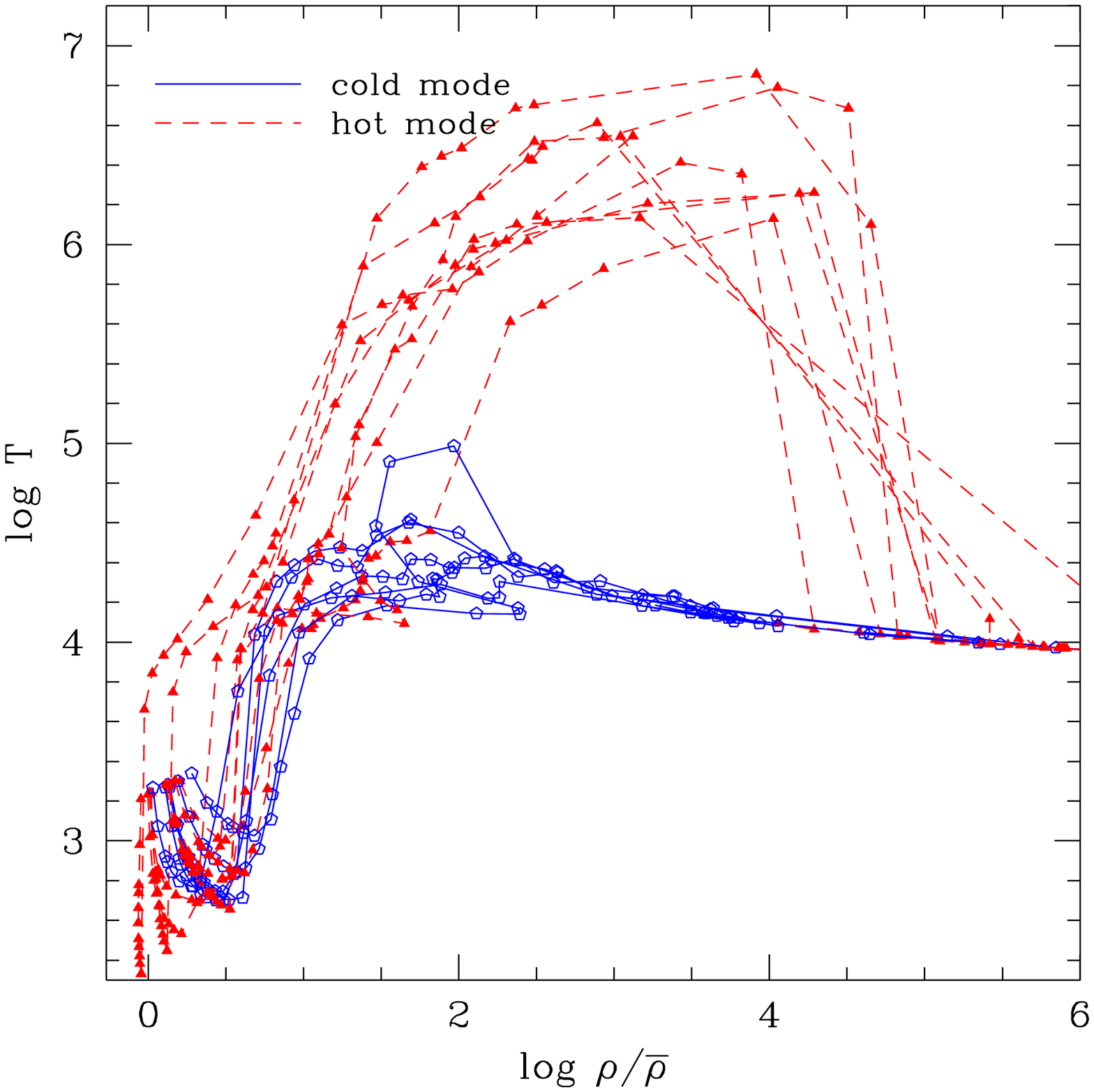}
}
\caption{
{\it Left}: Distribution of gas particles in the $\rho-T$ plane at $z=3$,
in the L22/128 simulation.  One can easily identify three major phases:
low density, low temperature gas in the photoionized IGM, shock heated
overdense gas, and high density, radiatively cooled gas within galaxies.
{\it Right}: Trajectories of 15 particles that accreted onto galaxies
shortly before $z=3$, illustrating the ``cold'' (solid lines, circles)
and ``hot'' (dashed lines, triangles) accretion modes.  Hot mode
particles are shock heated above $\sim 10^{5.5}\K$ before cooling,
while cold mode particles move directly from the diffuse IGM phase
to the dense, galactic phase without ever heating above $10^5\K$.
Trajectories start at $z=14.9$ and end at $z=3$.  Points mark the 
individual redshift outputs, which have typical time separations
of 0.05-0.1 Gyr.
}
\label{fig:rhot}
\end{figure*}

Figure~\ref{fig:rhot}a shows the distribution of gas particles
in the temperature-overdensity plane, from the L22/128 simulation at $z=3$.
As with previous studies (e.g., \citealt{dave99}), we identify three important
gas phases.  The narrow,
upward sloping locus with $\rho/\bar{\rho} < 10$ and $T<10^5\K$ consists
of low density, highly photoionized gas in the intergalactic medium (IGM),
which is responsible for the \lya\ forest.  
The tight temperature-density relation in this regime is maintained
by the competition between adiabatic cooling and photoionization
heating \citep{hui97}.  The plume of particles
with $\rho/\bar{\rho} \sim 10 - 10^4$ and $T \sim 10^5-10^7\K$
is comprised of shock heated gas in virialized halos and, at the lower
density end, around filaments.
The narrow, downward sloping locus at $\rho/\bar{\rho} > 10^3$
and $T \approx 10^4\K$ represents radiatively cooled, dense gas in galaxies.
Cooling times at these densities are short, so gas remains close to the
equilibrium temperature where photoionization heating balances radiative
cooling, which is a slowly decreasing function of density.  

According to the conventional sketch described in the introduction,
gas that ends
up in galaxies starts in the diffuse phase, enters the 
shock heated phase, then cools and condenses to reach the
galactic phase.  Figure~\ref{fig:rhot}b plots the 
$\rho-T$ trajectories of 15 randomly selected gas particles accreted
onto galaxies near $z=3$, starting near $z=15$.
Some of these particles follow just the
path described above.  However, about half of them start in the 
diffuse IGM phase and move directly to the dense galactic phase
without ever heating above $10^5\K$.  The virial temperature of the 
smallest resolved halo in the simulation is $3.0 \times 10^5\K$
at $z=3$, so these particles have never been close to the virial
temperature of any resolved halo.  
The separation between outputs in Figure~\ref{fig:rhot}b is typically
0.05-0.1 Gyr, and some of the low-$\tmax$ particles could shock
heat and cool rapidly in between two outputs.
However, the cooling
radiation arguments of \cite{fardal01} show that such ``missed cooling''
events cannot have a large impact.  The total energy radiated by gas
entering the simulated galaxies is of order the acquired 
gravitational potential energy, as expected, and most of this energy 
is radiated by
gas with $T < 3 \times 10^4\K$.  Gas shock heated to $10^6\K$, by
contrast, would radiate 90\% of its thermal energy by the time it
cools to $10^5\K$, so it would not have much energy left to radiate
at low temperature.  Counting luminosity as \citet{fardal01} do closes the
loophole of missing rapidly cooling particles because the
high luminosity of such particles would compensate for their rarity.

\begin{figure*}
\setlength{\epsfxsize}{0.75\textwidth} 
\centerline{\epsfbox{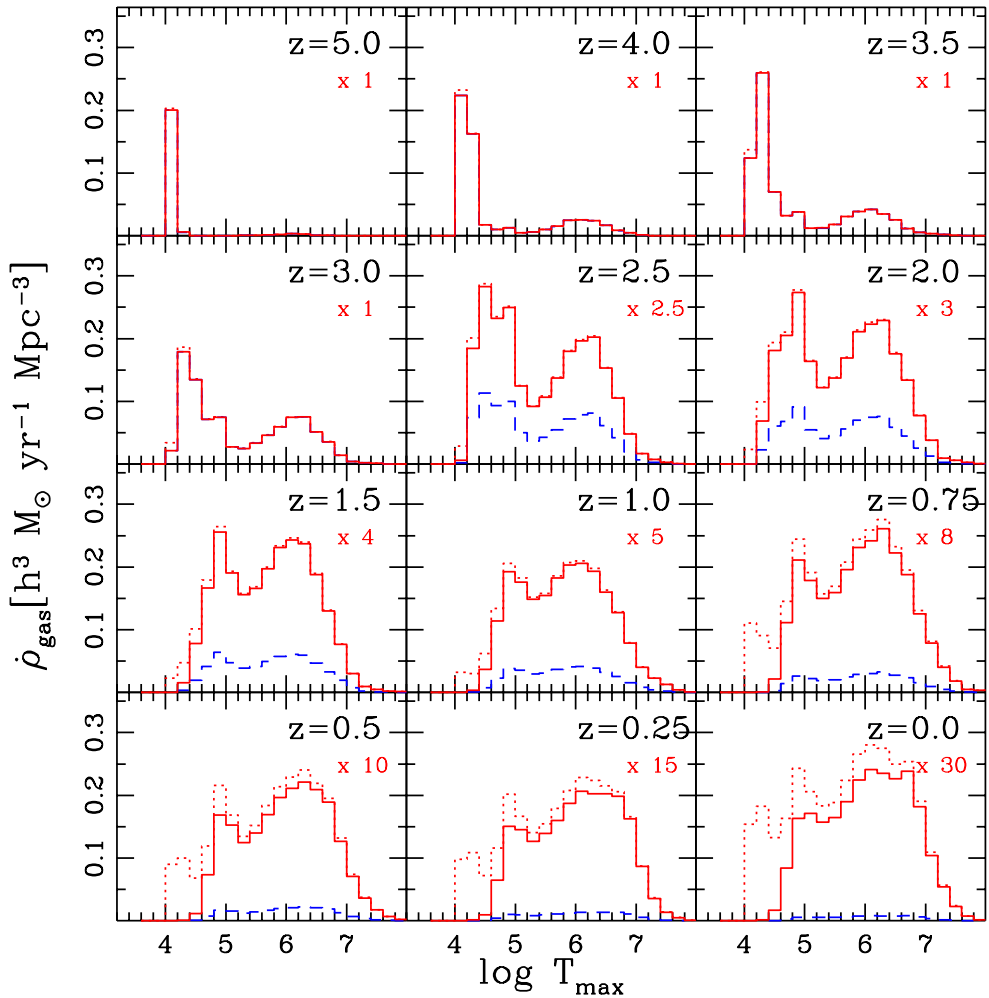}}
\caption{
Distribution of maximum temperatures of gas accreting onto galaxies.
For each particle that was smoothly accreted onto
a resolved galaxy between the previous output and the plotted redshift,
we trace back its history to determine the maximum temperature it had
at any previous time.  Dashed histograms show this distribution 
in units of $h^3M_{\odot}{\rm yr}^{-1}{\rm Mpc}^{-3}$ (comoving) per 0.2-dex
bin of $\log \tmax$.  Solid histograms have the same shape but
are multiplied by an arbitrary constant (as indicated in the panel)
to improve visibility.  Dotted histograms show the effect of including
accreted stars in the calculation. 
}
\label{fig:diff_acc_ns}
\end{figure*}

The two populations of trajectories in Figure~\ref{fig:rhot}b 
(also illustrated in Fig.~4 of \citealt{kay00}) represent
the processes we refer to as ``cold mode'' and ``hot mode'' gas accretion.
To quantify the global significance of these two modes,
we plot in Figure~\ref{fig:diff_acc_ns} the distribution of $\Tmax$ values
of accreted gas, computed as described in \S\ref{sec:acchistory}.
Each panel shows the $\Tmax$ histogram for particles accreted onto 
resolved galaxies between the previous output and the indicated
redshift.  At high redshifts, the histograms are clearly bimodal:
some gas enters galaxies after cooling from $T \sim 10^6 - 10^7\K$,
but a large fraction of the accreted gas was never hotter than
$T \sim 10^5\K$.  The bimodal nature of the $\Tmax$ histogram and
the comparable contributions of the two modes to mass accretion
by galaxies are the primary result of this paper.

The overall accretion rate declines towards low
redshift, though a significant amount of growth still occurs
because more time is available (see Fig.~\ref{fig:history} below).
The ``cold mode'' region of the histogram broadens
towards low redshift, and its centroid shifts towards higher $\Tmax$,
so by $z=0$ the distribution is no longer strongly bimodal.
As discussed in \S\ref{sec:acchistory}, we generally exclude accreted
stars from our statistics, since they reflect physical conditions 
that prevailed at earlier times.  Dotted histograms in 
Figure~\ref{fig:diff_acc_ns} show the effect of including accreted
stars, which is mainly to extend the low redshift histograms to
lower $\Tmax$ values.

\begin{figure*}
\setlength{\epsfxsize}{0.75\textwidth}
\centerline{\epsfbox{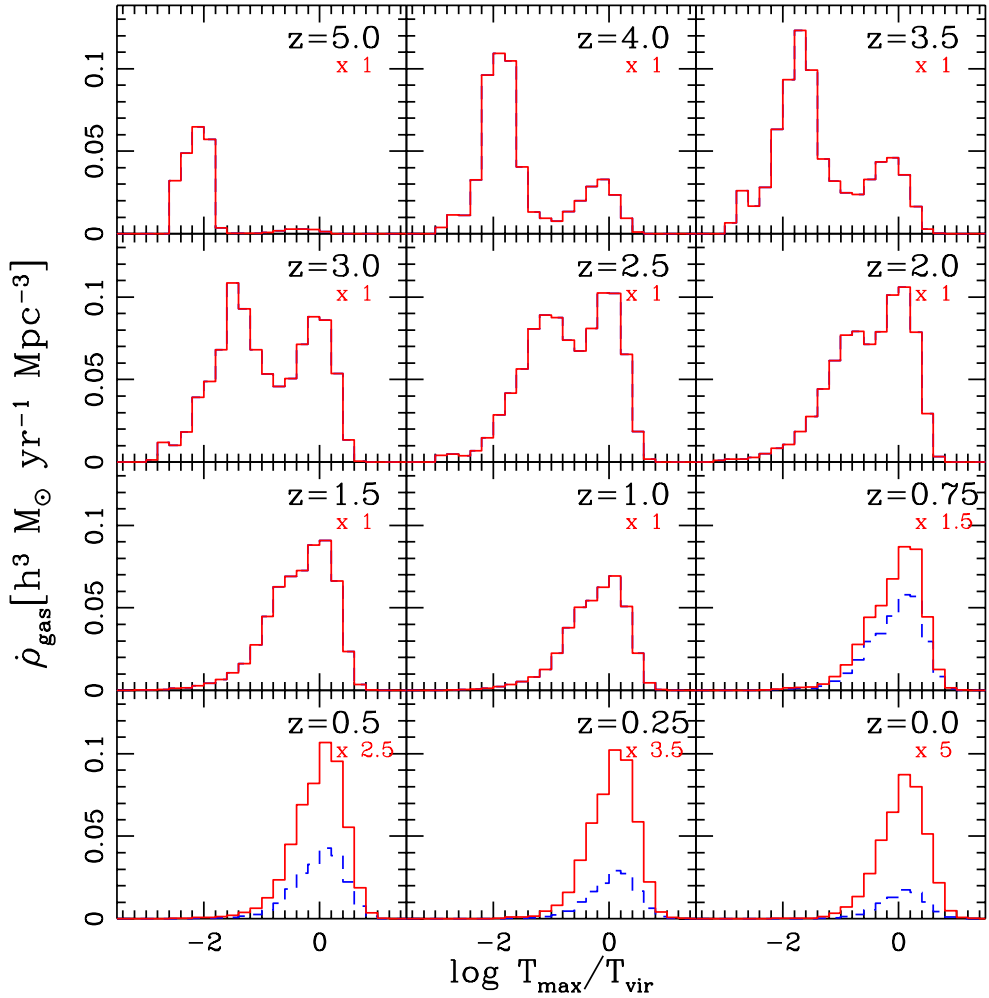}}
\caption{
Like Fig.~\ref{fig:diff_acc_ns}, but showing the distribution of
$\tmax/\tvir$, the ratio of a particle's maximum temperature to the
virial temperature of the accreting galaxy's dark matter halo.
}
\label{fig:tvir}
\end{figure*}

Figure~\ref{fig:tvir} is similar to Figure~\ref{fig:diff_acc_ns},
but here we plot the distribution of $\Tmax/\tvir$, where $\tvir$
is the virial temperature of the galaxy's parent dark matter halo,
identified as described in \S\ref{sec:skid}.  At $z \geq 2.5$,
there is again clear bimodality in the distribution, revealing
a distinction between gas that goes through a strong, virial-type
shock before cooling and gas that does not.  The high temperature
portion of the histogram peaks at $\Tmax \sim \tvir$ at all
redshifts.  However, the spread in $\Tmax/\tvir$ values is large,
roughly an order-of-magnitude, even for this hot mode accretion.
The spread presumably reflects departures from the spherical
geometry adopted in most analytic calculations, plus the effects
of post-shock adiabatic compression, which can heat gas to $\Tmax > \tvir$
before cooling sets in.  The peak of the low temperature portion of the
histogram moves steadily towards higher $\Tmax/\tvir$ with time.
In Figure~\ref{fig:diff_acc_ns} we saw that the characteristic
$\Tmax$ of cold mode accretion also increases with time, rising from
$\Tmax \sim 10^{4.2}\K$ at high redshift to $\Tmax \sim 10^5\K$ at
low redshift, but the shift is more rapid when we scale to $\Tvir$
because the virial temperatures of the smallest resolved halos
are themselves dropping with time (see \S\ref{sec:skid}),  
and these halos account for a substantial fraction of cold accretion.
The clear
bimodality of the histograms has therefore disappeared by $z\sim 1.5$
in Figure~\ref{fig:tvir}, and at lower redshift the cold mode appears
as a tail of the distribution towards low $\Tmax/\tvir$.

Figures~\ref{fig:diff_acc_ns} and~\ref{fig:tvir} show that the cold
and hot modes separate more cleanly in the distribution of physical
temperature $\Tmax$ rather than the scaled temperature $\Tmax/\tvir$.
This result is itself a useful clue to the physics of cold accretion.
If the characteristic temperature of the cold accretion mode were determined
by gravitationally induced shocks, then we would expect it to scale
with the characteristic dynamical temperature $\tvir$.  In this case,
the minimum in the $\tmax/\tvir$ histogram should stay roughly constant,
instead of rising steadily with time as it does in Figure~\ref{fig:tvir}.
The minimum of the $\tmax$ histogram, on the other hand, does stay nearly
constant at $\tmax \sim 10^{5.4}\K$, near a local minimum in the 
atomic cooling curve where free-free emission takes over from helium
line cooling (see, e.g., Fig. 1 of KWH).  This behaviour
could arise if a significant fraction of infalling gas starts at
low temperature ($T \la 10^4\K$) and gains energy through 
weak shocks or adiabatic
compression, in which case gas with short cooling
times could radiate energy without ever heating to high temperature,
while gas with long cooling times could not.  Clearly this is not the
full story, since the peak of the $\tmax$ histogram does move to higher
temperature over time, presumably reflecting the higher infall 
velocities associated with larger gravitationally induced structures.
We will discuss the
physics of cold accretion at greater length in \S\ref{sec:disc_cold}.  
For now, we use
the empirical evidence of Figure~\ref{fig:diff_acc_ns} to set the
dividing line between the cold and hot modes at $\tmax = 2.5\times 10^{5}\K$,
noting that the location of this division is likely a consequence
of atomic physics.  Gravitationally scaled temperatures
appear less effective at distinguishing hot and cold accretion.

\begin{figure*}
\setlength{\epsfxsize}{0.9\textwidth}
\centerline{\epsfbox[30 165 595 480]{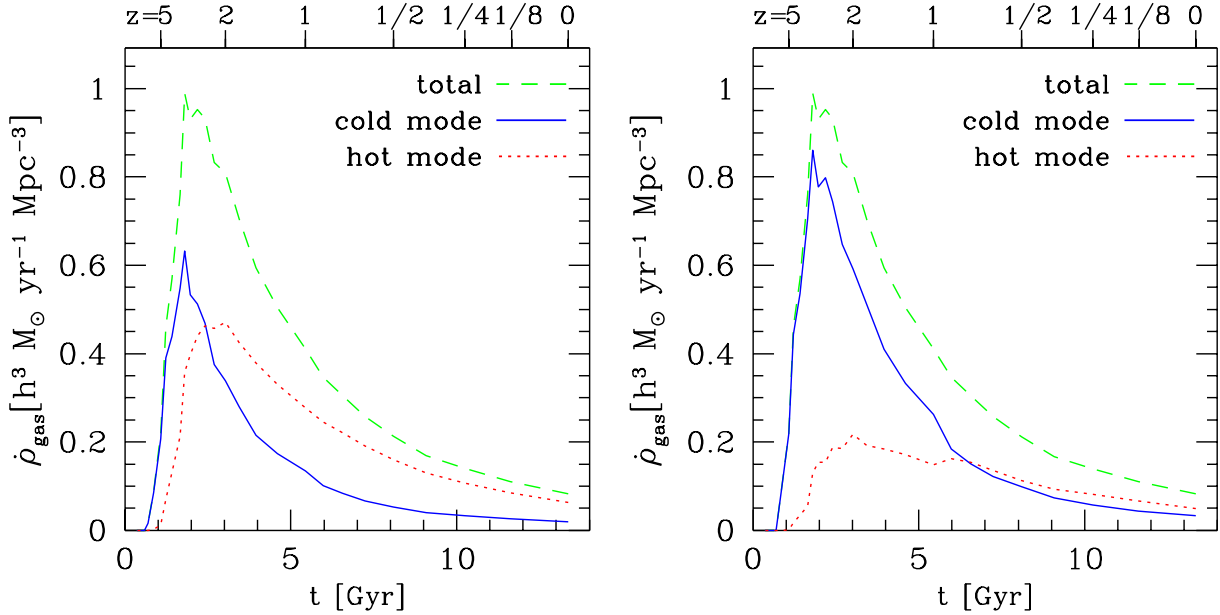}}
\caption{
Redshift history of the total smooth gas accretion rate (dashed line),
and the rates in cold mode and hot mode (solid and dotted lines,
respectively).  In the left panel, the division between hot and cold
modes is at $\tmax = 2.5\times 10^{5}\K$, while in the right panel it is at
$\tmax/\tvir=1$.
}
\label{fig:history}
\end{figure*}

Figure~\ref{fig:diff_acc_ns} shows that cold accretion dominates over
hot accretion at high redshift, then becomes steadily less important
at lower redshifts.  Figure~\ref{fig:history} quantifies this dependence
by showing the mean accretion rate in cold and hot modes as a function
of time and redshift.  The left hand panel adopts our standard division
between cold and hot mode, at $\tmax = 2.5\times 10^{5}\K$.  The cold 
accretion
rate and the total accretion rate rise rapidly together and peak
at $z \sim 3$, then decline towards low redshift.
Note that this Figure shows accretion onto galaxies above our 
resolution threshold; if we went further down the mass function,
the accretion rate would be flatter at $z>3$.
The hot accretion rate rises later, peaking at $z\sim 2$, and it
declines more gently thereafter.  Cold accretion dominates at $z \geq 3$,
while hot accretion dominates at $z \leq 2$.  The right hand panel
adopts a division at $\tmax/\tvir = 1$.  With this definition
(which is not, we think, the most appropriate one), cold accretion
dominates by a large factor at high redshift, and the two modes
are comparably important at $z<1$.

\begin{figure*}
\centerline{
\epsfxsize=0.8\textwidth
\epsfbox[35 280 570 680]{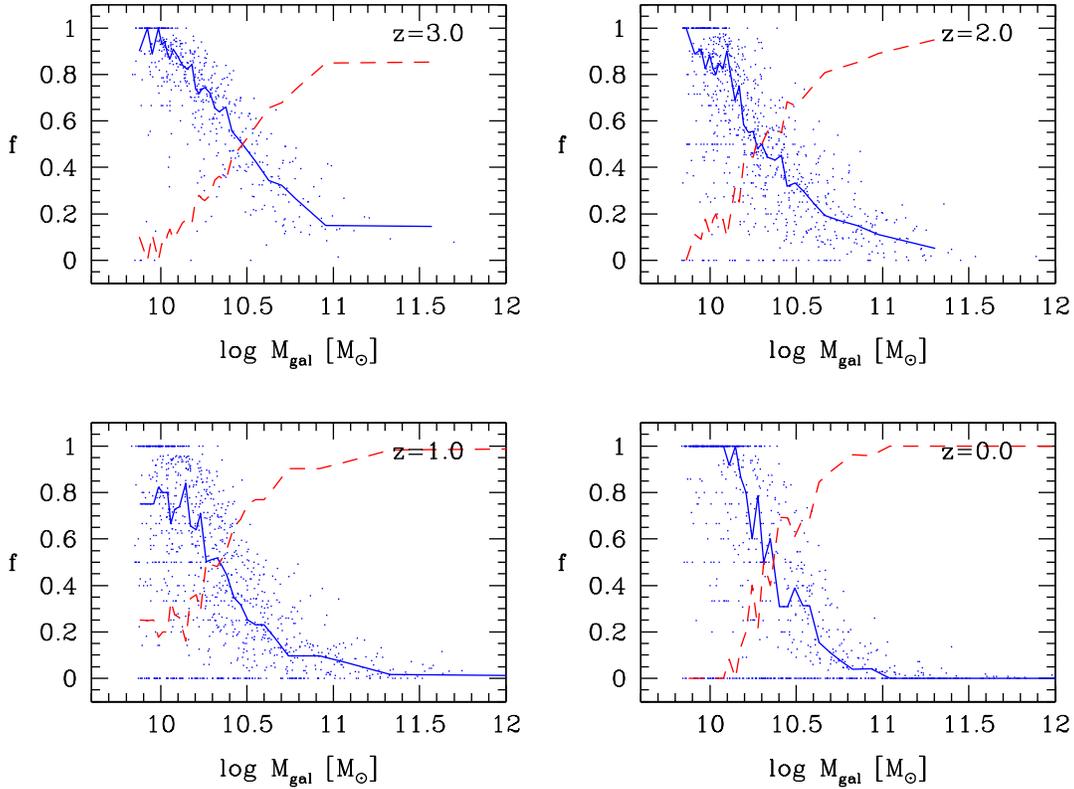}
}
\caption{
The cold accretion fraction as a function of galaxy baryonic mass
(cold gas + stars), at $z=3$, 2, 1, and 0.  Points show the cold
fractions of individual galaxies, and solid lines show the median
values in bins of baryonic mass.  Dashed lines show the median hot
fraction; solid and dashed curves sum to one by definition.
}

\label{fig:mass_ratio}
\end{figure*}

These global statistics, averaged over the full resolved galaxy population,
mask an important underlying trend, namely a strong dependence of the
cold/hot accretion fractions on galaxy mass.  
Figure~\ref{fig:mass_ratio} plots the cold accretion fraction against
galaxy baryonic mass at $z=3$, 2, 1, and 0.  Points show the cold
fractions for each resolved galaxy, the solid curve shows the median
value in bins with equal number of galaxies (usually $\sim 10-20$ per bin), 
and the dashed curve shows the median
hot accretion fraction.  The solid and dashed curves sum to one by
definition.  There is a strong and continuous trend of cold accretion
fraction with baryonic mass.  The transition mass where cold and hot
accretion are equally important is nearly constant at
$\mgal \sim 10^{10.3} \msun$ for $z \leq 2$, and slightly higher
($\mgal \sim 10^{10.5} \msun$) at $z=3$.
We will show in \S\ref{sec:numconv} that the value of this transition
mass is insensitive to the numerical resolution of the simulation.
The overall change of the cold mode accretion fraction with redshift,
seen in Figure~\ref{fig:history},
is largely a consequence of increasing galaxy masses at low redshift.

\begin{figure*}
\epsfxsize=0.8\textwidth
\centerline{\epsfbox[35 280 570 680]{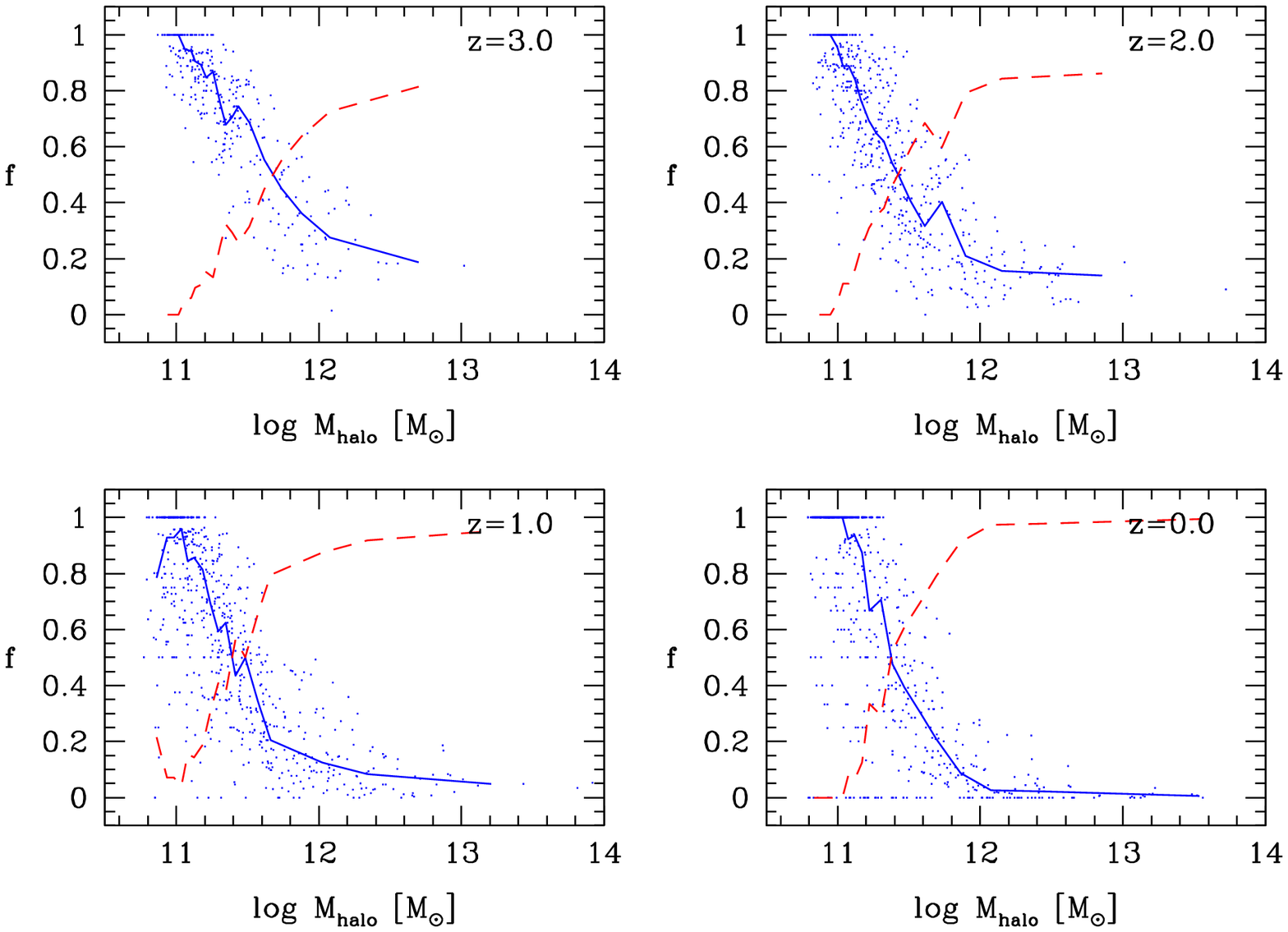}}
\caption{Like Figure~\ref{fig:mass_ratio}, but using parent
halo mass in place of galaxy baryonic mass.}

\label{fig:mass_ratio_halo}
\end{figure*}

We obtain similar results if we use, in place of a galaxy's baryonic mass,
the total mass of its parent halo, as shown in 
Figure~\ref{fig:mass_ratio_halo}. At high redshift, the trend with
halo mass is even cleaner than the trend with galaxy mass; comparing
Figures~\ref{fig:mass_ratio} and~\ref{fig:mass_ratio_halo} shows that
the low mass galaxies  that are dominated by hot mode are
mostly those that live in high mass halos (satellite
objects orbiting a larger central galaxy). The cold mode fraction
in massive halos is slightly larger at high redshifts,
an effect that is even more pronounced if we associate galaxies with 
FOF halos instead of SO halos. 
The transition halo mass at which cold and hot modes are equally
important is $\mhalo \sim 10^{11.4}\msun$ (slightly higher at $z=3$),
which is the value one would expect given our transition baryonic
mass of $\mgal \sim 10^{10.3} \msun$ if $\sim 2/3$ of the available
halo baryons typically end up in the central galaxy.
The $10^{11.4}\msun$ transition mass is about a factor of $~2-3$ higher than
the value found by \cite{birnboim03} based on 1-d numerical experiments,
a good level of agreement given the radically different 
calculational methods.  We will return to this point
in \S\ref{sec:disc_cold}.

\begin{figure*}
\epsfxsize=0.8\textwidth
\centerline{\epsfbox[60 260 580 680]{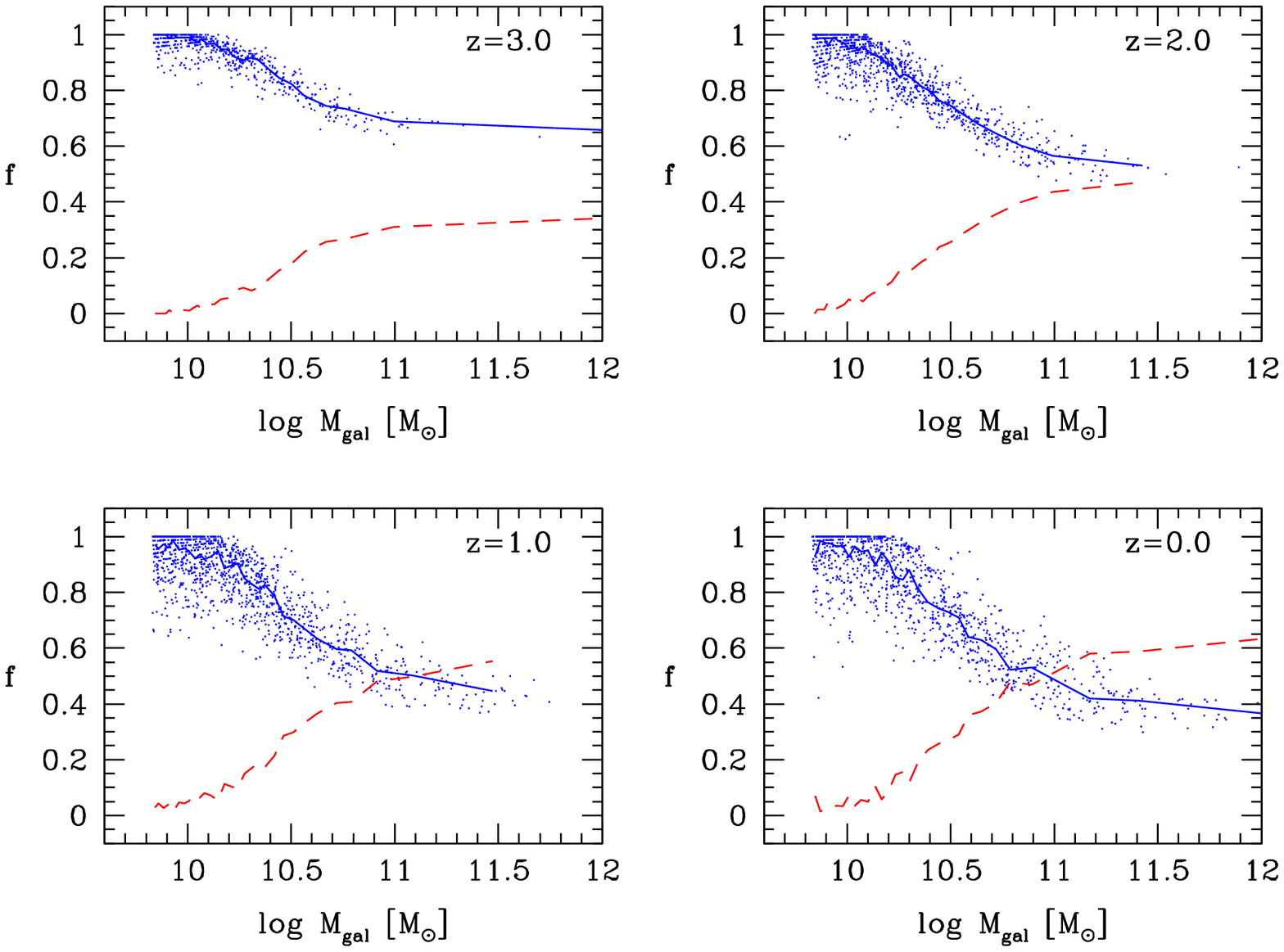}}
\caption{
Contribution of cold and hot accretion to total galaxy masses at 
$z=3$, 2, 1, and 0.  Points show the cold mode fraction of each galaxy,
i.e. the fraction of its mass with $\tmax < 2.5\times 10^{5}\K$, 
as a function of baryonic mass.  The solid line is the median
cold mode fraction, while the dashed line is the median hot
mode fraction; by definition, the two fractions sum to one.
}
\label{fig:cold_vs_mass}
\end{figure*}

Figure~\ref{fig:cold_vs_mass} characterises the overall contributions
of cold accretion to the simulated galaxy population by plotting the
fraction of each galaxy's baryonic mass that was first accreted
in cold mode against the galaxy's present mass.  Here we consider
all particles present in the galaxy, cold gas {\it and} stars, and
trace back each particle's history to determine its $\tmax$ value.
(Since each star particle comes from a unique gas particle, this
value is well defined even for stars.)
High mass galaxies have little or no cold gas accretion 
(Fig.~\ref{fig:mass_ratio}), but they are built largely from
mergers of lower mass systems, and these did accrete a significant
fraction of their gas in cold mode.  Thus, even for the most massive
galaxies present at $z=0$ (lower right panel), roughly 40\% of the
baryonic mass was never heated above our adopted threshold of
$2.5\times 10^{5}\K$.  For galaxies with $\mgal < 5\times 10^{10} M_\odot$,
more than half of the mass was accreted in cold mode, and for
$\mgal \la 2\times 10^{10}M_\odot$, nearly all of it.
If we look at higher redshifts, the cold mode fraction at
fixed mass is slightly higher, but the biggest difference
is that more of the galaxies are low mass.  Every galaxy present
in the simulation at $z=3$ was built primarily via cold accretion.
We conclude that, if our simulation predictions are accurate, 
cold accretion is a major ingredient in the recipe of galaxy formation.

\section{Numerical Issues} 
\label{sec:numerical}

\subsection{Sub-resolution merging \label{sec:submerge}}

One issue of both physical and numerical interest is whether the
``smooth'' accretion onto our resolved galaxies has a substantial
contribution from mergers with systems below our resolution
threshold, since we do not distinguish between gas particles
that enter individually and those that enter in unresolved 
groups (see \S\ref{sec:acchistory}).  If sub-resolution merging
dominates, then our computed $\tmax$ values could be spuriously
low because the gas does its cooling in unresolved systems,
and the cold accretion process found here might simply be analogous
to the mergers with low-$\tvir$ objects.
The best way to address this issue is to run a 
simulation that resolves galaxies all the way down to the limit where 
their formation is suppressed by the UV background,
a baryonic mass of about $\mgal =10^8M_{\odot}$
\citep{efstathiou92,quinn96,thoul96,gnedin00}.
We have one such
simulation, L5.5/128 with 64 times the mass resolution of L22/128,
which has been evolved to $z=3$.  
We first calculate the smooth accretion rates in L5.5/128 
for galaxies above the L22/128
resolution limit, counting as smooth accretion all the gas in galaxies
below this limit just as we do in our analysis of L22/128 itself.
We then calculate the fraction of this inferred smooth accretion
that is actually due to mergers with galaxies between the
$1.1 \times 10^8 M_\odot$ resolution threshold of L5.5/128 and the 
$6.8 \times 10^9 M_\odot$ threshold of L22/128.  We find that this
fraction is less than 7\% at $z=3$ and less than 3\% at $z=4$.
Our treatment of photoionization in these simulations
assumes ionization equilibrium (KWH).  If we included
non-equilibrium heating at the epoch of reionization
\citep{miralda94}, then our IGM temperature would be somewhat
higher, and galaxy formation would be physically suppressed at a higher
mass scale \citep{gnedin00}, further reducing the sub-resolution merger
contribution.

\begin{figure}
\centerline{
\epsfxsize=1.0\columnwidth
\epsfbox[20 120 590 720]{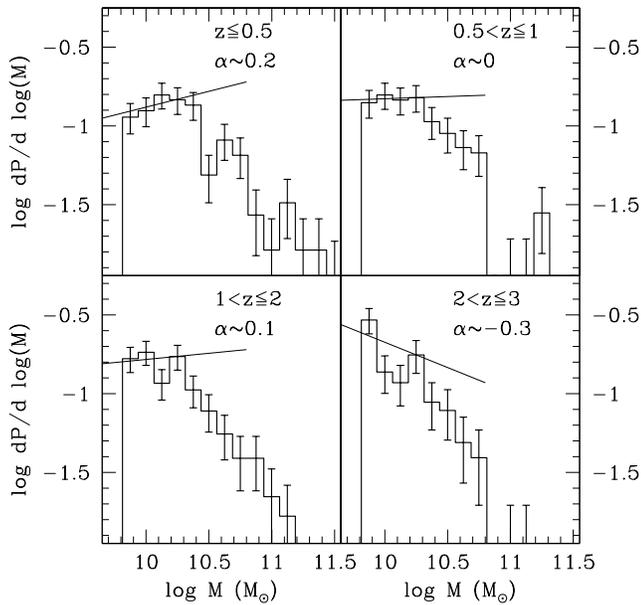}
}
\caption{
The mass spectrum of galaxies that merge with larger galaxies, in four
redshift intervals as indicated.  We fit power laws to the low ends of
these distributions as described in the text.  Straight lines show
these fits, which have logarithmic slopes $\alpha$ as listed.
}
\label{fig:merging_mf}
\end{figure}

At $z < 3$ we do not have such a high resolution simulation available, 
so we must use an indirect method to estimate sub-resolution merging.
Our approach is similar to that of MKHWD, based on extrapolating
the mass distribution
of resolved galaxies that merge with larger resolved galaxies.
This distribution is plotted in Figure~\ref{fig:merging_mf}, for the 
L22/128 simulation.
We consider all resolved
mergers that take place in four redshift intervals: $z\le 0.5$, $0.5<z\le1$,
$1<z\le2$, and $2<z\le3$.  We fit a 
power law $d P_M/d\log M \propto M^{\alpha}$ to the low mass
bins in each panel, where $d P_M/d\log M$ is
the probability per $d\log M$ 
for a galaxy of mass $M$ to merge with a larger galaxy.
We use the last five bins for the $z\leq 0.5$ distribution
and the last four bins for the other distributions.
We force the fits to go through the measured value in the fourth bin, 
so that we have just one free parameter, $\alpha$. 
Best-fit slopes range from 0.2 ($z \leq 0.5$) to $-0.3$ ($2 < z \leq 3$),
though the small number of bins and substantial Poisson error bars
make these estimates uncertain.
The mass contributed per $d\log M$ scales as 
$M^{\alpha+1}$, so high mass satellites dominate merger growth if
$\alpha > -1$, as we find at all redshifts.

Integrating our power law
fits from the resolution limit $\mgal = 6.8 \times 10^9 M_\odot$
down to $\mgal =10^8M_{\odot}$
yields an upper limit to the amount of sub-resolution merging,
since the merger mass distributions appear to be turning over steadily
and are likely to fall below our extrapolations.
With this estimate, the ratio of mass added to the galaxy
population by sub-resolution mergers to mass added by resolved 
mergers is approximately $0.1-0.15$ at $z=0-0.5$, $0.20-0.35$
at $z=0.5-1$, $0.25-0.45$ at $z=1-2$, and $0.5-2.0$ at $z=2-3$.
The quoted ranges reflect the $1\sigma$ statistical 
uncertainties in the fitted slopes.
Sub-resolution merging could thus make 
a significant contribution to {\it merger}
rates at higher redshifts, but because merger rates are much lower
than smooth accretion rates (see Figure~\ref{fig:SFR} below), it still 
does not substantially change the inferred accretion rates.
The exception is at $z=2-3$, where the maximal correction above
could raise the merger rate from 25\% of the smooth accretion rate
to 150\% of the smooth accretion rate.  In other words, up to half
of our estimated smooth accretion rate at $z=2-3$ 
could be due to sub-resolution merging.
However, at $z=3$ we have direct evidence from the comparison
of the L5.5/128 and L22/128 simulations that the unresolved
merger rate is only $\sim 7\%$ of the smooth accretion rate,
implying that our power law extrapolation greatly overestimates
the sub-resolution merging at high redshift, where the typical
galaxy mass is only a few times the resolution limit.
We hope to eventually redo our analysis
with higher resolution simulations that model larger volumes than 
L5.5/128 and continue all the way to $z=0$, but for now we conclude
that the quantity we estimate as the smooth accretion rate in 
the L22/128 simulation does represent truly smooth gas accretion,
with only a modest contribution from sub-resolution mergers.
We thus confirm the conclusion of MKHWD about the 
mechanisms of galaxy growth: typical galaxies gain most of their
mass through smooth accretion of gas rather than mergers with 
smaller galaxies.\footnote{This agreement is, however, somewhat 
fortuitous.  MKHWD applied the power law extrapolation method to
the L50/144nb simulation, but their Figure 9 incorrectly labels
the merger mass distribution as being per $dM$ instead of 
$d\log M$ (the quantity they actually calculated).
This translates to an error in the power law coefficient,
which caused MKHWD to underestimate the sub-resolution merging. 
If MKHWD had correctly applied their power law fit to their data,
they would have concluded 
that sub-resolution merging exceeds the true smooth accretion rates.
The higher resolution of the L22/128 simulation allows us to see
a turnover in the merger mass distribution, which is why we find
a small sub-resolution contribution from correct application
of the same method.}
Furthermore, the cold accretion process that is 
the focus of this paper is not a consequence of galaxies accreting
gas that has cooled in halos of low virial temperature.

\subsection{Time resolution\label{sec:tint}}

A second technical issue is the time interval between our analysis
outputs.  Since $\tmax$ is by definition the maximum temperature
of a particle during its pre-accretion history, our analysis
at discrete outputs necessarily yields a lower limit to the true $\tmax$
value of any particle.  The arguments of \cite{fardal01} imply that
we cannot be drastically underestimating $\tmax$ by missing rapidly
cooling particles, since these would then have high luminosities that
would cause the cooling radiation to emerge mainly at high temperatures,
which it does not.  However, we have also carried out a direct check
by repeating our analysis at several $z \geq 1$ outputs
using our standard time interval to determine which particles
are smoothly accreted but determining $\tmax$ values at
smaller intervals (by a factor $\sim 4$ at $z=3$,
$\sim 8$ at $z\sim 2$, and $\sim 20$ at $z=1$).
With this reanalysis, we find that
15-25\% of the smoothly accreted gas that was previously attributed to
cold mode accretion is actually hot mode accretion,
with the larger fraction applying at low redshift when the cold fraction
is itself small.
Further investigation shows that most of the ``missed'' hot particles
shock heat well inside the virial radius, where the density of the gas is
high and the cooling times are short,
so they are physically distinct from the ``classic'' hot mode
particles heated at the virial shock.
With higher time resolution, the $\tmax$ distribution of hot mode particles
also shifts towards slightly higher temperatures.
Note, however, that if
we use higher time resolution for all aspects of our analysis,
rather than just for computing $\tmax$ values, then we can 
actually get an increase in the cold mode fraction at higher redshift,
since the calculation now incorporates low mass galaxies that were
previously omitted because they were below the resolution limit
at the preceding redshift output.
Using the finer time resolution described above to identify galaxies and
compute $\tmax$ values yields a 
net increase in the cold mode fraction of 
30\% at $z\sim 1$, no significant change at $z \sim 3$, and
a small drop at higher redshifts.

\begin{table*}
\begin{tabular}{ccccc}
\hline Redshift&$\rho/\bar{\rho}=500$&$\rho/\bar{\rho}=1000$
&$\rho/\bar{\rho}=5000$&$\rho/\bar{\rho}=10000$ \\ 
\hline
$T=1\times10^6\K$\\
\hline
$z=1$&1030&513&103&51 \\ 
$z=2$&304&152&30&15 \\
$z=3$&128&64&12.8&6.4\\ 
$z=4$&65.6&33&6.6&3.3\\ 
\hline
$T=2\times10^6\K$\\ 
\hline 
$z=1$&2350&1175&235&118 \\
$z=2$&696&348&69.6&34.8 \\ 
$z=3$&294&147&29.4&14.7\\
$z=4$&150&75.2&15&7.5\\

\hline

\end{tabular}
\caption{Cooling times in Myr of gas at specified overdensity and 
redshift, for starting temperatures $T=10^6\K$ (top)
and $T=2\times 10^6\K$ (bottom). }

\end{table*}

We conclude that the finite time resolution in our analysis may
cause errors of $\sim 0.1-0.2$ in our estimates of global cold and hot 
accretion fractions.
This weak dependence on time resolution makes sense if we ask 
how far inside a halo a particle needs to be to get
shock heated and cool before the next output redshift, and 
thus be misidentified as cold accretion.
Our regular analysis interval is
$\Delta t\sim 1.0\,$Gyr at low redshifts and 
$\Delta t\sim 0.2\,$Gyr at high redshifts. 
Table~2 lists the time required for gas to cool from $10^6\K$ or 
$2\times 10^6\K$ (typical halo virial temperatures) down to $10^5\K$,
at various redshifts and overdensities.  We compute these times
by integrating over the actual cooling function. 
For our standard time interval, a particle heated to $2\times 10^6\K$
cannot cool to $T=10^5\K$ between two redshift outputs
unless its overdensity is greater than 500, corresponding to less
than 1/3 of the virial radius for an isothermal halo. For a particle heated 
to $T=10^6K$ the cooling time at overdensity 500 is comparable to our 
standard time interval, which means that the cooling time at the 
virial radius is much longer than the time interval between our simulation 
outputs.
For our finer time interval, a shock heated particle would only
cool between two outputs for an overdensity of several thousand,
requiring it to be very close to the galaxy's cold baryon component.

\subsection{Numerical Convergence\label{sec:numconv}}

\begin{figure*}
\centerline{
\epsfxsize=0.8\textwidth
\epsfbox[25 195 585 715]{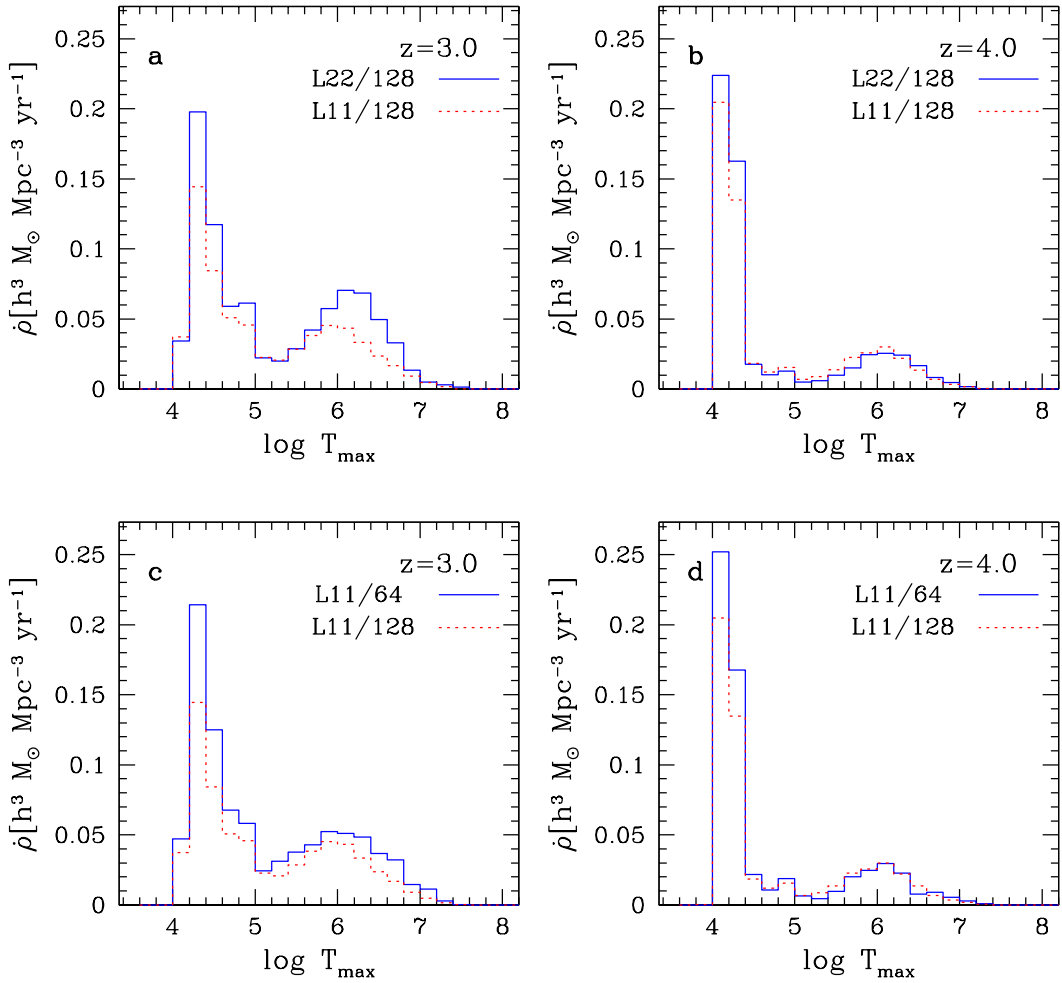}
}
\caption{
Dependence of the $\tmax$ distribution on numerical resolution.
Upper panels compare results from our standard, L22/128 simulation (solid
histogram)
to results from the higher resolution, L11/128 simulation (dotted histogram)
at $z=3$ (left) and $z=4$ (right).  Differences can be partly attributed
to resolution and partly to simulation volume.  Lower panels isolate the
resolution effect by comparing simulations L11/128 and L11/64, which have
the same initial conditions and volume and differ only in mass resolution.
}
\label{fig:resolution}
\end{figure*}

The most serious numerical concern is that our basic result,
the existence of a cold accretion mode that makes a major 
contribution to galaxy formation, is somehow an artifact
of the simulations' finite mass resolution.  
For example, numerical broadening of shocks might allow gas to
radiate energy while it is passing through the shock, instead of
first heating and then cooling \citep{hutt00}.
We can address this
issue by comparing simulations with different mass resolution 
to check the stability of the results.

The top panels of Figure~\ref{fig:resolution} compare the $\tmax$
distributions of gas accreted at $z=3$ and $z=4$ 
in the L22/128 simulation and the higher resolution, L11/128 simulation.
In L11/128, we only consider the accretion onto galaxies above the
$6.8 \times 10^9 M_\odot$ resolution limit of L22/128.  These objects
are resolved by at least 512 gas and star particles in L11/128.
Despite the factor of eight difference in mass resolution, the
two histograms agree nearly perfectly at $z=4$, and they show
only modest differences at $z=3$.  Since the simulation volumes are also
different, we expect some differences between these simulations
just because of the different structures they contain.
The lower panels of Figure~\ref{fig:resolution} compare the L11/128
and L11/64 simulations, which have identical initial conditions
(except for the additional high frequency modes in L11/128) and differ
only in mass resolution.  The change of simulation volume slightly
improves the agreement at $z=3$, but it appears that most of the
difference between L22/128 and L11/128 is an effect of resolution 
rather than simulation volume. 
Higher resolution lowers the overall
gas accretion rate by $\sim 30\%$. Some of the gas 
accreted in the lower resolution simulation is converted into stars within 
galaxies below the resolution limit of the lower resolution simulation, 
but that are resolved in the higher resolution simulation, before it is 
accreted onto a galaxy above the 512-particle threshold. 
This can explain a significant part of the differences in the cold mode 
accretion rates.
However, numerical resolution effects on the gas cooling are probably 
responsible for most of the differences in the hot mode accretion rates
and  the remaining cold mode differences. 
These numerical effects must depend on the galactic mass, since at $z=4$, where 
galaxies are significantly smaller on average, the differences are smaller.
Since the difference between L22/128 and L11/128 is fairly small and mainly 
attributable to mass resolution, we conclude that the volume of the L22/128 
simulation is large enough to give reasonably accurate statistics on the 
cold and hot accretion fractions, at least at $z=3$.

\begin{figure}
\centerline{
\epsfxsize=1.0\columnwidth
\epsfbox[25 155 365 710]{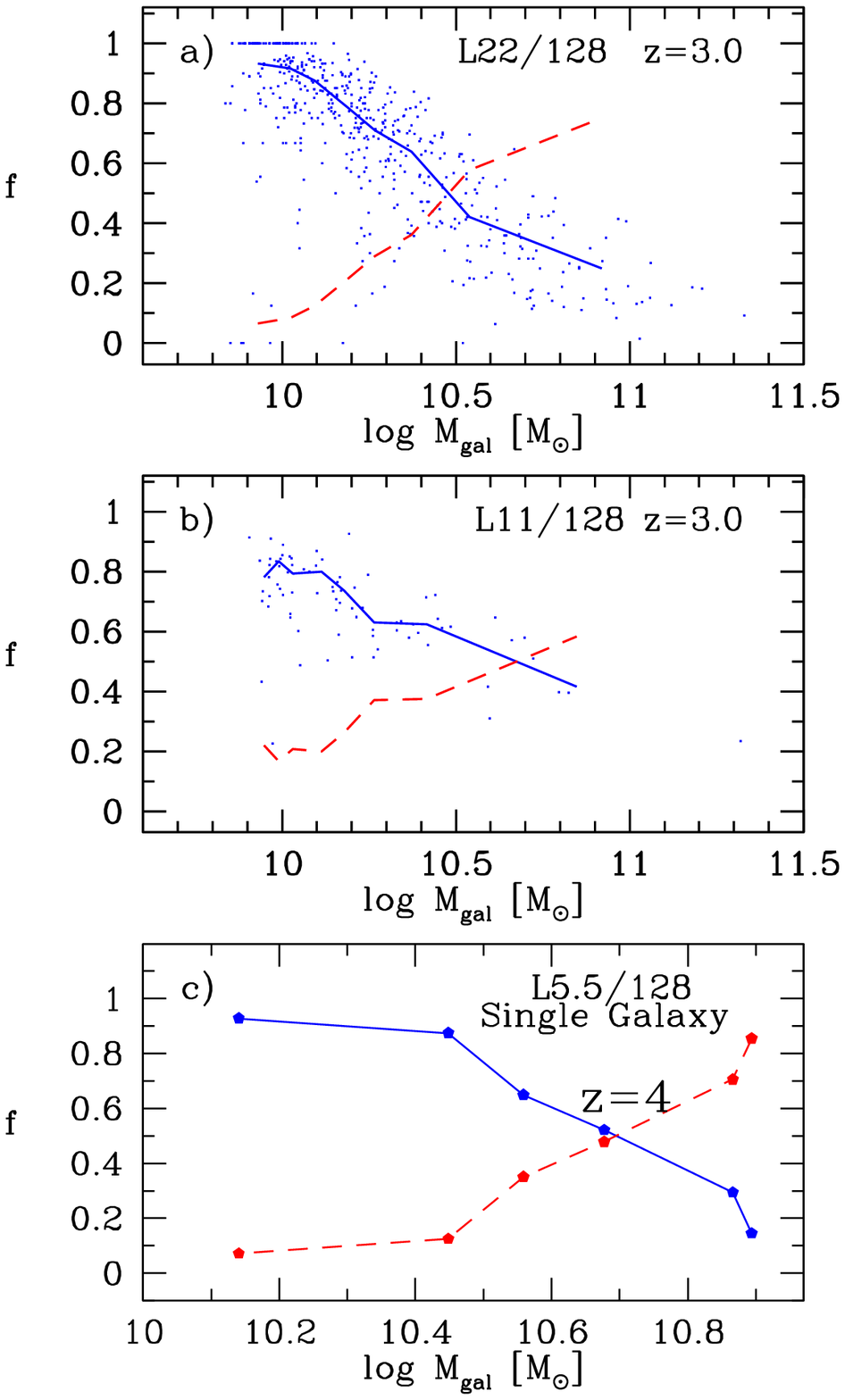}
}
\caption{
Resolution tests of the cold accretion fraction.
(a) Cold fraction vs. galaxy baryon mass for the L22/128 simulation
at $z=3$, repeated from Fig.~\ref{fig:mass_ratio}.
Points show individual galaxies, the solid line shows median cold 
fraction, and the dashed line shows median hot fraction.
(b) Same plot for the L11/128 simulation, at $z=3$.
(c) Cold fraction vs. galaxy baryonic mass for the single largest galaxy
in the L5.5/128 simulation at redshifts $z=5.5$, 5, 4.5, 4, 3.5, and 3
(each point is a redshift, with the highest redshift corresponding to 
the lowest mass).  Dashed line shows the hot fraction.
While the mass resolution of this simulation is 64 times higher than
that of L22/128, the crossover between cold mode domination and
hot mode domination occurs at nearly the same baryonic mass.
}
\label{fig:big_galaxy}
\end{figure}

Our most powerful test of numerical convergence comes from examining
the galaxy mass dependence of the cold accretion fraction, as shown
for the L22/128 simulation in Figure~\ref{fig:mass_ratio}.
Figure~\ref{fig:big_galaxy}a repeats the $z=3$ panel of 
Figure~\ref{fig:mass_ratio}, and Figure~\ref{fig:big_galaxy}b
shows the same result for the L11/128 simulation.  While the higher resolution
simulation contains few high mass galaxies because of its small volume,
the transition between cold and hot mode domination occurs at 
$\mgal=10^{10.7}\msun$, close to the transition in L22/128.
Our highest resolution simulation, L5.5/128, contains only nine galaxies
above the L22/128 mass resolution limit at $z=3$ (and only one at $z=4$),
so a comparison of similar form is not very useful.
However, we can check the mass dependence of
the cold accretion fraction by following the largest (and thus best
resolved) galaxy in the simulation volume. 
Figure~\ref{fig:big_galaxy}c plots the cold and hot accretion fractions
of this largest galaxy as it grows from
$\mgal=1.4 \times 10^{10}\msun$ ($8000 \msph$) at $z=6$ to 
$\mgal=7.4 \times 10^{10}\msun$ ($43,000 \msph$)
at $z=3$, the redshift of our last simulation output.
The transition between hot and 
cold mode dominance occurs close to the $z=4$ output, when the galaxy mass is
$\mgal=10^{10.7}\msun$.
This agrees well with the transition mass at $z=3$
found in L11/128 and L22/128
(and agrees even better with the $z=4$ transition mass, which 
is $10^{10.75}\msun$ for L22/128).
The second most massive galaxy in L5.5/128 does not reach
the cold/hot transition mass until $z=3$, the final output,
so it is not useful for this test.
The agreement in transition mass over simulations that span a
factor of 64 in mass resolution is the strongest single piece
of evidence that the importance of cold accretion in our simulations
is not an artifact of limited resolution.

\subsection{Influence of the UV Background\label{sec:UV}}

\begin{figure*}
\centerline{
\epsfxsize=0.8\textwidth
\epsfbox[25 475 600 715]{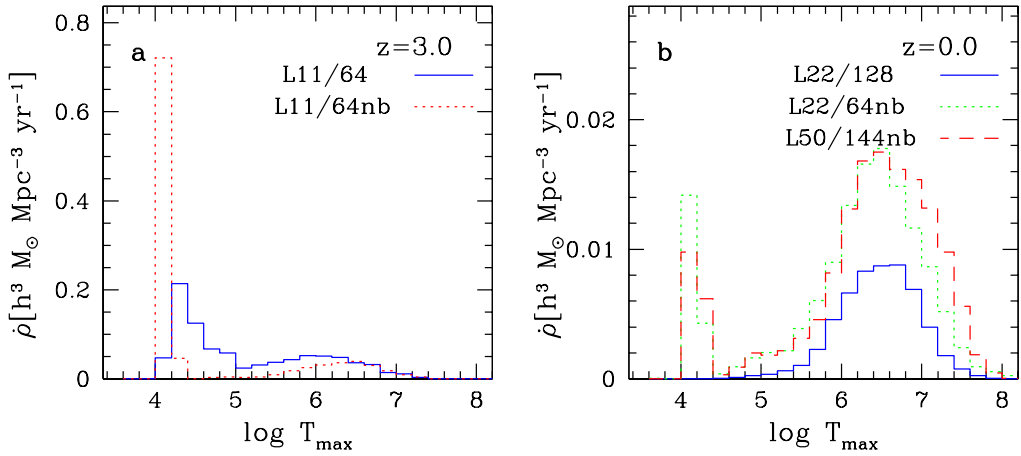}
}
\caption{
Influence of numerical resolution, simulation volume,
and the UV background on the
$\tmax$ distribution.  
(a) Comparison of L11/64 and L11/64nb, which differ only in the presence
of a UV background, at $z=3$; 
(d) Volume effect comparisons of L22/128, L22/64nb and L50/144nb at $z=0$.
}
\label{fig:UV_noUV}
\end{figure*}

Figure~\ref{fig:UV_noUV}a illustrates the impact
of the UV background on $\tmax$ distributions, comparing results from
L11/64 and L11/64nb at $z=3$.
The no-background simulation shows a somewhat reduced hot mode, 
and the cold accretion is sharply peaked at $\tmax \approx 10^4\K$. 
The UV background is the only difference between these two simulations. 
The different amplitude and shape of the histogram at low $\tmax$
is therefore driven by the difference in input physics,
most likely the fact that photoionization reduces the strength
of the collisional line peak in the cooling curve at moderate overdensity
(see KWH, Fig.~2).  The transition mass at which cold and hot 
accretion rates are equal is consistently higher in simulations with 
no UV background, increasing the global contribution of cold accretion.

L22/128 is the highest resolution simulation we have evolved to $z=0$.
All of our lower resolution simulations have been evolved without a UV
background, because photoionization at low mass resolution
spuriously suppresses galaxy formation \citep{weinberg97}.
Figure~\ref{fig:UV_noUV}b compares the $z=0$ $\tmax$ histograms from
L22/128, L22/64nb, and L50/144nb.
We now consider only the accretion onto galaxies above the 
64-particle threshold of the lower resolution simulations,
$\mgal = 5.4 \times 10^{10} M_\odot$, 
which is higher than the cold/hot transition mass of 
$\sim 3 \times 10^{10} M_\odot$ found for L22/128 at $z=0$
(see Fig.~\ref{fig:mass_ratio}).
As a result of this higher threshold mass, the low temperature
end of the $\tmax$ histogram for L22/128 is strongly suppressed
relative to the corresponding histogram in Figure~\ref{fig:diff_acc_ns}.
The no-background simulations, by contrast, show a significant
amount of cold accretion, with $\tmax$ values peaked just 
above $10^4\K$.  Although the simulations differ in both
numerical parameters and input physics, the similarity
to Figure~\ref{fig:UV_noUV}a implies that the UV background
is the cause of this difference.

The L22/64nb and L50/144nb simulations also show significantly
more hot accretion at $z=0$ than L22/128; a similar effect is
seen at $z=3$ (not plotted).  Here we suspect
that the coarse resolution of the no-background simulations
is to blame, since photoionization has minimal effect on cooling at
high temperatures and the differences are similar to those
seen in Figures~\ref{fig:resolution}a and~\ref{fig:resolution}c
(at $z=3$).  Finally, we note that 
the reasonably good  agreement between L22/64nb and L50/144nb suggests
that the $22\hmpc$ box is large enough to give representative
statistics for the cold/hot fraction even at $z=0$, except perhaps at very
large $\tmax$.

\subsection{Other simulations \label{sec:othersim}}

Within our own set of simulations, we find that the existence of an
important cold accretion mode is insensitive to mass resolution or
simulation volume, though the specifics of the $\tmax$ distribution
are significantly affected by the presence of a UV background.
SPH simulations of the formation of individual galaxies achieve
still higher resolution, with thousands of particles in the central
object, and they also show a large fraction of cold accretion
\citep{katz91,abadi03}.  Further evidence of the robustness of
the result comes from simulations of volumes like those considered here
performed with entirely independent SPH codes.  \cite{kay00}, 
studying a $2\times 32^3$ particle simulation of a $10\hmpc$ box
performed by the Virgo consortium, find that only $\sim 11\%$ of
the gas accreted by their simulated galaxies by $z=0$ was ever
hotter than $10^5\K$.  We have recently analysed a simulation,
kindly provided by L.\ Hernquist and V.\ Springel,
that uses Springel \& Hernquist's
(\citeyear{springel02}) entropy conserving formulation of SPH,
which makes the numerical treatment of shock heating significantly
different from that in our code.  This simulation, run Q3
of \cite{springel03b}, represents a $10\hmpc$ box with $2\times 144^3$ 
particles, and is evolved to $z=2.75$.  
The simulation also includes strong feedback from stellar winds,
using the methods described by \cite{springel03a}.  Despite these
differences in numerical method and input physics, the
\cite{springel03b} simulation produces a $\tmax$ histogram at high
redshifts that closely resembles those shown in 
Figure~\ref{fig:diff_acc_ns}, with a large fraction of gas
accretion occurring at low temperatures.
The transition between cold mode dominance and hot mode dominance
occurs at a halo mass similar to that found in 
Figure~\ref{fig:mass_ratio_halo}.  However, the strong feedback
makes galaxy baryonic masses significantly smaller in the
\cite{springel03b} simulation than in our simulation (numerical differences
may also play a role), so the baryonic mass at which this transition
occurs is lower than that in Figure~\ref{fig:mass_ratio} by
a factor $\sim 4$. In sum, galaxy masses and the amount of accretion are lower
 in the 
\citet{springel03b} simulation, but the existence of a cold accretion mode
and the relative amounts of cold and hot accretion at a given halo mass
appear entirely consistent with our results.

These internal and external comparisons strongly suggest that the bimodal
$\tmax$ histogram is a genuine physical result, at least given the
physical assumptions of these simulations.  Since the results quoted 
above are all based on SPH simulations, albeit with different
numerical implementations, it would be desirable to confirm the result
with Eulerian grid or adaptive mesh hydro simulations, which 
typically use shock capturing methods instead of artificial viscosity
to implement shock heating.  \citeauthor{cen99} (1999, Fig.~4) find
a broad temperature distribution for cooling radiation in their
Eulerian hydro simulation, suggesting that much of the gas in this
simulation also cools without ever reaching $T \geq 10^6\K$.  
A. Kravstov (private communication), in adaptive mesh simulations
of individual forming galaxies, finds that much of the gas penetrates
along cold filaments far inside the virial radius without being
shock heated.  These simulations cannot presently measure the $\tmax$
histogram itself because they do not track gas from cell to cell and 
therefore do not record the history of gas that ends up in galaxies,
but this measurement can be carried out in the future by adding
a population of ``tracer'' particles that follows the gas flow
(Kravtsov, private communication).

\section{The Environment Dependence of Accretion and Star Formation Rate}
\label{sec:environment}

\begin{figure}
\centerline{
\epsfxsize=1.0\columnwidth
\epsfbox[20 140 590 720]{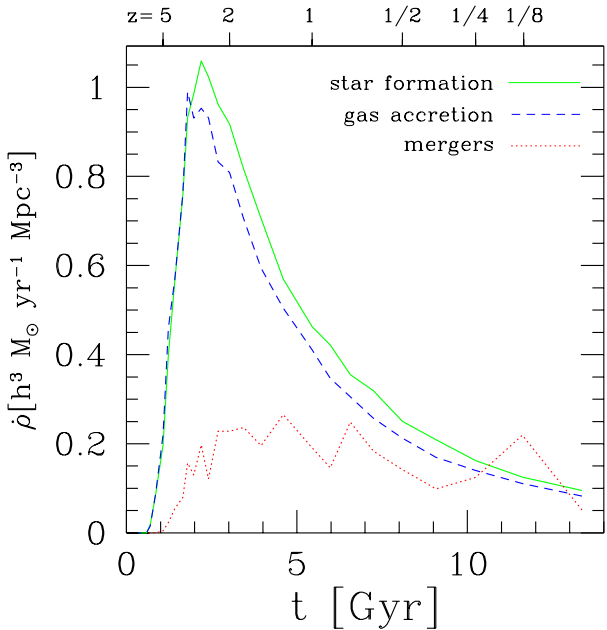}
}
\caption{Star formation rate per unit comoving volume 
(solid line) for all resolved
galaxies in the simulation compared to smooth gas accretion (dashed line) and
merger accretion rates (dotted line). The merger mass accretion rates 
include total mass gain in merger events (gas and stars).}
\label{fig:SFR}
\end{figure}

Figure~\ref{fig:SFR} plots the globally averaged star formation rate
(SFR) and the globally averaged contributions of smooth accretion and
mergers to galactic mass growth, as functions of time in the L22/128
simulation.  At high redshifts the smooth accretion rate dominates the
merging growth rate by a large factor,
while at $z < 1$ these two processes have  
comparable global rates. Of course, galaxies that gain most of their mass via 
mergers do exist in the simulation, but the growth of the typical galaxy is  
dominated by smooth accretion.
MKHWD found that the global star formation history in the L50/144nb 
simulation closely tracks the gas accretion history and does {\it not}
track the galaxy merger rate. 
Figure~\ref{fig:SFR} confirms this conclusion
using our higher resolution, L22/128 simulation --- the correlation between
the accretion rate and SFR curves is even tighter than the one found
by MKHWD.  The two curves are offset by $\sim 0.5$ Gyr, reflecting the average
time that it takes gas to be converted into stars after it is accreted
by a galaxy.  However, with our observationally motivated assumption
(e.g., \citealt{kennicutt98})
that the star formation rate is an increasing function of gas density,
any steadily accreting galaxy will form stars, and mergers are not needed
as triggering events.  
Indeed, the global star formation rate must roughly track the global
accretion rate for any star formation model in which galaxies do 
not build up large, quiescent gas supplies.
Mergers could play a role in causing rapid
bursts of star formation by driving galactic gas to higher densities,
and our simulations may underestimate this effect because of their 
limited resolution. However, such merger-induced acceleration would
have little effect on the globally averaged SFR because the merging
galaxies simply consume the same amount of gas on a somewhat shorter
timescale.

\begin{figure*}
\centerline{
\epsfxsize=0.8\textwidth
\epsfbox[35 280 560 680]{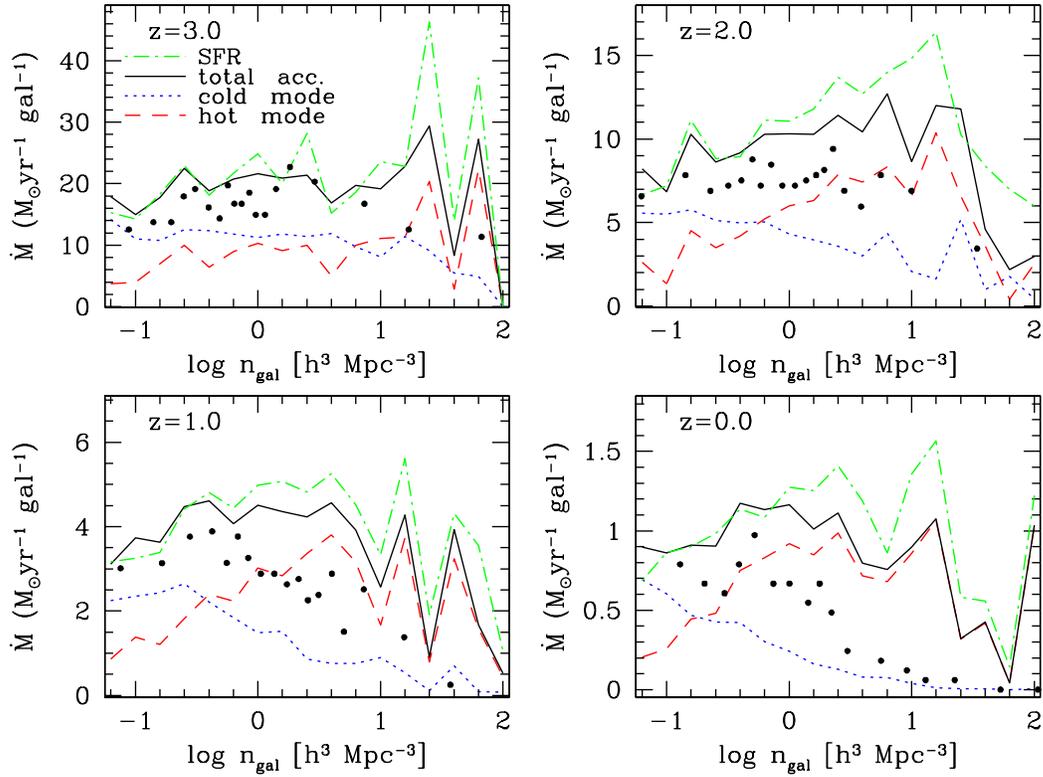}
}
\caption{
Dependence of accretion rates and star formation rates on local galaxy
number density (comoving), at $z=3$, 2, 1, and 0.  Solid lines show
the mean total accretion rates in bins of $\Delta \log n=0.2$, and
dotted and dashed lines show the mean cold and hot contributions,
respectively.
Points show the median total accretion rate in bins containing equal
galaxy numbers.  Dot-dashed lines show mean star formation rates, which
track the total accretion rate.
}
\label{fig:dens_ratio}
\end{figure*}

If our input physics is approximately correct, then understanding the
redshift and environmental dependence of the cold and hot accretion rates
is tantamount to understanding the redshift and environmental dependence
of galactic scale star formation.  
In this section we focus on the environmental dependence; we return to the
redshift dependence
in \S~\ref{sec:disc_sf} and \S~\ref{sec:disc_feedback}.
Figure~\ref{fig:dens_ratio} plots the mean SFR and accretion rate
per galaxy as a function of environment, specifically the comoving
space density of resolved galaxies computed using SPH-style kernel averaging
on a galaxy-by-galaxy
basis with a variable size spline kernel whose size is chosen to
enclose 15 neighbours (as in \citealt{weinberg04}).
The SFR and accretion rate track each other as a function of density,
just as the global rates track each other as a function of redshift.
At $z=3$, the average accretion rate is nearly independent of environment,
but at lower redshifts there is a break towards lower accretion rates
and star formation rates at $n_{\rm gal} \sim 1\; h^3\;{\rm Mpc}^{-3}$.
Points in Figure~\ref{fig:dens_ratio} show the median 
accretion rate, computed in bins containing equal numbers of galaxies
(while the mean is computed in bins of constant $\Delta \log n = 0.2$).
Comparing the medians and means, one can see that the accretion rate
in high density regions is dominated by the few galaxies with large
accretion rates; most galaxies in these regions
have very low accretion and correspondingly low star formation.

The dotted and dashed lines in 
Figure~\ref{fig:dens_ratio} show the separate contributions of cold
and hot accretion to the mean accretion rate.  There is a clear trend,
increasingly pronounced at lower redshift, of cold mode domination
in low density environments and hot mode domination in high density
environments.  The transition between the two modes shifts to progressively
lower comoving densities at lower redshifts.  Cold mode dominates in
all but the highest density environments at $z\sim 3$, but at $z=0$
it dominates only in the lowest density environments and is insignificant
at $n_{\rm gal} \ga 1\dunits$.  These trends can be largely explained
by the galaxy mass dependence of the cold/hot accretion fraction:
low density regions are populated by 
lower mass galaxies, which tend to have more cold accretion.
However, when we examine the trends separately for low and high mass
galaxies, we find that there is also a direct environmental effect,
in that low mass galaxies are often hot mode dominated in high density
environments, while they are always cold mode dominated in low density
environments.  This difference reflects the
higher characteristic {\it halo} masses in dense environments; as noted in
our discussion of Figures~\ref{fig:mass_ratio} 
and~\ref{fig:mass_ratio_halo}, low mass galaxies in high mass halos
tend to be hot mode dominated.
The rare high mass galaxies in low density environments
are still hot mode dominated, but they tend to have somewhat higher
cold accretion fractions than those in denser regions.

We can compare our $z=0$ predictions to the results of 
Gomez et al.\ (\citeyear{gomez03}, hereafter G03), who use
H$\alpha$ equivalent widths to infer the star formation rates
of a volume limited sample of galaxies from the SDSS.
To mimic their selection of galaxies above an absolute magnitude
limit $M_r^*+1$, we choose galaxies above a baryonic mass threshold
that yields the same galaxy number density, roughly the 120 most 
massive galaxies in our $22.222\hmpc$ box.  
Like G03, we calculate the surface density around
each galaxy by finding the distance $r_{10}$ to the tenth nearest
neighbour in projection, $\Sigma = 10/(\pi r_{10}^2)$, after eliminating
galaxies further than $\pm 1000\kms$ in redshift (making use of
our periodic boundary).  To improve our statistics, we combine
the results from the three orthogonal projections of the box.

\begin{figure*}
\centerline{
\epsfxsize=0.75\textwidth
\epsfbox[25 355 590 705]{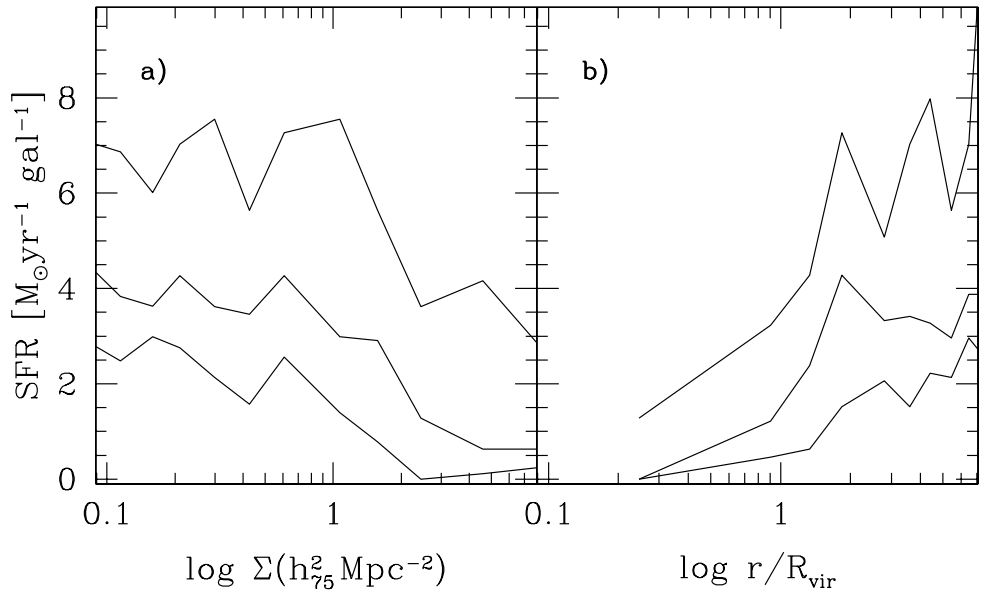}
}
\caption{(a) Star formation rate vs. galaxy surface density
at $z=0$.  Curves show the 25th-percentile, median, and 75th-percentile
of the SFR per galaxy in bins of surface density.
(b) Same as (a), but galaxies are binned as a function
of distance from the centre of the largest cluster in the L22/128
simulation, in units of the cluster's virial radius.
}
\label{fig:comp_sdss}
\end{figure*}

Figure~\ref{fig:comp_sdss}a plots the median star formation rate and the
25th and 75th percentiles of the SFR distribution in bins of galaxy
surface density containing equal number of galaxies, similar to Figure~4 
of G03.  
The simulation reproduces the principal qualitative result of G03,
namely a break in the SFR vs. $\Sigma$ relation at 
$\Sigma \sim 1 \surfdunits$, most prominent for the 75th-percentile curve,
with a slowly increasing SFR below the break and a rapidly declining
SFR at higher surface densities.  
G03 also examine the dependence of SFR on the distance to the centre of the
nearest group or cluster.  They emphasise that the break in SFR occurs
at $3-4$ times the cluster virial radius, implying that it is not an
effect associated with virialization or the intracluster medium.
We have just one prominent cluster in the L22/128 simulation, with
virial mass  $2.7\times 10^{14}\msun$.  
Figure~\ref{fig:comp_sdss}b plots SFR against projected 
distance to the centre of this cluster (defined by the position of
the most bound particle), in units
of the cluster virial radius $R_{\rm vir} = 1.1\hmpc$, for comparison
to G03's Figure~6.  The breaks in the simulated SFR curves also occur
beyond the cluster virial radius, at roughly $2R_{\rm vir}$.  
At least part of the drop at large radii is caused by the presence
of a large galactic group at distances between 1 and 2 $\rvir$, which 
dominates the statistics.
A more detailed comparison would require a larger simulation volume that would
allow us to mimic G03's group selection method more carefully.
Since the SFR tracks the gas accretion rate in our simulations,
a drop in SFR within high density regions at low redshifts is a direct 
consequence of a diminished gas supply and does not require cluster specific
processes that either rapidly consume the existing gas 
(e.g. galaxy harassment) or remove it from galaxies (e.g., ram pressure 
stripping).

G03 note that their SFR values should
be multiplied by a factor $\sim 5$ to account for the $3"$ aperture
of the SDSS fibers, but even so they are significantly lower than the
rates predicted by the simulation.  For example, the plateau of the
75th-percentile SFR curve is at $\sim 2\accgalunits$ in G03
(including the factor of five) and $\sim 5\accgalunits$ in
the simulation.  The 25th-percentile curve is consistent with zero SFR
at all $\Sigma$ in G03, while the simulation predicts 25th-percentile
rates $\sim 1.5\accgalunits$ in low density environments.
G03 include negative estimates of the 
SFR derived from  $H_{\alpha}$ flux in their median calculation, 
which slightly lowers their SFR values, but probably not by a large factor.
Limited mass resolution of our simulation may have some effect in
making the predicted SFR and accretion rates
artificially high (see \S~\ref{sec:numconv}).
In any event, we
are encouraged that the simulation reproduces one of the most striking
features of the observed correlation between SFR and environment, a break
at surface densities $\sim 1\surfdunits$ that lie beyond the virial
radii of rich galaxy groups, a result found independently
by G03 in the SDSS and by \cite{lewis02} in the 2dF Galaxy Redshift
Survey.  (The SDSS results of \citeauthor{kauffmann04} [\citeyear{kauffmann04}]
also appear consistent with this finding, although they do not present
them in this form.)  This break occurs in a density regime 
where the accretion rates are dominated by the hot mode
(see Fig.~\ref{fig:dens_ratio}), and galaxies above G03's $M^*_r+1$
threshold are mostly hot mode dominated in any case.
Thus, although these observations test
some aspects of the simulation predictions, they do not test the
existence of an important cold accretion mode or the transition of the dominant
accretion mode from cold to hot with increasing galaxy density.
Such effects should become more apparent if the galaxy sample is extended to
include lower luminosity systems.

\section{Discussion}
\label{sec:discussion}

\subsection{Physics of the Hot Mode \label{sec:disc_hot}}

In our simulations, hot accretion ($\tmax > 2.5\times 10^5\K$) dominates
over cold accretion in galaxies of baryonic mass
$\mgal \ga 10^{10.3}\msun$ (Fig.~\ref{fig:mass_ratio}) 
or in halos with 
$\mhalo \ga 10^{11.4}\msun$ (Fig.~\ref{fig:mass_ratio_halo}).
Roughly half of the total baryonic mass of high mass galaxies
($\mgal \ga 10^{11}\msun$) originates from hot accretion,
while lower mass galaxies are built predominantly by cold accretion
(Fig.~\ref{fig:cold_vs_mass}).
The hot accretion mode in the simulations corresponds reasonably well
to the accretion envisioned in the standard picture of galaxy formation
that underlies most semi-analytic models: gas shock heats to roughly
the halo virial temperature before cooling and settling into a galaxy
\citep{rees77,silk77,white78,white91}.

The actual distribution of $\tmax/\tvir$ for hot mode accretion is
broad, roughly an order of magnitude (Fig.~\ref{fig:tvir}).
This broad distribution reflects a combination of departures from 
spherical symmetry, hierarchical assembly, and post-virialization heating.
With aspherical infall, gas can hit an accretion shock with a range
of velocities and thus heat to different temperatures.  Some gas may
be shocked to $\tmax < \tvir$
in filamentary structures beyond the virial radius, then
enter the halo without experiencing a true virial shock.  When an
individual galaxy falls into a group or cluster,
some of its halo gas is heated to the higher virial temperature of the 
new halo and escapes the galaxy potential well, but other gas remains at the
galaxy's virial temperature and may be accreted before it is heated 
or stripped.  Since we calculate the ratio of $\tmax$ to the virial
temperature of a galaxy's host halo at the time of accretion, 
hot mode gas accreted in this way will also have low $\tmax/\tvir$.
Finally, gas that heats by 
an accretion shock near the virial radius can be
heated further by adiabatic compression as it flows inwards 
and increases its density.  We find that a significant fraction 
of the hot mode gas has $T$ significantly below $\tvir$ (but still 
much higher than the typical cold mode temperature) in the outer regions of 
the halo and does not reach $\tvir$ until it is inside $\sim 0.5 \rvir$.

Figure~\ref{fig:rhot_physical} shows the same $\rho-T$ trajectories
as Figure~\ref{fig:rhot}b, but now the horizontal axis represents
physical density (scaled to the mean baryon density at $z=3$) rather than
overdensity.  In this plane, adiabatic compression moves a particle
along a line $\log T \propto (2/3)\log \rho$.  In the diffuse IGM phase,
particles move to lower densities and temperatures as the universe
expands, along a locus of slope close to 0.6.  Hot mode particles then
experience sudden shock heating, which is often followed by 
an extended phase of heating along a nearly adiabatic trajectory, before
eventually cooling to $10^4\K$ and high overdensity.
The fact that shock heating is spread over several outputs
(which are typically spaced by $\sim 0.05-0.1$ Gyr) is a consequence
of the finite numerical width of the shocks.
We will discuss the behaviour of cold mode particles in this plot below.

We have also examined the distribution of cooling times
in halos at various redshifts.  
If we calculate the amount of hot gas that would cool between two
outputs in the absence of further heat input, we get a value that
exceeds the actual hot mode accretion rate by a factor $\sim 2$.
Therefore, adiabatic compression or further shock heating must
be sufficient to prevent roughly half of this gas from cooling.

\subsection{Physics of the Cold Mode \label{sec:disc_cold}}

\begin{figure}
\centerline{
\epsfxsize=1.0\columnwidth
\epsfbox[30 165 565 690]{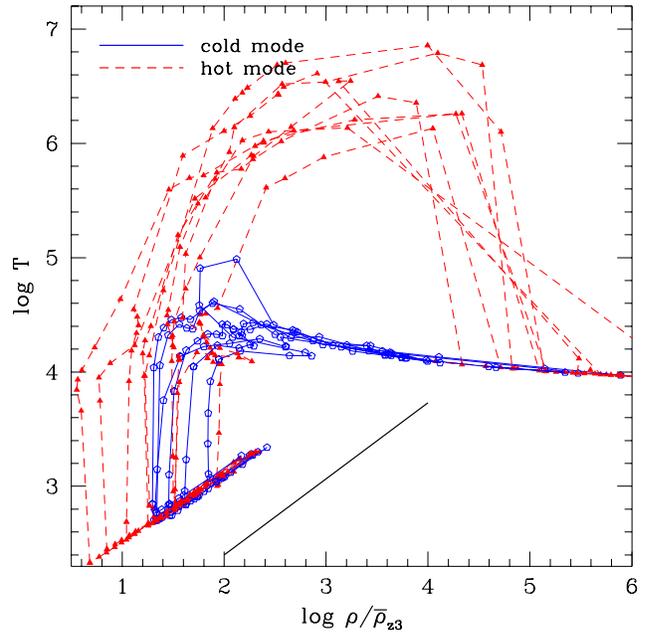}
}
\caption{Density-temperature trajectories for the same particles
shown in Fig.~\ref{fig:rhot}b, but with the horizontal axis representing
physical density (scaled to the cosmic mean density at $z=3$) instead
of overdensity.  The line segment at the bottom has the slope 
$d\log T /d\log\rho = 2/3$ expected for pure adiabatic evolution.
}
\label{fig:rhot_physical}
\end{figure}

Why does a substantial fraction of gas enter galaxies without
ever heating close to the virial temperature of the galaxy's parent
dark matter halo?  Our results do not provide a comprehensive answer
to this question, but they do offer a number of clues that allow us
to build a physical picture of how cold accretion occurs.
As discussed in \S\ref{sec:results}, the division between
hot and cold modes occurs at roughly constant $\tmax$ rather than
constant $\tmax/\tvir$, which suggests that it is determined mainly
by cooling physics rather than by shock heating.  
The strong dependence of the $\tmax$ distribution for
cold mode gas on the presence of a UV background
(Fig.~\ref{fig:UV_noUV}) further supports this inference.
The ratio
of cold to hot accretion is tightly correlated with galaxy mass
or halo mass, and the transition between cold mode dominance and
hot mode dominance occurs close to (though somewhat above)
the halo mass predicted by the
spherically symmetric models of \cite{birnboim03}.  The ratio of
cold to hot accretion also depends on the density of the environment,
at least at low redshift, and although this dependence largely
reflects the relative galaxy mass functions in high and low density
regions, low mass galaxies are dominated by hot accretion in high density
environments and by cold accretion in low density environments.

The trajectories of particles in the $\rho-T$ plane offer further insight
into the physics of cold accretion.  Figure~\ref{fig:rhot}b shows that
typical cold mode particles, after an initial phase of adiabatic cooling,
experience most of their heating at overdensities $\sim 3-30$, then
steadily increase their overdensity while cooling from slightly above
$10^4\K$ to $10^4\K$.  Figure~\ref{fig:rhot_physical}, 
plotted in physical density
rather than overdensity, shows that the heating phase of these trajectories
is very steep, implying that it is produced by shocks not by adiabatic 
compression.  The characteristic shock velocities must be 
$v \sim (kT/\mu m_p)^{1/2} \sim 10 - 35\kms$ ($T \sim 10^4-10^5\K$) and
are probably around filaments.

\begin{figure*}
\centerline{
\epsfxsize=0.8\textwidth
\epsfbox[30 165 595 410]{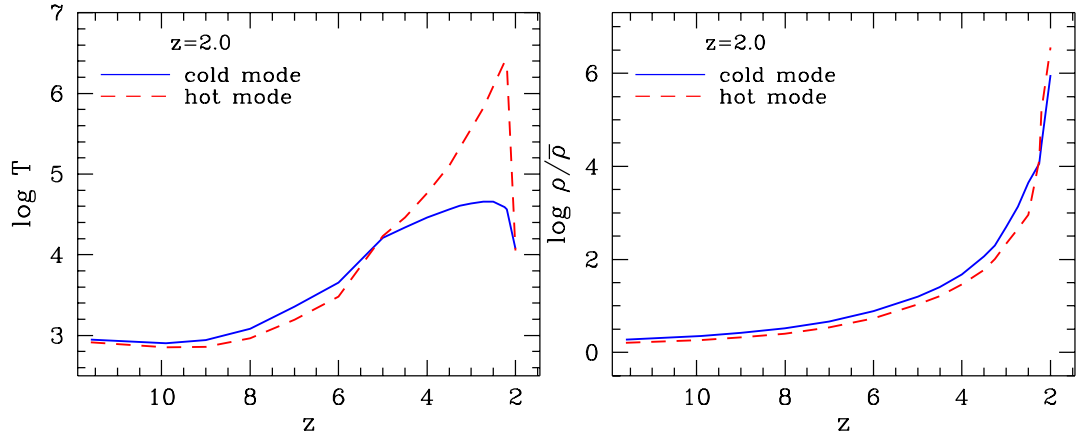}
}
\caption{The average temperature (left panel) and density (right panel)
evolution of cold mode (solid line) and hot mode (dashed lines) particles
that are accreted onto galaxies at $z=2$.
}
\label{fig:avgrhot}
\end{figure*}

To better understand the origin of the hot/cold distinction, it is 
useful to look for differences in the physical state of gas {\it before}
it is accreted in one of these two modes.  Figure~\ref{fig:avgrhot}
plots the average overdensity and temperature histories of gas that
is accreted onto galaxies at $z=2$.  Solid lines show histories for
cold mode particles and dashed lines for hot mode particles.
Averaging over many particles tends to smooth the density and temperature
evolution, since individual particles experience their rapid heating phases
at different times.  The important result
in Figure~\ref{fig:avgrhot} is that cold mode particles initially
have higher overdensities than hot mode particles, at least on average.
Because of the correlation between density and temperature in 
the photoionized IGM (Fig.~\ref{fig:rhot}), cold mode particles also
start at slightly higher temperature.  The differences are fairly
small, and become somewhat larger if we focus only on the lower-$\tmax$
cold mode particles.  The general appearance of
Figure~\ref{fig:avgrhot} is similar at other redshifts.

\begin{figure*}
\centerline{
\epsfxsize=0.76\textwidth
\epsfbox{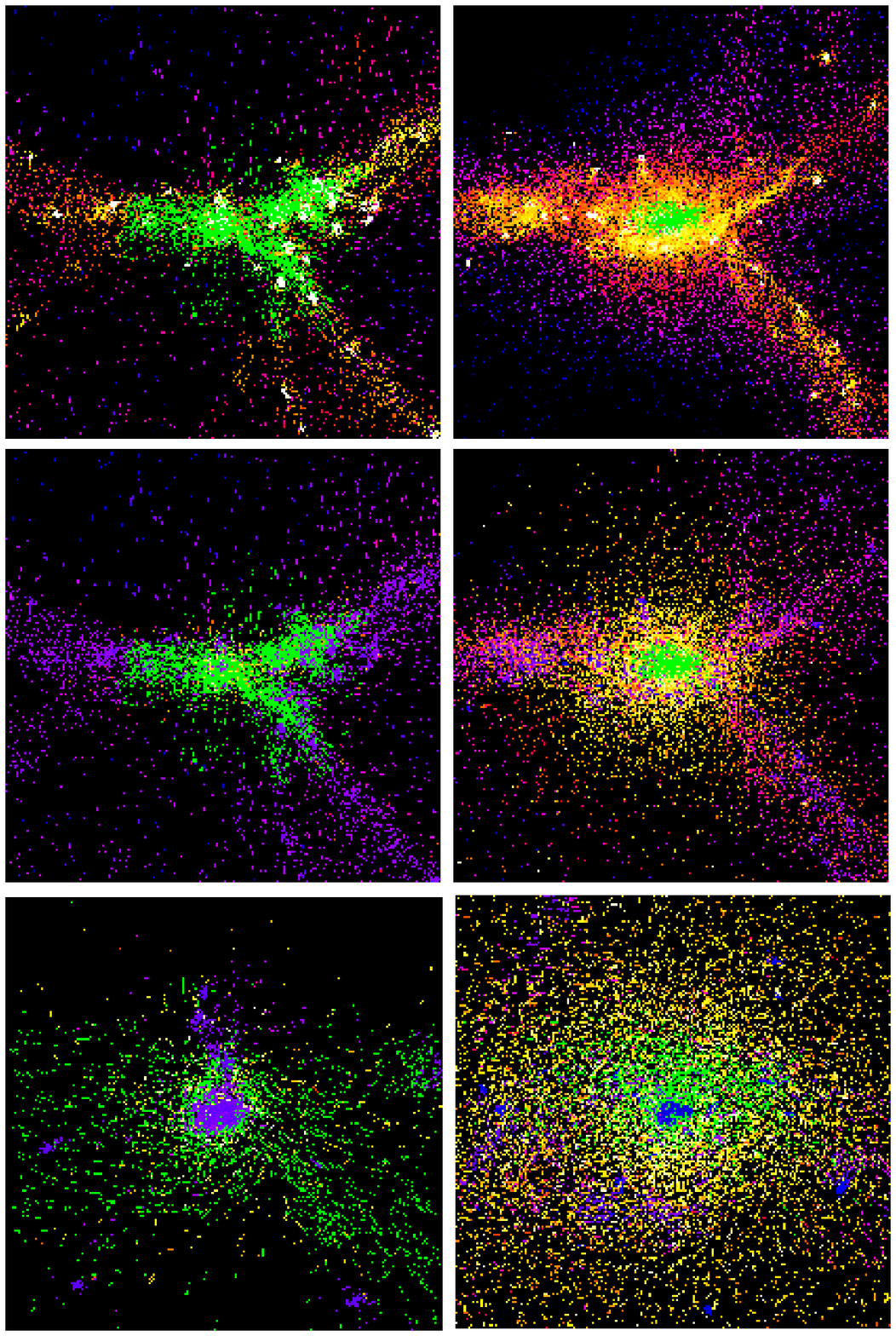}
}
\caption{
Accretion pattern around the largest galaxy in the L5.5/128 
simulation.  Left hand panels show redshift $z=5.52$, when the galaxy baryonic
mass is $1.6\times 10^{10}\msun$
and accretion is almost entirely dominated by cold mode.  Green 
particles in each panel represent gas that will be accreted in cold mode 
by $z=5$.  Right hand panels show redshift $z=3.24$, when the galaxy
mass is $7.35 \times 10^{10}\msun$ and accretion is almost entirely
dominated by hot mode.  Green particles in these panels show gas
that will be accreted in hot mode by $z=3$.
Other particles in the top panels are colour coded by overdensity,
from 0.67 (darkest blue) to $>2000$ (white), in a region $4\rvir$
on a side ($78 \hkpc$ and $202 \hkpc$ in physical units at $z=5.52$
and 3.24, respectively) and $2\rvir$ thick.
Other particles in the middle panels are colour coded by temperature in the 
same region. Temperature scale starts from the lowest temperature 
(darkest blue) $1500\K$ on the left and $3000\K$ 
on the right panels, to the highest temperature (white); $5\times 10^6\K$ 
on the  left and $10^7\K$ on the right panels.
Bottom panels have the same temperature colour-coding but zoom in 
to show a region $1\rvir$ on a side and $0.5 \rvir$ thick.
Vectors attached to particles show their projected velocities.
}
\label{fig:filament}
\end{figure*}

Since overdensities $\sim 3-30$ are associated with filamentary structures,
while halo overdensities are $\ga 100$, these $\rho-T$ trajectories
suggest that cold mode particles experience much of their
heating in filamentary structures, while hot mode particles are heated
at halo virial shocks.  Figure~\ref{fig:filament} provides anecdotal
evidence to support this idea. 
The left panels show an example of a cold mode dominated halo
($\mhalo=2.6 \times 10^{11} \msun$) containing a well resolved galaxy
($\mgal = 1.6 \times 10^{10} M_\odot \approx 10,000 m_{\rm SPH}$) 
in the L5.5/128 simulation at $z=5.52$.  
The right panels show this same halo and its surroundings at $z=3.24$,
when it is well into the hot mode dominated regime
($\mhalo=1.26 \times 10^{12} \msun$, $\mgal=7.35 \times 10^{10} \msun$). 
Particles are colour coded by overdensity in the upper panels and
by temperature in the middle panels, all of which are $4\rvir$ on 
a side and $2\rvir$ thick.  The lower panels show a zoomed region of the 
temperature plot, $1\rvir$ on a side and $0.5 \rvir$ thick.  Lines 
attached to particles show their projected velocities.  In the left 
hand panels, green particles represent
cold mode gas that will be accreted by the central galaxy by $z=5$, 130 Myr
later.  In the right hand panels, they show hot mode gas that will
be accreted by $z=3$, 190 Myr later.  

The $z=3.24$ picture corresponds reasonably well to the conventional
notion of galactic gas accretion, with accreting material coming
from the inner regions of a hot, quasi-spherical halo.  However,
even at this redshift
cold filaments penetrate fairly far inside the halo, where they heat up to
higher temperatures.
The cold mode accretion at $z=5.52$, by contrast, is clearly directed
along the intersecting filaments, and this coherent filamentary
flow allows the galaxy to accrete gas from larger distances.
A.\ Kravtsov (private communication) finds similar behaviour in adaptive
mesh simulations of an individual galaxy: the forming galaxy is 
surrounded by a hot gas halo, but cold filaments with coherent flows
penetrate far inside this halo.  The cluster scale simulations
of \cite{nagai03} do not include gas cooling, but they still show
filamentary structures in which the gas entropy is far below that of the
surrounding, spherically distributed gas.

\begin{figure*}
\centerline{
\epsfxsize=0.8\textwidth
\epsfbox[30 155 590 410]{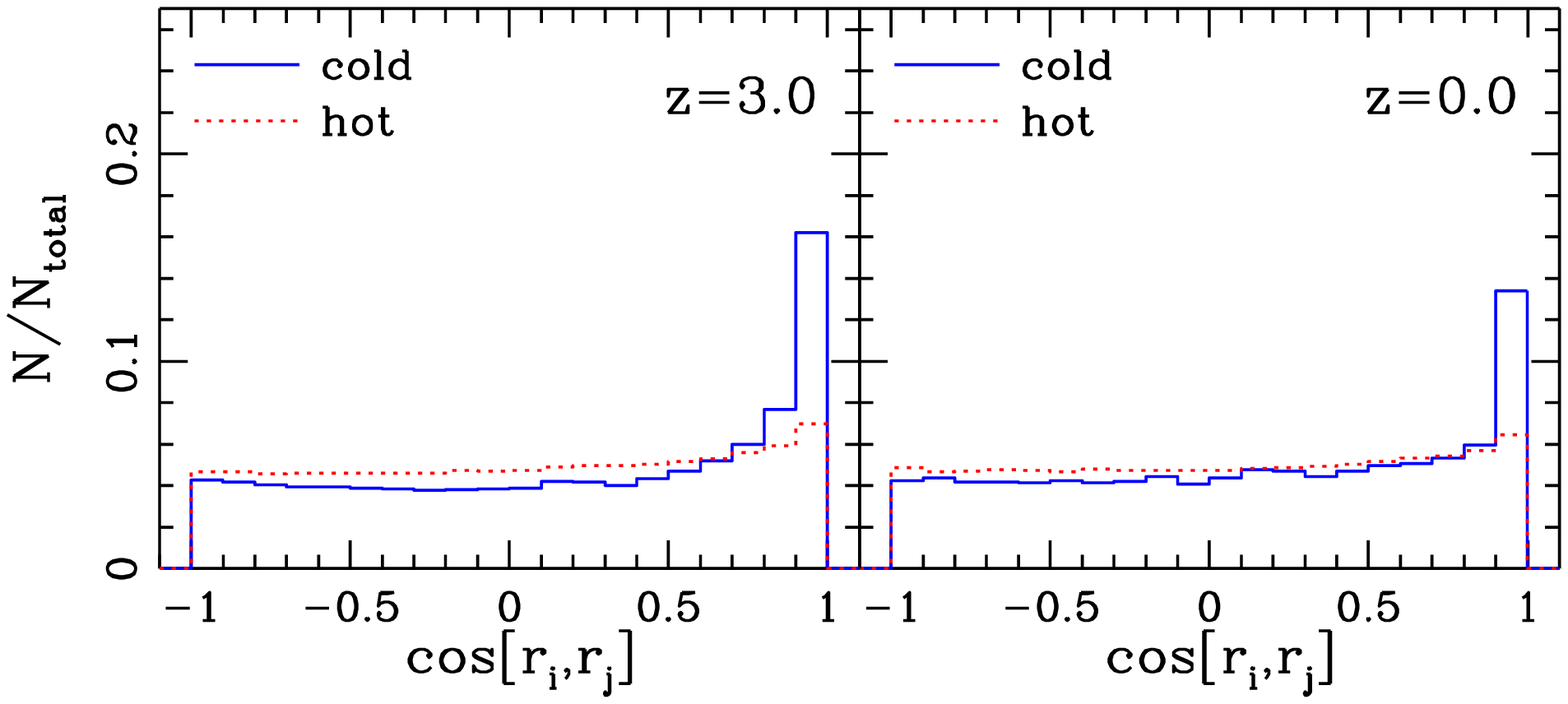}
}
\caption{The distribution of the normalised scalar products of the radius
vectors of accreting particles (see text) at the indicated redshifts.
The dotted histogram is hot mode and the solid histogram is cold mode.}
\label{fig:scalar_product}
\end{figure*}

To quantify the evidence for filamentary cold accretion, we applied
the following test.
First, at the redshift output before the accretion output we find the radial
vectors ${\bf r}^g_i$ connecting the centre of each resolved galaxy $g$
to the positions of all particles $i$ that it will accrete.
Next we determine the normalised
scalar product $\hat{{\bf r}}^g_i \cdot \hat{{\bf r}}^g_j = \cosij$
for pairs of particles that accrete onto the same galaxy $g$.
We compute the distribution of $\cosij$ separately for all pairs of
hot mode particles and all pairs of cold mode particles, summing the
distributions for all galaxies $g$ and normalising them to unity.
Figure~\ref{fig:scalar_product} shows the results at $z=3$ and $z=0$.

Hot mode particles show a nearly uniform distribution of $\cosij$, which
is expected for a statistically isotropic accretion pattern.
Cold mode particles, by contrast, show a strong peak at $\cosij \approx 1$,
indicating accretion from preferred directions, i.e., along filamentary
structures.  The size of this peak underestimates the fraction of 
accretion that is filamentary, since particle pairs accreting from
different filaments that feed the same galaxy produce $\cosij$ peaks
at other angles, and averaging over the full galaxy population turns
these multiple peaks into a uniform spread.  Also, filaments are not
perfectly straight, and they have a finite geometrical cross section,
both of which tend to spread the $\cosij$ values.
Nonetheless, this test shows a clear statistical difference in the
geometry of cold and hot accretion, consistent with the visual 
impression of Figure~\ref{fig:filament}.  If we remake these plots
separately for halos above and below $M = 2\times 10^{11}\msun$,
we find that the directional signal in the cold accretion mode
is stronger for high mass halos and weaker for low mass halos,
while hot accretion remains approximately isotropic in both regimes.
Some of the difference between low and high mass halos could be
numerical in origin, but we believe that it primarily reflects
a more isotropic nature of cold accretion in low mass systems.

From these clues, we can build the following picture of cold accretion.
As argued by \cite{binney77} and \cite{birnboim03},
accreting gas with short cooling times does not form
a classic virial shock --- its effective equation of state is too soft,
and it passes through the virial radius essentially in free fall.
In the spherically symmetric models of \cite{birnboim03}, there is
a sharp transition between low mass halos that do not have a virial 
shock and high mass halos that do, with the approximate condition
being that a virial shock arises if the post-shock cooling time would be
larger than the dynamical time.  
The important new ingredient highlighted by the simulations is the
filamentary nature of cold accretion.  Departures from spherical 
symmetry allow galaxies to have a mix of cold and hot accretion at
one time, so the sharp transition of the \cite{birnboim03} models
is replaced by a steady change from cold mode domination at low mass
to hot mode domination at high mass.  
A low mass galaxy has no virial shock anywhere, so it can
accrete cold gas quasi-spherically, though filaments still channel
much of the flow.  An intermediate mass galaxy has a virial shock
everywhere except at the points where dense filaments of cold gas
penetrate the virial radius.  These galaxies have a mix of cold, filamentary
accretion and hot, quasi-spherical accretion.  Gas in denser regions of
the IGM is more likely to join a filament before accreting onto a galaxy,
explaining the difference of density histories in Figure~\ref{fig:avgrhot}.
Finally, high mass galaxies have a virial shock everywhere and have only
hot accretion.  

The characteristic halo mass at which we find equal cold and hot
accretion rates, $M\sim 3\times 10^{11} M_\odot$, is a factor
$\sim 2-3$ larger than the transition mass calculated by
\cite{birnboim03}.  However, \cite{dekel05} show that the derived
transition mass depends on the precise criterion they use, 
in particular the radius at which they require a shock to form.
For low metallicity gas, they find a transition mass 
$\sim 1.5 \times 10^{11} M_\odot$ for a shock at $0.1 \Rvir$
and $\sim 10^{12} M_\odot$ for a shock at $\Rvir$, though the
addition of a UV background (as used here) would lower cooling
rates and probably reduce these masses.  Given the radical 
difference in calculational methods, the agreement between the
SPH simulations and the spherically symmetric models is remarkably
good, pointing to similar underlying physics.

Semi-analytic models of galaxy formation distinguish between accretion
with long post-shock cooling times, in which the cooling radius is 
smaller than the halo virial radius, and accretion with rapid 
post-shock cooling, in which the cooling radius exceeds the virial
radius \citep{white91}.  In both cases, the cooling time is 
calculated assuming shock heating to the virial temperature,
but in the latter case the gas is assumed to accrete onto the
central galaxy at the free-fall rate instead of first forming
a hot atmosphere in quasi-hydrostatic equilibrium.  We have calculated
the halo mass that marks the transition between these two cases
for our adopted cosmology and cooling function, including the
effects of the UV background.  We calculate cooling radii assuming
that halo gas follows an NFW density profile \citep{nfw96}
reduced by $\Omega_b/\Omega_m$, and we take typical concentrations as
a function of redshift from \cite{bullock01}.
If we define the cooling radius to be the radius at which the cooling
time equals $t(z)$, the age of the universe at redshift $z$,
then we find a transition mass 
$\sim 7 \times 10^{10} M_\odot$ at $z \ga 1.0$, roughly 2 to 4 times
lower than the transition mass found in our simulations.
At $z < 1.0$ there is no transition mass at all; in the presence of
a UV background, the cooling radius never exceeds the virial radius
at the lower densities that prevail in low redshift halos.
If we use either the freefall time or average dynamical time of
the halo to define the cooling radius then the transition mass
at high redshift is reduced by another factor of two to four,
and there is no transition mass at $z \la 3$ and $z \la 2.5$.
Thus, while the cooling radius transition in existing 
semi-analytic models bears a qualitative resemblance to 
the cold-hot transition identified in this paper, 
a standard calculation of this transition yields a result
that is quantitatively different and underestimates
the significance of cold accretion (see the Appendix for
details of our calculation).
The transition mass goes
up if one ignores the UV background (thus raising the cooling rates)
or assumes an $r^{-2}$ gas density profile (thus raising the
density at the virial radius), but the first of these assumptions 
ignores a physical process present in the simulation,
and dark matter profiles, at least, are much better described
by NFW profiles than by isothermal spheres.
The cooling radius approach
also fails to capture the filamentary nature of cold accretion,
and because the cooling radius is calculated assuming virial
temperature gas, it implicitly predicts that most of the emission
from even rapidly cooling gas will be at the X-ray wavelengths
characteristic of these temperatures, while in the simulations
this gas radiates much of its energy in atomic line transitions,
hydrogen Ly$\alpha$ in particular \citep{fardal01}. 
The exact amount  of radiated Ly$\alpha$ emission, however,
needs further investigation, since it 
could be partially affected by the numerical broadening of 
shocks.

One question remains: how does cold mode gas actually enter a galaxy?
Our ability to answer this question is limited because our simulations
do not have good enough resolution at the relevant scales, but we know that
gravitational infall to galaxy mass potential wells produces typical
inflow velocities of $\sim 100-300\kms$, and something must happen to 
the corresponding kinetic energy.  (The simulation code conserves energy,
so it cannot be lost to numerical effects.)
We can identify three possibilities, all of which may
operate to some degree.  First, gas could penetrate inside the virial
radius but be stopped by a strong shock close to the galaxy disk,
heating to the virial temperature and converting its infall energy to
thermal energy.  The post-shock cooling time is then very short because
the gas density near the disk is high.  Some particles in our simulations
exhibit exactly this behaviour, and A. Kravtsov's adaptive mesh simulations
also show shock heating far inside the virial radius at the terminations of
cold filamentary flows (A.\ Kravtsov, private communication).
The lower left panel of Figure~\ref{fig:filament} shows some examples of such 
particles, with temperatures near $\tvir$ in a quasi- spherical region of 
several kpc around the central galaxy.

We formally count such particles as hot mode accretion, since they have
a high $\tmax$, but physically they are more analogous to cold mode 
particles, since they do not shock heat near the virial radius.
Because of our finite time resolution, we miss the heating events for some
of these particles and actually count them as cold mode.
However, we think it is unlikely that {\it most} of the particles
we identify as cold mode experience this kind of strong shock heating
near the galaxy disk, in part because improving the time resolution
makes only a small difference to the fraction of cold mode particles
(\S\ref{sec:tint}) and in part because of the cooling radiation
arguments of \cite{fardal01}: the total amount of energy radiated
by accreting gas is of order the gravitational binding
energy, and a large fraction of this energy emerges from low temperature
gas. However, it is possible that high density accretion shocks near the 
forming disk are more common in the real universe and
are missed in our simulations because of their finite spatial
resolution. 

A second possibility is that cold mode gas smoothly merges onto the
galaxy disk like a stream of cars entering an expressway, converting
its infall velocity to rotational velocity.  This idea may seem
outlandish at first, but the disk acquires its angular momentum
from the accreting gas, so the required alignment of the disk with the 
accretion flow may arise naturally.  The large scale filamentary structure 
around galaxies persists for long periods of time, especially at high
redshifts where the galaxies are highly biased with respect to the dark matter.
A close examination of galaxies in our highest resolution
simulation provides some evidence for this expressway mechanism,
as one can see visually in the lower left panel of Figure~\ref{fig:filament},
but further analysis will be needed to assess its overall importance.
In any event, the virial theorem suggests that this mechanism could
provide only part of the energetic solution, since the galaxy's rotational
energy should be smaller than its gravitational binding energy.
(Note, however, that the virial theorem cannot be applied with precision
to non-isolated, time-dependent systems like forming galaxies, and the
bookkeeping is further complicated by the fact that dark matter dominates
the gravitational potential.)

\begin{figure*}
\centerline{
\epsfxsize=0.8\textwidth
\epsfbox[30 155 590 410]{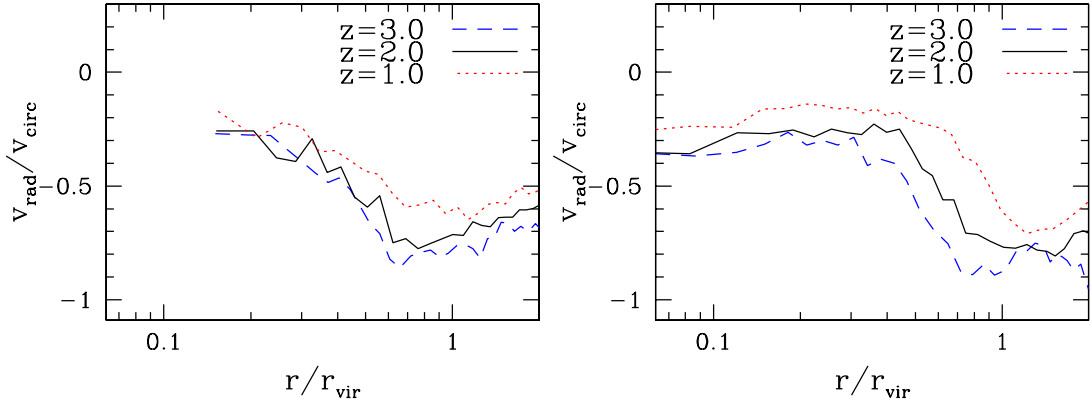}
}
\caption{Left: The median radial velocity of cold mode gas 
before it is accreted as a function of galactocentric radius.
Radii are scaled to the halo virial radius, $r_{\rm vir}$, and
velocities to the halo circular velocity, $v_{\rm circ}$.
Results are shown at $z=1.0$ (accreted by $z=0.0$, dotted), $z=2$ (accreted by 
$z=1.0$, solid), and $z=3$ (accreted by $z=2.0$, dashed).  We only consider 
gas accreted onto the central galaxies of their parent dark matter halos.
Right: Same for hot mode particles.}
\label{fig:velocities}
\end{figure*}

The third possibility is that infalling cold mode gas is slowed either
adiabatically (i.e., by pressure gradients) or in a series of shocks
that are individually too weak to cause strong heating.  
The left panel of Figure~\ref{fig:velocities} plots the average radial infall 
velocity of cold mode gas, scaled to the halo circular velocity,
as a function of $r/r_{\rm vir}$. We only consider cold mode gas that will 
be accreted onto the central galaxies by the next indicated
redshift. We define the central galaxy as the most massive galaxy in the halo, 
which is almost always the galaxy closest to the halo center.
We show results at $z=3$ (dashed line, accreted by $z=2$), $z=2$ (solid line,
accreted by $z=1$) and $z=1$ (dotted line, accreted by $z=0$).
Near the virial radius, the
average infall velocity is similar to the halo circular velocity,
but starting at $r \sim 0.5\Rvir$ the gas decelerates smoothly,
and by $r \sim 0.1\Rvir$ the average infall velocity is only 10-20\%
of the circular velocity.  The mean infall pattern is similar 
at all three redshifts, with the trend that the normalised velocities are
lower at lower redshifts.  There is a large scatter about the mean velocity,
so some accreting particles must either shock heat near the disk or
join the rotational flow as discussed above. 
However, based on Figure~\ref{fig:velocities}, we tentatively conclude 
that deceleration
by pressure gradients or weak shocks is the primary mechanism by
which cold mode gas loses its infall velocity, at least in the simulations,
radiating its dissipated energy as it slows. 
Artificial cooling of gas in numerically broadened shocks 
(see \S\ref{sec:numconv}) could have some effect on this result,
at least in its quantitative details.

In contrast, the right panel of Figure~\ref{fig:velocities} shows the 
situation for hot mode particles. We see that hot mode accretion 
more closely resembles the standard picture where the gas slows
rapidly at a shock near the virial radius 
However, the zone where the gas slows down is
significantly inside the virial radius at $z=3$ (at about 0.6$\rvir$) and 
moves steadily towards $\rvir$ by $z=0$.  Furthermore, this zone is fairly 
broad, and the
post-shock gas still has a mean radial velocity that is a significant
fraction of $\vcirc$, especially at high $z$.
These complications reflect the departures from spherical symmetry
and equilibrium discussed in \S\ref{sec:disc_hot}. Although hot accretion 
shows no significant filamentary signal in Fig~\ref{fig:scalar_product},
from visual inspection we see that even in some of the biggest halos at high
redshift a large fraction of the accreted gas comes from filaments. Filaments
in these halos are hot (with temperatures well above the cold mode limit)
but with temperatures much lower than $\tvir$. These filaments and mergers 
can deliver hot gas (as well as cold) well inside
$\rvir$, with significant radial velocity.  Halos at low redshift
have better developed virial shocks and clearer separations from
their filamentary surroundings, and the standard picture becomes
a better approximation.

Further numerical investigations should significantly improve our
physical understanding of cold accretion.  \cite{birnboim03}
and \cite{dekel05} argue that the physical criterion that determines
whether a virial shock forms in their spherically symmetric calculations
is the ratio of the cooling time to the ``compression time'' under
post-shock conditions.  In future work, we will investigate whether
this condition works as a {\it local} criterion for strong shock heating
in our simulations.  Our comparison of simulations with and without
a UV background already points to an important role for cooling rates, and
comparison of simulations with primordial and metal-line cooling and 
with different choices of $\Omega_b$ and $\Omega_m$ will help unravel
the factors that determine the cold/hot transition mass.
Most valuable will be very high resolution simulations that can accurately
follow cold accretion streams all the way into galaxy disks,
preferably performed with both SPH and adaptive mesh methods.

\subsection{Connection to Global Star Formation and Galaxy Structure}
\label{sec:disc_sf}

As shown in Figure~\ref{fig:SFR}, the cosmic star formation rate drops
by a factor of $\sim 11$ from $z=3$ to $z=0$.  Unfortunately,
this star formation history cannot be directly compared to observations
because no one of our current simulations has the dynamic range to include all
the galaxies that contribute significantly to the global star formation
rate at all redshifts.  In a forthcoming paper (Fardal et al.,
in preparation), we carefully combine the results from several simulations
that span a range of volume sizes and resolutions to determine a global
rate that can fairly be compared with the observational results.
We find that the
global SFR is approximately constant from $z=6$ to $z=2$, then
drops by a factor of 12 from $z=2$ to $z=0$.  The simulation predictions are
consistent with the observations within their scatter, but this is not
a stringent test because the observational estimates themselves
span a wide range of values at any one redshift.

Since the history of the cosmic star formation rate in our simulations
closely tracks the history of gas accretion, as shown in Figure~\ref{fig:SFR},
to understand the evolution of the cosmic star formation rate it is
sufficient to understand the evolution of the smooth gas accretion rate.
From Figure~\ref{fig:history} we see that both the cold
and hot accretion rates decline at $z<2$, with the cold mode decrease
starting earlier and occurring more rapidly.  
As already discussed in \S\ref{sec:results}, the transition in the
global accretion rate from 
cold mode dominated at high redshift to hot mode dominated at low redshift
simply reflects the increasing mass scale of galaxies and halos,
since low mass systems are cold mode dominated and high mass systems are 
hot mode dominated at every redshift.  However, the drivers of the
declining accretion rate are somewhat different in the two cases.

\begin{figure}
\centerline{
\epsfxsize=1.0\columnwidth
\epsfbox[20 140 590 710]{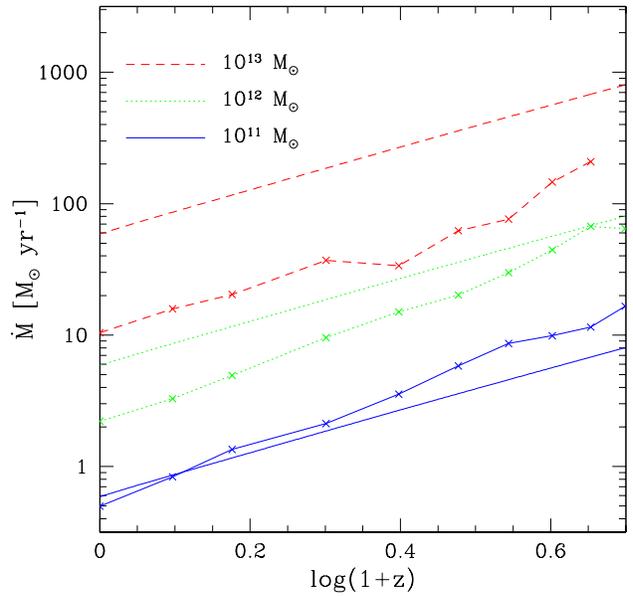}
}
\caption{
The average gas accretion rate onto the central galaxies of halos with 
$\log \mvir/\msun = 11 \pm 0.1$ (solid lines),
$\log \mvir/\msun = 12 \pm 0.15$ (dotted lines), and
$\log \mvir/\msun = 13 \pm 0.15$ (dashed lines),
as a function of redshift.  Lines marked with crosses show results
from the L22/128 simulation, while smooth, unmarked lines show 
the analytical infall models described in the text.
}
\label{fig:rates}
\end{figure}

First let us concentrate on the drop in the global cold mode accretion rate.
In principle this could reflect a decrease in the space density 
of cold mode galaxies or a decrease in the amount of cold mode accretion per 
galaxy, or a combination of the two.  
The rapid rise seen in Figure~\ref{fig:history} reflects the increasing
space densities of galaxies above our mass resolution threshold; most
of these galaxies are cold mode dominated at high redshift.  However,
while the space density of cold mode galaxies peaks at $z\sim 2$, it
declines only very slowly thereafter.  Most of the decline in cold
mode accretion must therefore be driven by the declining accretion
rate per galaxy, as shown directly in 
Figure~\ref{fig:rates}, which plots the average
gas accretion rate onto the central galaxy in halos of different
masses as a function of redshift.
For halos in the mass range
$\log M_{\rm vir}=11\pm 0.1 \msun$, which is cold
mode dominated at every redshift, the mean accretion rate per
galaxy drops by a factor of $\sim 12$ between $z=2$ and $z=0$,
accounting for the drop in the global cold mode accretion rate.

The smooth solid curve in Figure~\ref{fig:rates} shows the prediction of
a simple analytical model along the lines of \cite{white91}: the accretion
rate is just $\dot{M} = f_g \dot{M}_{\rm vir}$ where $f_g$ is the universal
gas fraction (i.e., the universal baryon fraction minus the fraction
of baryonic mass in stars) and $\mvir=(4\pi/3)\rvir^3\rhovir$ is
the virial mass.
The change in virial mass $\dot{M}_{\rm vir}$ is determined through
the cosmological growth of the virial radius assuming an isothermal profile,
fixed circular velocity and the cosmological evolution of $\rhovir$.
For $10^{11}M_\odot$ halos, this model explains
the simulated accretion rates quite well.
One can view the steady decline of $\dot{M}$ as a consequence of the
declining density at the virial radius, which is tied to the declining
mean density of the Universe,
or of the corresponding increase in
characteristic dynamical times --- the time for a halo to double its
mass scales with the age of the Universe.

At high redshifts the analytical model underpredicts the accretion
rate by a factor $\sim 2$, for at least two reasons.
First, much of the accretion is filamentary, allowing halos to draw
material from beyond the virial radius and accrete more efficiently
than a spherical model predicts. 
Second, halos in this mass range are only $\sim 2$ times our resolution
limit, and we are biased towards including 
halos that have higher accretion rates because we only count halos
that have already formed a resolved central galaxy. 
This effect operates at all redshifts, but it is  more important
at high redshifts when galaxies have had less time to grow, causing
us to omit lower $\dot{M}$ galaxies from our average. 
In addition, our limited mass resolution may play some role in overestimating 
accretion rates. 
When we repeat the $\dot{M}$ analysis for the higher resolution, L11/128
simulation at $z=3$ and $z=4$, the mean accretion rate is lower and agrees
even better with the analytical model, though the rate is still higher by
$\sim 20\%$ at $z=3$ and $\sim 40\%$ at $z=4$. 

Crosses connected by dotted and dashed lines in Figure~\ref{fig:rates}
show the mean accretion rates for the central objects
of more massive halos, with 
$\log M_{\rm vir}/\msun = 12\pm 0.15$ and
$\log M_{\rm vir}/\msun = 13\pm 0.15$, respectively.
Halos of $10^{13}\msun$ are dominated by hot mode accretion at
all of the plotted redshifts, and halos of $10^{12}\msun$ are
dominated by hot mode at all but the highest redshifts
(see Figure~\ref{fig:mass_ratio_halo}).  Unmarked lines show the corresponding
predictions of the infall model described above; 
these correspond to the $10^{11}\msun$
model curve (solid line) multiplied by factors of 10 and 100.
In these higher mass halos, gas must cool before it can accrete 
onto the central galaxies, so it is not surprising that the pure
infall model now overpredicts the galaxy accretion rates.
However, the infall model still captures the redshift dependence of
these rates remarkably well, especially if one ignores the 
highest redshifts, where the number of $10^{13}\msun$ halos is
small and the $10^{12}\msun$ halos have a significant cold mode
contribution.  It is as if cooling just places a constant
``tax'' on the newly available halo gas, allowing an almost redshift
independent fraction of the gas to cool at a given halo mass.
This fraction decreases from $\sim 0.45$ for $10^{12} \msun$ halos
to $\sim 0.20$ for $10^{13} \msun$ halos, as one might expect given
the higher virial temperatures and correspondingly longer cooling times
in more massive halos.

To model the cooling effects in simpler terms, we have followed an
analytical prescription along the lines of
\cite{white91}, assuming spherical halos with isothermal
density and temperature profiles. We suppose that all gas that is able to cool
accretes instantaneously onto the central galaxy. 
The accretion rate is, therefore,
$\dot M = 4\pi r_{\rm cool}^2 \rho_{\rm cool} dr_{\rm cool}/dt$, where
$\rho_{\rm cool}$ is the gas density at cooling radius and the cooling radius,
$r_{\rm cool}$, is the radius where $t_{\rm cool}=t_H$ and
$t_H$ is the Hubble time. Therefore, the increase of the cooling
radius in time defines the accretion rate.
This model predicts accretion rates about 30\%-60\% lower
than we find for $10^{12}$ and $10^{13}\msun$ halos, but it
reproduces the evolution of these rates fairly well.

At least part of the difference in magnitude of accretion rates may be a result
of our limited numerical resolution.
\citet{springel02} find that the accretion rates can be overestimated in
formulations of
SPH like those that we use here when there are not enough gas particles in the
galaxy halo.  We have repeated the above analysis
for the L11/128 and L5.5/128 simulations (at $z=3$ and $z=4$) and find that the
accretion rates drop by
40\%-50\% for $10^{12}\msun$ halos. However, each of these
simulations has only one halo in this mass range, so this result is
only suggestive.
Further work using simulations with higher dynamic range will be needed
to fully elucidate the mechanisms that drive the evolution of hot mode
accretion.

With these results in mind, we can build the following interpretation of
the accretion rate evolution in Figure~\ref{fig:history}, and consequently
of the evolution of the cosmic SFR in the simulations.
The accretion rate climbs very rapidly starting at $z\sim 5$ as more galaxies
form above the resolution threshold --- if we had higher resolution and
a lower threshold, the accretion rate at these redshifts would be higher
and would evolve more slowly.  The number of galaxies in the low mass regime
where cold accretion dominates reaches a maximum at $z\sim 2$ and 
declines slowly thereafter.  The rapid drop in the global cold accretion
rate, starting at $z\sim 4$, 
is driven mainly by the decreasing accretion rate per galaxy 
shown in Figure~\ref{fig:rates}, which reflects the increasing infall
timescale in a lower density Universe.  The mass threshold for hot
accretion is higher, so the hot accretion starts to climb later and peaks
at a lower redshift, $z\sim 2$.  Furthermore, the number density of hot
mode halos stays roughly constant at $z<2$, and the average mass of these
halos increases with time.  (Note that higher mass halos always have 
higher mean accretion rates in both the simulation and the analytic model,
even though the ``cooling tax'' makes the increase
slower than $\dot{M} \propto M$.). The drop in global hot accretion is,
therefore, also a consequence of the fast drop in the 
accretion rate per galaxy. 
This drop is slightly slower than for the cold mode owing to an increase in 
the characteristic mass of hot mode halos with time and to a slight 
increase in their number density.

In broader terms, 
most accreting galaxies at high redshift are low mass, do not have
virial shocks, and connect directly to their large scale environment
by filamentary ``umbilical cords.'' At intermediate redshifts, the
galaxies experiencing the most accretion are intermediate mass systems
with virial shocks penetrated by cool filaments, accreting in a mix
of hot and cold mode.  At low redshift, there are many high mass galaxies
with no cold accretion, and many low mass galaxies reside in high mass
halos where they experience primarily hot accretion.
The filamentary structures
that play such a vital role in cold mode accretion grow in scale as the 
Universe evolves, particularly 
around high density regions.
They eventually become too large to
feed individual galaxies and instead channel their flows into the hot
IGM of groups and clusters.  
The combination of galaxy mass scales and large scale structure geometry
drives the predicted dependence of the SFR on environment,
allowing it to extend beyond the virial radii of large groups and clusters.

One of the most striking features of the galaxy population to emerge from
the SDSS is a rapid transition in typical galaxy 
properties at a stellar mass $M_* \sim 3\times 10^{10} M_\odot$
\citep{kauffmann03,kan04}.  
Galaxies below this mass tend to be actively star
forming with lower surface densities, high gas fractions, and late type 
morphologies, while
galaxies above this mass tend to have old stellar populations, high
surface density, low gas fraction, and early type morphology.  
The observed transition mass is remarkably close to the one at which 
we find a transition from cold mode domination to hot mode domination,
so it is tempting to see the two as connected.  It is not clear just
how the second transition would bring about the first.  However, if a 
galaxy accretes much of its mass along filamentary structures,
this process is bound to influence its angular momentum,
perhaps radically modifying the traditional picture of disk formation
in which the disk's specific angular momentum distribution is closely
related to that of its parent dark matter halo 
\citep{hoyle49,peebles69,fall80,mo98,bullock01,vandenbosch01}.
In particular, cold accretion might be crucial to disk formation,
with hot accretion contributing mainly to spheroid growth.
The lower efficiency of hot accretion will also cause high mass galaxies
to have less star formation and older stellar populations, though it
is not clear that this effect in itself will produce a transition as
strong as the observed one.  If, as we speculate in the next section,
hot accretion is suppressed in the real universe by physical processes
not represented in our simulations, then the transition would be much sharper.
Furthermore, galaxies substantially above the cold-to-hot transition
mass would then have to be built by mergers, explaining their
predominantly elliptical morphology.  A tight connection between cold
accretion and disk formation would also help explain the observed
morphology-density relation, since the cold accretion fraction depends
strongly on environment.  For the moment, these remarks
are largely speculation, but we hope that future investigations of 
higher resolution simulations will deepen our understanding of 
the cold and hot accretion processes and their connection to galaxy
morphologies and stellar populations.

When the baryons within a dark matter halo cool to form the central galaxy
in the standard hot mode picture, the dark matter density near the galaxy 
increases through adiabatic contraction \citep{blumenthal86}.  This
results in an increase in the galaxy's circular velocity, which can lead
to conflicts with observations. For example,
it is difficult for models to simultaneously match the galaxy
luminosity function and the zero-point of the Tully-Fisher relation 
(e.g. \citealt{vandenbosch03}).
However, if the galaxy assembles through cold mode accretion, adiabatic 
contraction might not occur or be greatly reduced since the dark matter
and baryonic components assemble nearly simultaneously.  This might help
alleviate  some of the tension with observations, especially since
most disk galaxies should be cold mode dominated given that on average
they have lower masses (e.g. \citealt{bell03}).

The most direct observational tests of cold accretion would come
from detecting the cooling radiation associated with gas infall onto
galaxies.  As emphasised by \cite{fardal01}, the existence of an
important cold accretion mode allows a significant fraction of the
gravitational energy acquired by accreting gas to emerge in the
\lya\ line, producing luminosities that are detectable with large
telescopes.  The main challenge to this test is separating the
contribution of cooling radiation from the contribution of star 
formation, since galaxies with high accretion rates should also
have high star formation rates.  Because the \lya\ photons from
cooling radiation are generally produced at larger galactocentric
radii, it should be possible to distinguish the two contributions,
but accurate predictions of the angular and frequency distribution
of \lya\ cooling radiation require radiative transfer calculations
that have not yet been applied to 3-d simulations.
Filamentary gas flows that penetrate close to the galaxy disk
and then shock should produce X-ray ``hot spots'' in the inner
regions of some galaxy halos.  Further investigation will be
required to assess the observability of this phenomenon.
High column densities of neutral hydrogen near the disk might
absorb much of the X-ray emission and re-radiate it in \lya\
\citep{birnboim03}.

\subsection{Accretion and Feedback}
\label{sec:disc_feedback}

Although our simulations appear reasonably consistent with the 
observationally inferred history of cosmic star formation
(Fardal et al., in preparation), they overpredict the $z=0$
baryonic mass function relative to the observational estimates of,
e.g., \cite{cole01} or \cite{bell03}.  Globally, the predicted
fraction of baryons in the form of stars and cold gas exceeds the
observationally inferred fraction by a factor $\sim 2-3$.  
Of course, the integral of the mean cosmic star formation rate
must equal the mean stellar mass density (when corrected for the stellar 
mass loss and recycling of this gas), so the 
agreement with one and disagreement with the other implies that
at least some of the observational estimates of these quantities are 
themselves in conflict.
However, the SFR estimates have substantial systematic uncertainties,
while the stellar mass densities appear reasonably secure {\it if}
the assumed stellar IMFs are correct, so we think it more likely that
the simulations are indeed producing overly massive galaxies.
A related problem is that the simulations predict ongoing accretion
and star formation in high mass galaxies, while in the real universe
these systems are predominantly ellipticals exhibiting little or no
recent star formation.  Similar difficulties with the luminosities
and colours of high mass galaxies appear in semi-analytic models of
galaxy formation, unless they are specifically modified to suppress them
(see, e.g., \citealt{benson03}).

Simply distinguishing between hot and cold accretion does not, of course,
change the simulation predictions.  However, the conflicts mentioned above
suggest that these simulations are still missing some physical processes
that play a significant role in galaxy formation, and these processes
might have different effects on hot and cold accretion.  One can
imagine, for instance, that hot mode accretion might be
more easily suppressed by some forms of feedback because the hot
mode gas is lower density and more isotropically distributed.
In the local universe, at least, it appears that hot accretion {\it is}
suppressed in groups and clusters, where X-ray emissivities imply
short cooling times but X-ray spectra show that the predicted gas
at $T \la 1$ keV is not present \citep{peterson03}.
The leading hypothesis for explaining this conflict is that the
central regions of the intracluster medium are heated by
recurrent AGN activity, thermal conduction from the outer regions,
or both (e.g., \citealt{binney95,ciotti01,narayan01,ruszkowski02}).
The same processes might operate in lower mass halos at moderate redshifts.
\citet{maller.bullock04} argue that a proper, multi-phase
treatment of hot halo gas reduces predicted accretion rates by a 
substantial factor, even without conduction or extra feedback, 
because the cooling gas forms dense clouds and the remaining
gas has lower densities and longer cooling times.
This effect is omitted in standard analytic treatments and would
be missed by simulations (like ours) that do not resolve the
$\sim 10^6\msun$ scale of clouds formed by thermal instability.
Whatever the mechanism,
preferential suppression of hot mode accretion would lead to 
a sharper cutoff at the high end of the luminosity function,
a more rapid decline in the cosmic SFR at $z \la 1$, and older stellar
populations in massive galaxies.  All of these changes would 
improve the agreement between our SPH simulations and observations,
and they would also improve the agreement for most semi-analytic models.
\cite{binney04} has made similar points, drawing on our preliminary results
\citep{katz03} and the 1-d calculations of \cite{binney77} and
\cite{birnboim03} for theoretical motivation
(see also \citealt{dekel05}).

In our simulations, supernova feedback has little impact on galaxy
masses because the supernova energy is usually deposited in a dense
medium, where it radiates away before it can drive a galactic wind.
Simple models of a multi-phase interstellar medium do not change
this result \citep{springel03a}, but collective effects may allow real
galaxies to drive winds more easily, and a number of groups have added
{\it ad hoc} wind models to simulations 
(e.g., \citealt{theuns02,abadi03,springel03a}).
The interplay of supernova-driven winds
with accretion modes might be different from that
of conduction or AGN feedback, because once a wind is launched
the main challenge is getting it out of the 
galaxy halo, so that the gas does not simply return and form stars
on a short timescale.  In a galaxy fed by filamentary cold accretion, 
supernova-driven gas might encounter little resistance because
there is no surrounding halo medium (see left panels of 
Fig~\ref{fig:filament}).
In a galaxy dominated by hot accretion, on the other hand,
the disk and bulge are surrounded by a shock-heated gas halo, and a wind 
may be trapped at small radius.  Thus, one can easily imagine that
expulsion of cooled gas by supernova feedback, in contrast to the suppression
of cooling by AGN feedback or conduction, might operate more efficiently
in low mass galaxies.  This might provide a way of reducing the masses
of low mass galaxies and flattening the faint end of the
galaxy luminosity function, which would again improve agreement
between the simulations and data.  

The conventional understanding of
supernova feedback also suggests that it will be more effective in
low mass systems because gas is more easily driven out of shallow
potential wells \citep{dekel86}.  We have focused instead on the
role of the surrounding hot gas, which we suspect is even
more important, since simulations of the feedback process show that only
low mass galaxies ($\mgal \la 10^9 M_\odot$) blow out significant
amounts of gas if they are surrounded by conventional hot halos 
\citep{ferrara00}.
The form of the observed mass-metallicity relation
can be understood if supernova-driven winds are efficient below
$\mgal \sim 3\times 10^{10}M_\odot$ and inefficient at higher
masses \citep{tremonti04}. 
This transition mass is close to the mass where cold accretion gives way 
to hot accretion
in our simulated galaxies, so it might emerge naturally if trapping   
by the hot gas halo is the correct explanation for it.

The simulation predictions would also change in interesting ways if
we assumed that cold accretion is associated with the conventional
stellar IMF (e.g., \citealt{miller79,kennicutt83})
but that hot accretion is associated with {\it either} a bottom-heavy IMF 
rich in brown dwarfs {\it or} a top-heavy IMF
truncated below $\sim 2M_\odot$.  In the first case, massive galaxies would
have high stellar mass-to-light ratios and therefore lower luminosities.
In the second case, the massive galaxies would have lower stellar masses
(assuming that much of the recycled gas from evolved stars
is driven from the galaxy), and they would have higher UV luminosities
during their star-forming phases.
Allowing IMF variations introduces a frightening level of freedom in
galaxy formation models, but with two physically distinct accretion
processes one should at least consider the possibility that the 
associated star formation is different.

For the moment, these proposals are largely speculation.
We are carrying out further investigations to see how the predicted properties
of the galaxy population would change if we assume preferential suppression 
of hot mode accretion or different stellar IMFs for the two modes.
Such adjustments may make distinctive observational predictions
that could compensate for the extra freedom they introduce,
and their physical underpinnings can
be investigated with higher resolution simulations focused on these questions.
Given the variety of possible assumptions, semi-analytic calculations
might be a sensible approach to exploring models that treat
cold and hot accretion differently.

\section{Conclusions} 
\label{sec:conclusions}

In SPH simulations of galaxy formation, a substantial fraction of the
gas accreted by galaxies is never heated close to the virial temperature
of the galaxy potential wells.  The importance of this cold accretion 
mode, relative to the hot accretion mode envisioned in the traditional
picture of galaxy formation, depends strongly on galaxy mass.  Most
galaxies below a baryonic mass $\mgal \sim 10^{10.3}\msun$ or dark halo
mass $\mhalo \sim 10^{11.4} \msun$ accrete primarily in cold mode,
while more massive galaxies accrete primarily in hot mode.  Since
high mass galaxies are built from lower mass systems, even galaxies at
the top end of the luminosity function today acquired a significant
fraction of their mass via cold accretion.  Globally, the increasing
mass scale of galaxies means that cold accretion 
dominates at high redshift and hot accretion at low redshift.
The ratio of cold to hot accretion is also environment dependent,
mostly because higher mass galaxies are more common in dense environments,
and partly because low mass systems in dense environments have a
larger hot accretion fraction than their isolated counterparts. 
Hot mode accretion is quasi-spherical, but cold mode accretion is 
often directed along filamentary channels, increasing its efficiency.

A number of arguments suggest that these are physical results, not
numerical artifacts.  A quantitatively important cold 
mode appears in simulations covering a factor of 512 in mass resolution,
though the lowest resolution simulations do not include a photoionizing
UV background and therefore predict a noticeably different temperature
distribution for cold mode gas.  The transition mass of 
$\mgal \sim 10^{10.3}\msun$ is stable in simulations covering a factor
of 64 in mass resolution.  Cold accretion is seen in other SPH 
simulations, including one with a substantially different implementation
of shock heating.  Cold filamentary flows are found in adaptive mesh
simulations, though these cannot yet test whether gas gets into galaxies
without experiencing strong shock heating.  Finally, the transition mass
in our simulations agrees (to within a factor of $\sim 2-3$)
with that found by \cite{birnboim03} using
entirely independent analytic methods and high-resolution 1-d calculations.
Despite this litany of tests, we cannot rule out the possibility 
that cold accretion is an artifact of finite resolution of shocks in
SPH simulations.  Confirmation with still higher resolution simulations
and with hydrodynamic methods that directly impose shock jump conditions
at velocity discontinuities is therefore desirable.
The transition mass separating hot and cold accretion is quantitatively
different from the transition mass separating rapid and slow 
post-shock cooling in semi-analytic models of galaxy formation,
at least as calculated by our implementation, which we believe to
be the standard one (see \S\ref{sec:disc_cold} and the Appendix).
However, semi-analytic models could be easily revised to incorporate
a cold mode of gas accretion calibrated on hydrodynamic simulations
like those analyzed here.

Cold accretion could have important implications for the cosmic star formation
history and for the correlations of galaxy star formation rates, stellar
populations, and morphology with environment.  If cold and hot accretion
are affected differently by stellar or AGN feedback, or if they are
associated with different stellar IMFs,
then the simulation predictions
would change in interesting ways, perhaps resolving conflicts with the
observed colours of ellipticals and with the observationally inferred
galaxy baryonic mass function.  Our predicted transition mass is close
to the observed transition mass at which typical galaxy properties 
change markedly \citep{kauffmann03}, and we have speculated on possible
connections between the two.  
In particular, if AGN feedback, multi-phase cloud production,
or some other mechanism strongly suppresses hot accretion in the 
real universe, then galaxies above the transition mass would have little 
recent accretion and star formation, and they would have to be 
built by mergers of lower mass systems, which would produce
spheroidal morphologies.
Many questions remain about both the
underlying physics and the observational implications of cold accretion,
but we hope that higher resolution simulations and comparisons to
data will answer these questions over the next few years and solidify
our understanding of how galaxies get their gas.  

\section*{Acknowledgements}
We thank L.\ Hernquist and V.\ Springel for allowing us to analyse
their Q3 simulation, as discussed in \S\ref{sec:othersim}.
We thank Y.\ Birnboim, A.\ Dekel, A.\ Kravtsov, A. Maller,
and H. J. \ Mo for helpful discussions,
C.\ Murali for providing the merger code used in our analysis,
and an anonymous referee for constructive comments.
This project was supported by
NASA ATP grants NAG5-12038 and NAG5-13308, 
NASA LTSA grant NAG5-3525, and NSF grants
AST-9988146 \& AST-0205969.

\appendix
\section{The Rapid Cooling Transition in Semi-Analytic Calculations}

Semi-analytic models of galaxy formation, which trace back to the 
formalism introduced by \cite{white91}, calculate the growth of galaxies
fed by the accretion of gas cooling within dark matter halos.
While cooling rates are calculated assuming that the gas initially
heats to the halo virial temperature, these calculations do 
distinguish between a regime of slow cooling, in which accretion 
from a quasi-hydrostatic gas halo is regulated by the cooling rate,
and a regime of rapid cooling, in which no quasi-hydrostatic atmosphere
forms and gas accretion is regulated by the infall rate.
Given the qualitative resemblance of this criterion to the
one introduced by \cite{birnboim03}, it is interesting to ask
whether the predicted transition between infall and cooling 
dominated accretion is quantitatively similar to the transition
between cold and hot accretion found in our simulations.
In this Appendix, we describe our calculation of the infall-cooling
transition mass, discussed earlier in \S\ref{sec:disc_cold}.

We assume that galaxy halos initially contain the universal baryonic
fraction of gas, $\Omega_b/\Omega_m$,
with a constant gas temperature equal to $\tvir$. 
We model the initial gas density distribution as an NFW profile \citep{nfw96}, 
and, for simplicity, assume a constant concentration for the NFW profiles, 
which is typical for $10^{11}M_{\odot}$ halos \citep{bullock01b}: 
$c=\Rvir/r_s=15/(1+z)$. $\Rvir$ is the virial radius and $r_s$ is the 
scale radius of the NFW halo.
For a given halo virial mass, $\mvir$, $\Rvir$ is defined as the 
radius within which the enclosed density equals $\Delta_{\rm vir}$ 
times the mean mass density of the universe at that redshift, i.e.
\begin{equation}
\Rvir=\left({{3 \over 4\pi} {\mvir \over 
       \Delta_{\rm vir}\bar{\rho}_m}}\right)^{1/3} ~.
\end{equation}
Here we use the approximation for $\Delta_{\rm vir}$ from \citet{bryan98},
which gives similar results to \citet{kitayama96}.
For comparison we also consider an alternative model where the gas density
follows that of a singular isothermal sphere (SIS) with $T_{\rm gas}=\tvir$.

Following \cite{white91}, we define the local cooling time of the halo gas,
$\tcool$, as
\begin{equation}
\tcool(r)={3 \over 2} {k_B T \rho_g(r) \over f_g\mu m_p n_H^2(r)\Lambda(T)}~,
\end{equation}
where $k_B$ is the Boltzmann constant, $\mu$ is mean ``molecular" weight, which
we assume to be that of a fully ionized gas, $f_g$ is the fraction of 
halo gas available for cooling, $m_p$ is the proton mass, $n_H$ is the 
hydrogen 
density, $\rho_g$ is the total gas density, and $\Lambda(T)$ is the cooling
function.  Like in the simulation, we assume that the gas is primordial with a 
hydrogen mass fraction of 0.76.
The cooling time determines the cooling radius $\rcool$ at which
gas can radiate its thermal energy in a specified amount of time.
\citet{white91} define the cooling radius by setting $\tcool(\rcool)$
equal to the Hubble time $t(z)$, the age of the Universe at redshift $z$.
However, different groups
use different characteristic times in their semi-analytic calculations. 
Therefore, to cover a range of possibilities we compare the cooling
time to three time scales: the Hubble time $t(z)$, 
the dynamical time defined as
\begin{equation}
\tdyn=\sqrt{3\pi \over 16 G \bar{\rho}(z)}~,
\end{equation}
where $\bar{\rho}$ is the average enclosed density,
and the free fall time, which is $\tff=\tdyn/\sqrt{2}$.
Therefore, the cooling radius is defined as the radius where
$\tcool(\rcool)=t(z)$, $\tdyn(z)$, or $\tff(z)$.
In \citet{white91}, infall dominated accretion takes place when 
$\rcool > \Rvir$, and cooling dominated accretion takes place
when $\rcool > \Rvir$.
The transition halo mass between the two regimes is thus the halo mass
for which $\rcool=\Rvir$.

To fairly compare this mass with the transition mass between 
cold and hot mode accretion in our simulations,
we use the same cosmology (see \S~\ref{sec:param}) and the same cooling
function (see KWH96) as in our simulations, neglecting only inverse Compton
cooling.
Since some gas has already transformed into stars, the fraction of the gas 
available for cooling, $f_g$, is less than one and decreases with redshift.
We approximate $f_g=1-0.28/(1+z)$, which
roughly matches the global fraction of the baryons in the simulation that
are not in stars or cold galactic gas at a given redshift.

The UV background affects gas cooling in our simulations after $z=6$, so
to make a fair comparison it is necessary to include it in our calculation. 
As mentioned in \S~\ref{sec:UV}, for typical densities at the virial radius,
the UV background can significantly alter the hydrogen and helium line 
cooling peaks in the cooling function, making the cooling times at the virial
radius significantly longer. 

\begin{figure}
\epsfxsize=1.0\columnwidth
\epsfbox[20 140 580 700]{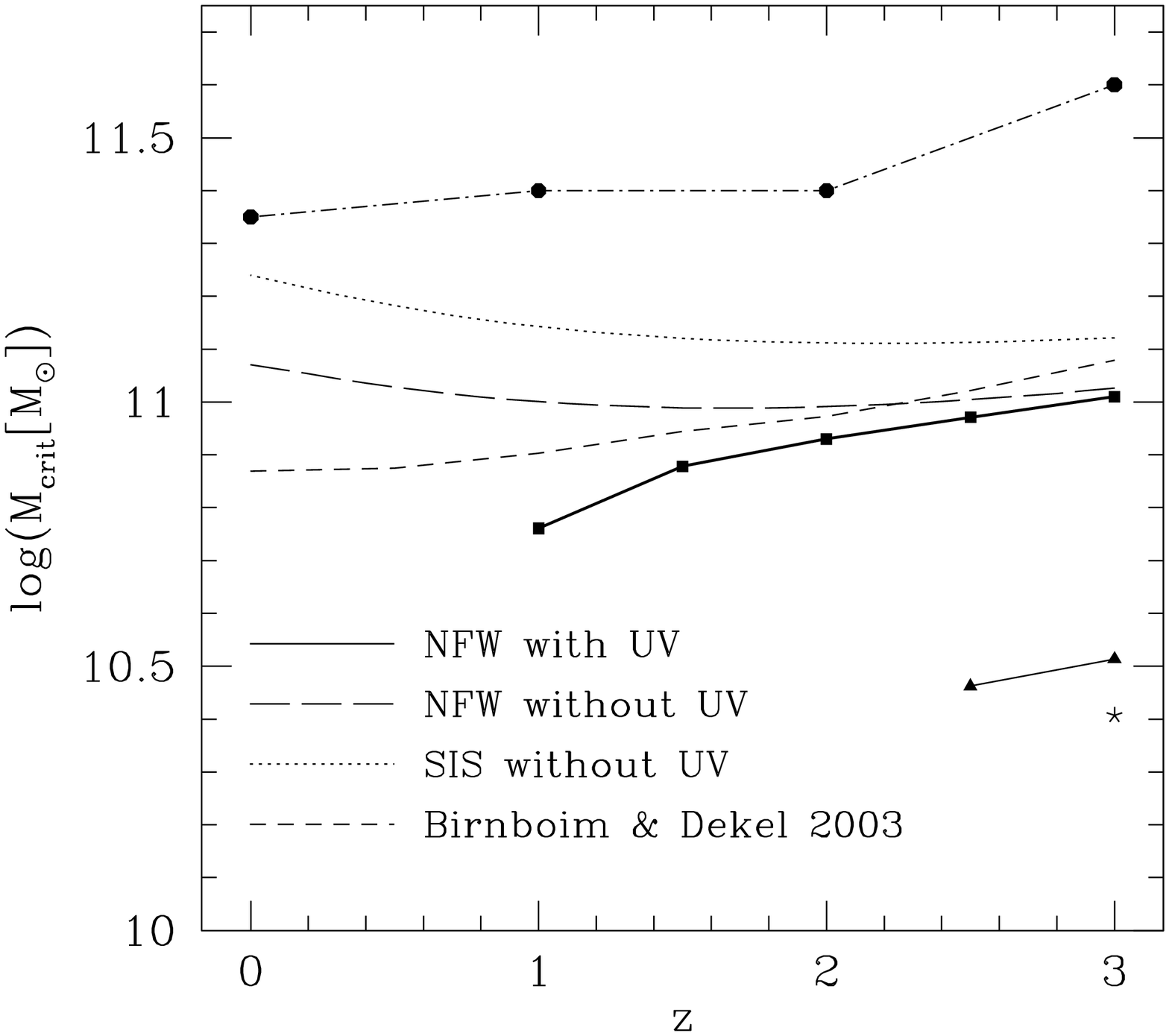}
\caption{Transition halo mass between infall and cooling dominated accretion,
calculated in a manner similar to that of \citet{white91}. 
Squares, triangles, and the star at $z=3$ assume an NFW
gas density profile with a UV background and a cooling radius definition 
$\tcool(\rcool)=t(z)$, $\tdyn(z)$, and $\tff(z)$, respectively.
With a UV background, all halos are cooling dominated below
$z=1$, 2.5, or 3, respectively, for these three cases.
The dashed line assumes an NFW profile but no
UV background, and the dotted line assumes an SIS profile with no
UV background, both for $\tcool(\rcool)=t(z)$.
Filled circles show the transition halo mass at which we find equal
amounts of cold and hot accretion in the L22/128 simulation,
which includes a UV background. For comparison we also show results from
the model of \citet{birnboim03} (see text).
}
\label{fig:append}
\end{figure}

In Figure~\ref{fig:append}, solid lines show the transition mass between 
infall and cooling dominated accretion for an NFW density profile 
in the presence of the UV background. 
Squares correspond to a definition of $\rcool$
with $\tcool=t(z)$, triangles to 
$\tcool=\tdyn(z)$, and the star to $\tcool=\tff(z)$. 
For direct comparison we plot the mass where the transition between cold
and hot mode occurs in our simulations as the dot-dashed line with circles.
At $z\ge 1$ the transition mass in the simulation is a factor of $2-4$ higher
than the highest of the semi-analytic transition masses, defined for
$\tcool(\rvir)=t(z)$.  At $z<1$, there is no halo mass for which
$\tcool(\rvir)\leq t(z)$, so the semi-analytic calculation predicts
no infall dominated halos at all.
If we define the transition with respect to $\tdyn$ or $\tff$, the
semi-analytic transition mass drops by another factor of $3-4$ at 
high redshift, and infall dominated halos disappear after 
$z=2.5$ ($\tdyn$) or $z=3$ ($\tff$).

The long-dashed curve in Figure~\ref{fig:append} shows the effect
of omitting the UV background, which increases cooling rates and 
therefore increases the transition mass.  
With no UV background, there are infall dominated halos all the
way to $z=0$, though the transition mass is still 
lower than the cold/hot transition mass in our simulations with 
a UV background.  Furthermore, as discussed in \S\ref{sec:UV}, 
the increased cooling in the absence of a UV background leads to
a substantially higher transition mass in our simulations themselves.
Since our simulations without a UV background are lower resolution
or run only to $z=3$, we do not have a very accurate measurement
of this increase, but it appears to be at least a factor of four.
The dotted curve in Figure~\ref{fig:append} shows the effect of
assuming an SIS gas profile instead of an NFW profile, again with
no UV background.  Since an SIS profile has a higher gas density
at the virial radius than an NFW profile, the transition mass increases.
However, while gas and dark matter profiles inside halos will generally
be different to some degree, the NFW assumption seems likely to be
more realistic near the virial radius.
For comparison we also plot the model of
\citet{birnboim03}(zero metallicity line from the upper panel of their 
Figure 11), which is motivated by their 1D simulations. 
Their cosmology is different from ours, but the baryonic mass fraction,
one of the main factors determining the transition mass, is similar.
\citet{birnboim03} do not include a UV background in their
calculations, and their model agrees fairly well with the
``NFW without UV'' line in Figure~\ref{fig:append}, though the redshift
dependence is somewhat different.
\citet{dekel05} discuss the dependence of the derived transition
mass on details of the calculational assumptions, gas metallicity,
and cosmological parameters.

We conclude that the conventional definition of the transition
between infall and cooling dominated accretion does not correspond
quantitatively to the transition between cold and hot mode accretion
found in our simulations, if one makes the same physical assumptions
in the semi-analytic calculation.
However, it may be possible to alter the semi-analytic calculation,
for example by defining the transition at some fraction of the 
virial radius instead of at $\rvir$ itself,
so that the correspondence is better.
It would be particularly interesting to see whether a definition of
transition mass calibrated to match hydrodynamic simulations for
one set of cosmological parameters and cooling rates then reproduces
the results of simulations that assume different parameters or
different cooling rates (e.g., because of different metallicity
or a change of the UV background).  A successful analytic model of
this sort would provide physical insight into the distinction between
cold and hot accretion, and it would be a useful practical tool
for investigating the effects of parameter variations.

\begin{thebibliography}{}

\bibitem[Abadi, Moore \& Bower(1999)]{abadi99}
Abadi, M.G., Moore, B., \& Bower, R.  G. 1999 MNRAS, 308, 947

\bibitem[Abadi et al.(2003)]{abadi03}
Abadi, M. G., Navarro, J. F,
Steinmetz, M., Eke, V. R 2003, \apj, 591, 499

\bibitem[Avila-Reese et~al.(1998)]{avila98}
Avila-Reese, V., Firmani, C., \& Hernandez, X. 1998 ApJ, 505, 37

\bibitem[Balogh et al.(1998)]{balogh98} 
Balogh, M. L., Schade, D.,
Morris, S. L., Yee, H. K. C., Carlberg, R. G., \& Ellingson, E. 1998,
ApJL, 504, 75

\bibitem[Balogh et al.(2001)]{balogh01} 
Balogh, M. L., Pearce, F. R.,
Bower, R. G., Kay, S. T. 2001 MNRAS, 326, 1228

\bibitem[Barnes \& Hut(1986)]{barnes86}
Barnes, J.E., \& Hut, P. 1986, Nature, 324, 446

\bibitem[Baugh et al.(1998)]{baugh98} Baugh,
C. M., Cole, S., Frenk, C. S., \& Lacey, C. G. 1998, ApJ, 498, 504

\bibitem[Baugh et al.(2004)]{baugh.etal04} Baugh,
C. M., Lacey, C. G., Frenk, C. S., Granato, G.L., Silva, L., Bressan, A.,
Benson, A.J. \& Cole, S. 2004, MNRAS submitted (astro-ph/0406069)

\bibitem[Bell et al.(2003)] {bell03} Bell, E. F., McIntosh, D. H.,
Katz, N., Weinberg, M. D. 2003, ApJS, 149, 289

\bibitem[Bennett et al.(2003)]{bennett03}
Bennett, C.~L.~et al.\ 2003, \apjs, 148, 1 

\bibitem[Benson et al.(2000)]{benson00} Benson, A. J., Bower, R. G.,
Frenk, C. S., White, S. D. M. 2000, MNRAS, 314, 557.

\bibitem[\protect\astroncite{Benson et al.}{2001}]{benson01}
Benson, A.J., Pearce, F.R., Frenk, C. S., Baugh, C. M. \& Jenkins,
A. 2001, MNRAS, 320, 261

\bibitem[Benson et al.(2003)]{benson03}
Benson, A.~J., Bower, 
R.~G., Frenk, C.~S., Lacey, C.~G., Baugh, C.~M., \& Cole, S.\ 2003, \apj, 
599, 38 

\bibitem[\protect\astroncite{{Binney}}{1977}]{binney77} {Binney},
J. 1977, MNRAS, 215, 483

\bibitem[Binney(2004)]{binney04}
Binney, J. 2004, \mnras, 347, 1093

\bibitem[Binney \& Tabor(1995)]{binney95}
Binney, J., \& Tabor, G. 1995, \mnras, 276, 663

\bibitem[Birnboim \& Dekel(2003)]{birnboim03} Birnboim, Y., Dekel,
A. 2003, \mnras, 345, 349

\bibitem[Blanton et al.(2000)]{blanton00} Blanton, M., Cen, R.,
Ostriker, J. P., Strauss, M.A., \& Tegmark, M. 2000, ApJ, 531, 1

\bibitem[Blanton et al.(2003)]{blanton03}
Blanton, M.\ R.\ et al.\ 2003, \apj, 592, 819

\bibitem[Blumenthal et al.(1986)]{blumenthal86} Blumenthal, G. R., Faber, 
S. M., Flores, R., Primack, J. R. 1986, \apj, 301, 27

\bibitem[Bryan \& Norman(1998)]{bryan98} Bryan, G. L., Norman, M. L. \apj, 495,
80

\bibitem[Bullock et al.(2001a)]{bullock01}
Bullock, J. S., Dekel, A., Kolatt, T. S., Kravtsov, A., Klypin, A. A.,
Porciani, C., \& Primack, J.R., 2001a, \apj, 555, 240

\bibitem[Bullock et al.(2001b)]{bullock01b}
Bullock, J.S., Kolatt, T. S., Sigad, Y., Somerville, R. S., Kravtsov, A. V.,
Klypin, A. A., Primack, J. R., Dekel, A. 2001b, \mnras, 321, 559

\bibitem[Burles \& Tytler(1998)]{burles98}
Burles, S., \& Tytler, D. 1998, \apj, 507, 732

\bibitem[Cen \& Ostriker(1999)]{cen99} Cen, R., \& Ostriker,
J. P. 1999, ApJ, 514, 1

\bibitem[Ciotti \& Ostriker(2001)]{ciotti01}
Ciotti, L., \& Ostriker, J. P. 2001, \apj, 551, 131

\bibitem[Cole et al.(1994)]{cole94} Cole, S., Aragon-Salamanca, A.,
Frenk, C. S., Navarro, J. F., \& Zepf, S. E.  1994, MNRAS, 271, 781

\bibitem[Cole et al.(2000)]{cole00} Cole, S., Lacey, C. G., Baugh,
C. M., Frenk, C. S. 2000, MNRAS, 319, 168

\bibitem[Cole et al.(2001)]{cole01} Cole, S. et al. 2001, MNRAS, 326,
255

\bibitem[Colless et al.(2001)]{colless01}
Colless, M.~et al.\ 2001, \mnras, 328, 1039 

\bibitem[Dav\'e, Dubinski, \& Hernquist(1997)]{dave97} Dav\'e, R.,
Dubinski, J., \& Hernquist, L. 1997, New Astron, 2, 227

\bibitem[Dav\'e et al.(1999)]{dave99}
Dav\'e, R., Hernquist, L., Katz, N., \& Weinberg, D. H. 1999, \apj, 511, 521

\bibitem[de Bernardis et al.(2002)]{debernardis02} de Bernardis, P. et
al. 2002 ApJ, 564, 559

\bibitem[Davis et al.(1985)]{davis85} Davis, M, Efstathiou, G., Frenk,
C. S., White, S. D. M. 1985, ApJ, 292, 371

\bibitem[Dekel \& Silk(1986)]{dekel86} Dekel, A., \& Silk, J.  1986,
ApJ, 303, 39
 
\bibitem[Dekel \& Birnboim(2005)]{dekel05} Dekel, A., \& Birnboim, Y. 2005,
\mnras, submitted, astro-ph/0412300

\bibitem[Dressler (1980)]{dressler80} Dressler, A. 1980, ApJ, 236, 351

\bibitem[Dressler et al.(1997)]{dressler97} Dressler, A. et al. 1997,
ApJ, 490, 577

\bibitem[Efstathiou(1992)]{efstathiou92}
Efstathiou, G.\ 1992, \mnras, 256, 43P

\bibitem[\protect\astroncite{{Fall} and
{Efstathiou}}{1980}]{fall80}
{Fall}, S.~M. and {Efstathiou}, G. 1980,MNRAS, 193, 189.

\bibitem[Eke, Cole \& Frenk(1996)]{eke96}
Eke, V. R., Cole, S. \&
Frenk, C. S. 1996, MNRAS, 181, 375

\bibitem[Evrard, Summers, \& Davis(1994)]{evrard94}
Evrard, A.E., Summers, F.J., \& Davis, M. 1994, \apj, 422, 11

\bibitem[Fardal et al.(2001)]{fardal01} Fardal, M.\ A., Katz, N.,
Gardner, J.\ P., Hernquist, L., Weinberg, D.\ H.\ \& Dav{\'e}, R.\
2001, ApJ, 562, 605

\bibitem[Ferrara \& Tolstoy(2000)]{ferrara00}
Ferrara, A.~\& Tolstoy, E.\ 2000, \mnras, 313, 291 

\bibitem[Gelb \& Bertschinger(1994)]{gelb94}
Gelb, J. M., \& Bertschinger, E. 1994, \apj, 436, 467

\bibitem[Gingold \& Mohaghan(1997)]{gingold77} Gingold, R. A. \&
Monaghan, J. J. 1977, MNRAS, 181, 375

\bibitem[Gnedin(2000)]{gnedin00} Gnedin, N. 2000, ApJ, 542, 535

\bibitem[Gomez et al.(2003)] {gomez03} Gomez, P. L, et al. 2003, ApJ,
584, 210

\bibitem[Gunn \& Gott(1972)] {gunn72} Gunn, J. E., \& Gott, J. R. 1972, 
ApJ, 176,1

\bibitem[Haardt \& Madau(1996)]{haardt96} Haardt, F., Madau, P. 1996,
ApJ, 461, 20

\bibitem[Haiman, Spaans, \& Quataert(2000)]{haiman00}
Haiman, Z., Spaans, M., \& Quataert, E. 2000, \apjl, 537, L5

\bibitem[Hashimoto et al.(1998)]{hashimoto98} Hashimoto, Y., Oemler, A. J.,
Lin, H. \& Tucker, D. L. 1998, ApJ, 499, 589

\bibitem[Helly et al.(2003)]{helly03}
Helly, J.~C., Cole, S.,  
Frenk, C.~S., Baugh, C.~M., Benson, A., Lacey, C., \& Pearce, F.~R.\ 2003,
\mnras, 338, 913

\bibitem[Hernquist(1987)]{hernquist87}
Hernquist, L. 1987, ApJS, 64, 715

\bibitem[Hernquist \& Katz(1989)]{hernquist89}
Hernquist, L., \& Katz, N. 1989, \apjs, 70, 419

\bibitem[Hoyle(1949)]{hoyle49}
Hoyle, F. 1949, in Problems of Cosmological Aerodynamics,
ed. J. M. Burgers \& H.C. van de Hulst (Internat. Union Theoret. Appl.
Mech. \& Inter. Astr. Union), p. 195

\bibitem[Hubble(1936)]{hubble36}
Hubble, E.P. 1936, The Realm of the Nebulae (Oxford University Press: Oxford),
79

\bibitem[Hui \& Gnedin(1997)]{hui97} Hui, L., Gnedin, N. 1997 MNRAS, 292, 27

\bibitem[Hutchings \& Thomas(2000)]{hutt00} Hutchings, R. M., Thomas, P. A.
 2000, MNRAS, 319, 721

\bibitem[Kannappan(2004)]{kan04} Kannappan, S. 2004, ApJL, 611, 89

\bibitem[Katz \& Gunn(1991)]{katz91} Katz, N., \& Gunn, J. E. 1991,
ApJ, 377, 365

\bibitem[Katz, Hernquist, \& Weinberg(1992)]{katz92} Katz, N.,
Hernquist, L., \& Weinberg, D. H. 1992, ApJ, 399, L109

\bibitem[Katz et al.(2003)]{katz03}
Katz, N., Keres, D., Dav\'e, R., \& Weinberg, D. H.\ 2003, in
The IGM/Galaxy Connection: The Distribution of Baryons at z=0,
eds. J. L. Rosenberg \& M. E. Putman, Kluwer, Dordrecht, p. 185,
astro-ph/0209279

\bibitem[Katz et al.(1994)]{katz94} Katz, N., Quinn, T., Bertschinger, E., 
\& Gelb, J. M. 1994,
\mnras, 270, L71

\bibitem[Katz, Weinberg, \& Hernquist(1996)]{katz96} Katz, N.,
Weinberg D.H., \& Hernquist, L. 1996, ApJ Supp., 105, 19 (KWH)

\bibitem[Katz \& White(1993)]{katz93} Katz, N., \& White,
S. D. M. 1993, ApJ, 412, 455

\bibitem[Kauffmann, White, \& Guiderdoni(1993)]{kauffmann93} Kauffmann,
G., White, S. D. M., \& Guiderdoni, B. 1993, MNRAS, 264, 201

\bibitem[Kauffmann et al.(2003)]{kauffmann03} Kauffmann, G. et
al. 2003, MNRAS, 341, 54

\bibitem[Kauffmann et al.(2004)]{kauffmann04} Kauffmann, G., 
White, S.~D.~M., Heckman, T.~M., Menard, B., 
Brinchmann, J., Charlot, S., Tremonti, C., \& Brinkmann, J. 2004, 
\mnras, 353, 713

\bibitem[Kay et al.(2000)]{kay00} Kay, S.~T., Pearce, F.~R., Jenkins,
A., Frenk, C.~S., White, S.~D.~M., Thomas, P.~A., \& Couchman,
H.~M.~P.\ 2000, MNRAS, 316, 374

\bibitem[Kennicutt(1983)]{kennicutt83}
Kennicutt, R.~C.\ 1983, \apj, 272, 54

\bibitem[Kennicutt(1998)]{kennicutt98}
Kennicutt, R. C. 1998, \apj, 498, 541 

\bibitem[Kitayama \& Suto (1996)]{kitayama96} Kitayama, T., Suto,
Y. 1996, 469, 480

\bibitem[Kodama et al. (2001)]{kodama01} Kodama, T., Smail, I., Nakata,
F., Okamura, S., Bower, R.G. 2001, ApJL, 562, 9

\bibitem[Larson, Tinsley \& Caldwell (1980)]{larson80} Larson, R. B.,
Tinsley, B. M., \& Caldwell, C. N. 1980 ApJ, 460, 1

\bibitem[Lacey \& Cole(1993)]{lacey93} Lacey, C.\& Cole, S. 1993, 
MNRAS, 262, 627

\bibitem[Lewis et al.(2002)]{lewis02}
Lewis, I.~et al.\ 2002, \mnras, 334, 673

\bibitem[Lucy(1977)]{lucy77}
Lucy, L. 1977, \aj, 82, 1013

\bibitem[Madau et al.(1996)]{madau96} Madau, P., Ferguson, H. C.,
Dickinson, M. E., Giavalisco, M., Steidel, C. C., \& Fruchter,
A. 1996, MNRAS, 283, 1388

\bibitem[Maller \& Bullock (2004)]{maller.bullock04}
Maller, A. H., \& Bullock, J. S. 2004
\mnras, submitted, astro-ph/0406632


\bibitem[Miller \& Scalo(1979)]{miller79} 
Miller, G. E.,\& Scalo, J. M. 1979, ApJS, 41, 513

\bibitem[Miralda-Escud\'e \& Rees(1994)]{miralda94} 
Miralda-Escud\'e J., \& Rees, M. J. 1994, \mnras, 266, 343

\bibitem[Mo, Mao, \& White(1998)]{mo98} Mo, H. J., Mao, S., \& White,
S. D. M. 1998, MNRAS, 295, 319

\bibitem[Moore et al.(1996)]{moore96}
Moore, B., Katz, N.,
Lake, G., Dressler, A., \& Oemler, A.\ 1996, \nat, 379, 613

\bibitem[Moore, Lake, \& Katz(1998)]{moore98}
Moore, B., Lake, G., \& Katz, N.\ 1998, \apj, 495, 139

\bibitem[Murali et al.(2002)]{murali02} Murali, C., Katz, N.,
Hernquist, L., Weinberg, D. H., \& Dav\'e, R. 2002, ApJ, 571, 1 (MKHWD)

\bibitem[Nagai \& Kravtsov(2003)]{nagai03} Nagai, D.~\& Kravtsov,
A.~V.\ 2003, ApJ, 587, 514

\bibitem[Nagamine et al.(2001)]{nagamine01} Nagamine, K.,
Fukugita, M., Cen, R., \& Ostriker, J. P. 2001, MNRAS, 327, 10

\bibitem[Narayan \& Medvedev(2001)]{narayan01} Narayan, R., Medvedev,
M. 2001, ApJ, 562, L129

\bibitem[Navarro, Frenk, \& White(1996)]{nfw96} Navarro, J. F., Frenk, C. S., 
White, S. D. M. 1996, \apj, 462, 563

\bibitem[Pearce et al.(1999)]{pearce99} Pearce, F. R., Jenkins, A.,
Frenk, C. S., Colberg, J. M., White, S. D. M., Thomas, P. A.,
Couchman, H. M. P., Peacock, J. A., \& Efstathiou, G. 1999, ApJ, 521,
L99

\bibitem[Peebles(1969)]{peebles69}
Peebles, P.\ J.\ E.\ 1969, \apj, 155, 393

\bibitem[Peterson et al.(2003)]{peterson03}
Peterson, J.~R., Kahn,
S.~M., Paerels, F.~B.~S., Kaastra, J.~S., Tamura, T., Bleeker, J.~A.~M.,
Ferrigno, C., \& Jernigan, J.~G.\ 2003, \apj, 590, 207

\bibitem[Postman \& Geller(1984)]{postman84} Postman, M., Geller,
M. J. 1984, ApJ, 281, 95

\bibitem[Quinn, Katz, \& Efstathiou(1996)]{quinn96}
Quinn, T., Katz, N., \& Efstathiou, G.\ 1996, \mnras, 278, L49

\bibitem[\protect\astroncite{{Rees} \&  
{Ostriker}}{1977}]{rees77} Rees, M.J., \& Ostriker,
J.P. 1977 MNRAS, 179, 541.

\bibitem[Ruszkowski \& Begelman(2002)]{ruszkowski02}
Ruszkowski, M.~\& Begelman, M.~C.\ 2002, \apj, 581, 223

\bibitem[Sawicki et al.(1997)]{sawicki97} Sawicki, M. J., Lin, H., \&
Yee, H. K. C. 1997, AJ, 113, 1

\bibitem[Schechter(1976)]{schechter76}
Schechter, P. 1976, \apj, 203, 297

\bibitem[Schmidt(1959)]{schmidt59}
Schmidt, M. 1959, \apj, 129, 243

\bibitem[Seljak \& Zaldariaga(1996)]{seljak96} Seljak, U., Zaldariaga, M. 
1996, ApJ, 469, 437

\bibitem[\protect\astroncite{{Silk}}{1977}]{silk77} Silk, J.I. 1977
ApJ, 211, 638.

\bibitem[Somerville \& Primack(1999)]{somerville99} Somerville, R. S.,
\& Primack, J. R. 1999, MNRAS, 310, 1087

\bibitem[Spergel et al.(2003)]{spergel03} Spergel, D. N. et
al. 2003, \apjs, 148, 175

\bibitem[Springel \& Hernquist(2002)]{springel02}
Springel, V. \& Hernquist, L. 2002, \mnras, 333, 649

\bibitem[Springel \& Hernquist(2003a)]{springel03a}
Springel, V. \& Hernquist, L. 2003a, MNRAS 339, 289

\bibitem[Springel \& Hernquist(2003b)]{springel03b}
Springel, V. \& Hernquist, L. 2003b, MNRAS 339, 312

\bibitem[Steidel et al.(1999)]{steidel99} Steidel, C. C., Adelberger,
K. L., Giavalisco, M., Dickinson, M., Pettini, M. 1999, ApJ, 519, 1


\bibitem[Theuns et al.(2002)]{theuns02}
Theuns, T., Viel, M.,
Kay, S., Schaye, J., Carswell, R.~F., \& Tzanavaris, P.\ 2002, \apjl, 578,
L5

\bibitem[Thoul \& Weinberg(1996)]{thoul96} Thoul, A. A.,
Weinberg, D. H. 1996 ApJ, 465, 608

\bibitem[Tremonti et al.(2004)]{tremonti04}
Tremonti, C. et al.\ 2004, \apj, in press, astro-ph/0405537

\bibitem[van den Bosch(2001)]{vandenbosch01}
van den Bosch, F.~C.\ 2001, \mnras, 327, 1334

\bibitem[van den Bosch, Mo \& Yang(2003)]{vandenbosch03} van den Bosch, F. C.,
Mo, H. J., Yang, X. 2003, \mnras 345, 923

\bibitem[Weinberg et al.(2004)]{weinberg04}
Weinberg, D. H., Dav\'e, R., Katz, N., \& Hernquist, L. 2004, \apj, 601, 1

\bibitem[Weinberg, Hernquist, \& Katz(1997)]{weinberg97}
Weinberg, D. H., Hernquist, L., \& Katz, N. 1997, \apj, 477, 8

\bibitem[White \& Frenk(1991)]{white91} White, S. D. M., \& Frenk,
C. S. 1991, ApJ, 379, 52

\bibitem[White, Efstathiou, \& Frenk(1993)]{white93}
White, S. D. M., Efstathiou, G. P., \& Frenk, C. S. 1993, \mnras, 262, 1023

\bibitem[White \& Rees(1978)]{white78} White, S. D. M., \& Rees,
M. J. 1978, MNRAS, 183, 341

\bibitem[York et al.(2000)]{york00}
York, D. G. et al.\ 2000, \aj, 120, 1579 

\bibitem[Yoshida et al.(2002)]{yoshida02} Yoshida, N., Stoehr,
F., Springel, V. \& White, S.D.M. 2002, MNRAS, 335, 762.

\bibitem[Zaldarriaga, Seljak \& Bertschinger(1998)]{zaldarriaga98} 
Zaldarriaga, M., Seljak, U., Bertschinger, E. 1998, ApJ, 494, 491 

\bibitem[Zheng et al.(1997)]{zheng97} Zheng, W., Kriss, G. A., Telfer,
R. C., Grimes, J. P., \& Davidsen, A. F. 1997, ApJ, 475, 469

\end{thebibliography}
\end{document}